\documentclass[12pt]{article}

\usepackage{graphicx}

\begin{document}
        
\title{\bf PAIR PRODUCTION OF SCALAR TOP QUARKS IN POLARIZED 
         PHOTON-PHOTON COLLISIONS AT ILC}

\author{A.~Bartl$^{a,b}$, ~~ W.~Majerotto$^{c}$, \\
K.~M$\ddot o$nig$^{d}$, ~~ A.N.~Skachkova$^{e}$, 
~~  N.B.~Skachkov$^{e}$ }

\maketitle

\begin{center}
{\normalsize \it $^a$ University of Vienna,
 Faculty of Physics, 1090 Vienna,
 Boltzmanngasse 5, Austria. \\
 $^b$ AHEP Group, Instituto de Fisica
 Corpuscular - C.S.I.C.,  Universidad de Valencia, 
 Edificio Institutos de Investigacion,
 Apt. 22085, E-46071 Valencia, Spain \\
 $^c$ Institute for High Energy Physics 
 (HEPHY Vienna), Nikolsdorfergasse 18, 
 A-1050 Vienna, Austria.\\
 $^d$ DESY, Platanenallee 6,  
 D-15738 Zeuthen, Germany. \\
 $^e$ JINR, Joliot-Curie 6, 141980 Dubna,
 Moscow region, Russia. \\}
\end{center}

\date{}	 

\bigskip 
\begin{abstract}
\noindent 

 We study  pair production of scalar top quarks
 (stop, $\tilde t_{1}$) in polarized photon-photon collisions 
 with the subsequent decay of the top squarks into $b$-quarks and 
 charginos  $\tilde t_{1} \to b \tilde \chi_{1}^{\pm}$. 
We simulate this process  by using PYTHIA6.4 for an electron beam energy  $2E^{e}_{beam}=\sqrt {s_{ee}}=1000$ GeV. A set of criteria for physical variables is proposed  which leads to  a good separation of stop signal events from  top quark pair production 
being the main background.  These  criteria allow us 
 to reconstruct the mass of the top squark 
  provided  that the neutralino mass is known.

\end{abstract}


\section{Introduction.}

 ~~~~The scalar top quark, the bosonic
 partner of the top quark,  is expected to be the lightest colored
 supersymmetric (SUSY) \cite{SUSY} particle.
$\tilde t_{L}$ and $\tilde t_{R}$, the supersymmetric partners of the left-handed and right-handed  top quarks, mix and the
 resulting two mass eigenstates $\tilde t_{1}$ and
$\tilde t_{2}$, can  have a large mass splitting. It is even possible that the lighter eigenstate $\tilde t_{1}$ could be 
 lighter than the top quark itself \cite{JEllis}, \cite{STOP_SUSY}.

Searches for top squarks were performed at  LEP and Tevatron and   will continue  at  LHC and ILC  \cite{ILCRDR1}, \cite{ILCRDR2}.  
At  ILC it is planned to have the option of a photon collider (PLC),
 as originally planned for TESLA \cite{TESLA}.
 This will be achieved by using  backscattered
 photon beams by Compton scattering of laser photon
 beams with electron  beams   \cite{Ginzburg1} - \cite{Telnov95},
  (for  recent review  on this  subject see 
 \cite{F.Bechtel}).

It has been stressed that the polarization effects  in the interactions of backscattered laser photons \cite{Ginzburg1111}--\cite{Telnov95}
 provide additional opportunities for studying the properties of the produced   particles  (see also \cite{TESLA} and \cite{ILCRDR1},
  \cite{ILCRDR2}). In the following we study the reaction 
\begin{equation}
  \gamma + \gamma \to \tilde t_{1} + \bar{\tilde t_{1}} . 
\end{equation} 
 
  Among the possible  $\tilde t_{1}$-decay channels within
 the MSSM (see \cite{A. Bartl} for details),  
 we focus on the decay $\tilde t_{1} \to b \tilde \chi_{1}^{\pm}$
  followed by the two-body chargino decay 
 $\tilde \chi_{1}^{\pm} \to  \tilde \chi_{1}^{0} W^{\pm}$,
where one of the W's decays hadronically,  $W \to q_{i} {\bar q_{j}} $, and the other one leptonically, $W \to \mu\nu_{\mu}$ \cite{Paris2004}
 \footnote{ 
     The process $e^{+}e^{-} \to \tilde t_{1}  \bar{\tilde t_{1}}$
            with  the subsequent decay  channels
            $\tilde t \to  c \tilde \chi^{0}$ and 
	   $\tilde t_{1} \to b \tilde \chi_{1}^{\pm}$  were
	   considered in \cite{H.Nowak1} -\cite{H.Nowak3}
	   and  \cite{ee-hep2}, \cite{ee-hep1}, respectively.}.
  The final state of this signal process, shown in the left
plot of Fig.1, contains  two $b$-quarks and two  quarks (originating from thedecay of  one W boson),  a hard muon  plus a  neutrino 
(from the decay of the other W) and two neutralinos:
 \begin{equation}
  \gamma\gamma \to \tilde t_{1} \bar{\tilde t_{1}} \to
  b\bar{b}\tilde\chi^{+}_{1}\tilde\chi^{-}_{1} \to
   b \bar{b}W^{+}W^{-}\tilde\chi^{0}_{1}\tilde\chi^{0}_{1} \to 
   b\bar{b}q_{i}\bar{q_{j}}\mu\nu_{\mu}\tilde \chi^{0}_{1}\tilde\chi^{0}_{1}. 
 \end{equation}
 The main background process is top quark pair production 
 with the subsequent decay $t \to bW^{\pm}$ (for W's we 
 use the same decay channels as in the stop case): 
  \begin{equation}
  \gamma\gamma \to  t \bar{ t} \to  b \bar{b}W^{+}W^{-} \to 
   b\bar{b}q_{i}\bar{q_{j}}\mu\nu_{\mu}.
 \end{equation}
 The only difference between the final states of  stop and top 
 production (shown in the right diagram of Fig.1)  is that the
 stop pair production has two neutralinos which are undetectable.
 Thus,  both processes  have the same signature: two $b$-jets, two jets  from W decay and a  muon. In the following we 
 show that the physical variables constructed out of the final state
may us allow to  reconstruct  the scalar top quark mass. 
 In the present paper we consider only  top pair 
 production as background.
 
      \begin{figure}[!ht]
     \begin{center}
    \begin{tabular}{cc}
\mbox{a)\includegraphics[width=7.8cm, height=5.28cm]{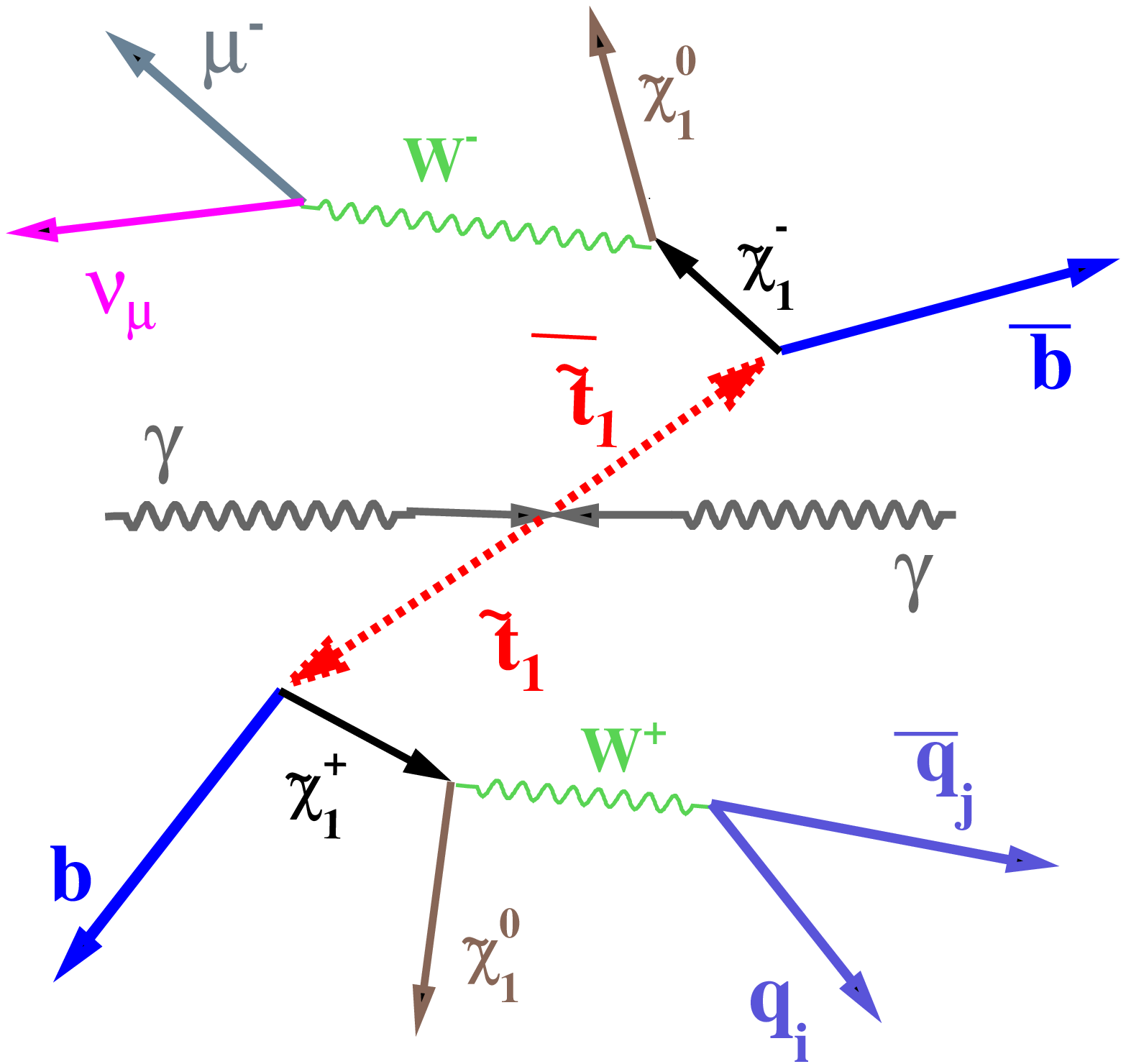}}
\mbox{b)\includegraphics[width=7.8cm, height=5.28cm]{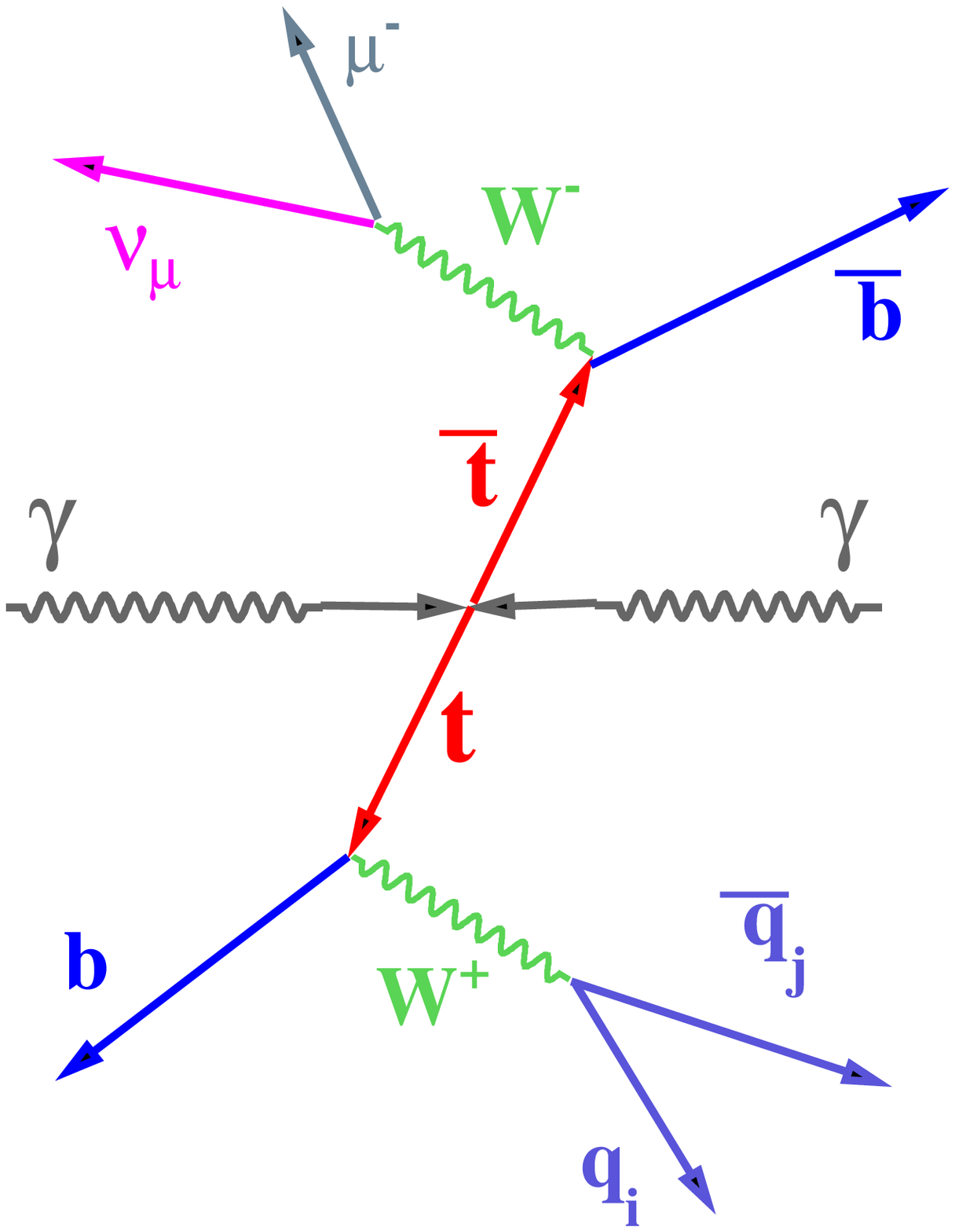}} \\   
    \end{tabular}
     \caption{\it \small Left is the stop signal event diagram, 
                         Right is the top background diagram. }    
     \end{center}        
     \end{figure}
     
     We analyse   the processes (2) and (3) 
    with the help  of Monte Carlo  samples of the
  corresponding events. Two  programs PYTHIA6.4 
  \cite{T. Sjostrand} and CIRCE2  \cite{T.Ohl} were 
  used. To simulate stop pair production process
  (1),  we used the  PYTHIA6.4 event generator in which 
  the  formula for the cross section
  of the stop pair production  in $e^{+}e^{-}$ 
  annihilation  was replaced by the formula for 
  two scalar particles (s) production
  $\gamma\gamma \to s \bar{s}$ from 
  \cite{Ginzburg88}, \cite{Ginzburg92}, 
 \cite{Ginzburg7}, (see \cite{S.Berge2}
  for the NLO corrections and  \cite{S.Berge1} for more
  details about  differential cross sections), 
  which takes into account various photon polarization
  states.  The top background was also simulated with 
  PYTHIA6.4. The program CIRCE2 was used to generate the
  momentum spectra of the backscattered photons involved
  in the process (1). The energy of the electron beams
  was chosen to be $E_{beam}^{e}$ = 500 GeV (i.e. the total
  $e^{-}e^{-}$ energy is $E_{e-e-}^{tot}=\sqrt {s_{ee}}=1000$ GeV).  

    In Section 2 we give the set of MSSM parameters used in 
  our study. 
 
    In Section 3 the important backscattered photon beam  
  characteristics, namely, momentum spectra and luminosity,
  are considered  for the case of  polarized 
  photon production in  Compton scattering of 
  polarized laser photons and  polarized electrons.  

     In Section 4 we discuss  some  general characteristics 
  of the signal process
  $\gamma \gamma \to \tilde t_{1} \tilde {\bar{t}}_{1}$
  and the main background  $\gamma \gamma \to t\bar{t}$.
  The subsections include kinematical distributions 
  for the produced stop quarks, for the jets 
  originating from W  boson decay and for  $b$-jets. We  compare them in  detail with those of  top pair production.  
   Subsection 4.2 also deals with the reconstruction of the 
  invariant mass of the two-quark system stemming from  
  the W boson decay as well as  with the reconstruction of 
  the invariant mass of the corresponding two-jet 
  system. 
 Subsection 4.3 contains the energy and transverse momentum spectra
 and some angular distributions of $b$-quarks and the corresponding 
 $b$-jets.
    In Subsection 4.4 we demonstrate how to discriminate between
  the signal muons produced in  W boson decays and
  those stemming from hadron decays in the same events.
  
  In Section 5 we show the distributions of the 
  global variables as missing energy, total 
  visible  (i.e. detectable) energy, the 
  scalar sum of  the transverse momenta of all
  visible particles in the event and the invariant
  mass of the  final-state hadronic jets plus
  the signal muon.  Two further global variables, 
  the invariant mass of  all final-state hadronic jets and the missing mass, are also introduced here. It is shown that  
  they are  very useful for the separation of 
  background top events.
  
 In Section 6 we  propose  three cuts which provide
  a good  signal-to-background  ratio (S/B).

  Section 7  is devoted to the  mass 
  reconstruction of the scalar top quark 
  based on the distribution of the  invariant mass  
  of one $b$-jet and  
  the other two $non-b$-jets  (from W decay), 
  provided that  the neutralino mass is known.
  
 In Section 8 we show   the distributions  of the invariant
 variables described in Section 7 for
  a stop mass $M_{\widetilde t_1} = 200$ GeV.
   
   Section 9 contains some  conclusions.

\section{ MSSM parameters and cross section.}
   The scalar top quark system is described by the mass matrix 
   (in the $\tilde t_{L}-\tilde {t}_{R}$ basis) \cite{JEllis},
   \cite{Gunion}
\begin{equation}    
   \left(\begin{array}{cc} M^2_{\tilde t_{LL}} & M^2_{\tilde t_{LR}} \\    
           M^2_{\tilde t_{RL}} & M^2_{\tilde t_{RR}} \end{array}\right) 
\end{equation}	   
with
\begin{equation}  
 M^2_{\tilde t_{LL}} = M^2_{\tilde Q} + (\frac{1}{2} - \frac{2}{3} sin^2 \Theta_W) cos2 \beta M^2_Z + M^2_t, 
\end{equation}
\begin{equation} 
M^2_{\tilde t_{RR}} = M^2_{\tilde U} +  \frac{2}{3} 
sin^2 \Theta_W cos2 \beta M^2_Z + M^2_t, 
\end{equation}
\begin{equation}  
M^2_{\tilde t_{RL}} = (M^2_{\tilde t_{LR}})^* =  M_t(A_t - \mu ^* cot \beta). 
\end{equation}
   The mass eigenvalues are given by
\begin{equation} 
 M^2_{\tilde t_{1,2}} = \frac{1}{2}\left[ (M^2_{\tilde t_{LL}} + 
  M^2_{\tilde t_{RR}}) \mp \sqrt{(M^2_{\tilde t_{LL}} + 
  M^2_{\tilde t_{RR}}) + 4 |M^2_{\tilde t_{LR}}|} \right]
\end{equation}
with the mixing angle
\begin{equation}
cos \Theta_{\tilde t} = \frac {-M^2_{\tilde t_{LR}}} {\sqrt {
|M^2_{\tilde t_{LR}}|^2 + (M^2_{\tilde t_{1}} -  M^2_{\tilde t_{LL}} )^2}}
\end{equation}
\begin{equation}
sin \Theta_{\tilde t} = \frac {M^2_{\tilde t_{LL}} - M^2_{\tilde t_{1}} } {\sqrt {
|M^2_{\tilde t_{LR}}|^2 + (M^2_{\tilde t_{1}} -  M^2_{\tilde t_{LL}} )^2}}
\end{equation}

   In the following we shall consider only one particular 
  choice of the MSSM parameters that are defined, in the 
  notations of PYTHIA6,  in the following  way: 
 
 \noindent 
 $\bullet$ $ M_{\widetilde{Q}}   = 270$ GeV; \\
  $\bullet$ $ M_{\widetilde{U}} = 270$ GeV;  \\
  $\bullet$ $ A_t = -500$ GeV (top  trilinear coupling); \\
  $\bullet$ $ tan \beta = 5$;\\
 $\bullet$ $ \mu =  -370 $ GeV;  \\
 $\bullet$ $ M_{1} = 80 $ GeV; \\
 $\bullet$ $ M_{2} = 160 $ GeV.\\ 
 
  Note that in PYTHIA6 $M_{\widetilde{Q}} $ corresponds 
  to $ M_{\widetilde{t}_L}$ (left squark mass for the
  third generation) and  $M_{\widetilde{U}}$ 
  corresponds to $M_{\widetilde{t}_R}$. These
  parameters give $M_{\widetilde{t}_1}=167.9$ GeV, 
  $M_{\chi^{+}_{1}}=159.2$ GeV 
  and  $M_{\chi^{0}_{1}}=80.9$ GeV. This parameter
  point is compatible with all experimental
  data. We have chosen this value of $ M_{\widetilde{t}_1}$ to be rather close to the mass of the top quark 
 $M_{t}=170.9 \pm 1.8$  GeV  
  \cite{Schieferdecker}.
  Therefore, one expects a rather  large  contribution 
  from the top background, which means that the choice 
  of this value of the stop mass makes the analysis 
  most difficult. Finding a suitable set of 
  cuts separating stop and top events  is therefore crucial.
  
\section{ Photon beam characteristics.}
~~~~~Let us mention two main features of photon-photon collisions. 
The first one is that the   monochromaticity of the backscattered
photon beam is considerably increased if the mean helicities $\lambda_{e}$ and $P_{c}$ of the electron beam and the laser photon beam are chosen such that $2\lambda_{e}P_{c} \approx -1$, as has been shown  in \cite{Ginzburg1111}-\cite{Ginzburg111111}
    \footnote{A laser beam polarization of $100\%$ 
              can be assumed. An electron polarization of 
	      $85\%$ is expected at the ILC.}.
In this case the relative number   of hard  photons becomes nearly twice as large  in  the region of the photon   energy fractions 
  $y_i=E_i^{\gamma}/E^{e}_{beam}\approx 0.7-0.85,$ i = 1, 2,
where  $E_{1,2}^{\gamma}$ are the energies of the two backscattered photon beams. Thereby the luminosity in collisions of
  these photons increases by a factor of 3-4.  The 
  growth of backscatterd photon energy spectra in the 
  region of large $y_i$ with the increase of 
  ($-2\lambda_{e}P_{c}$) is illustrated in Fig.3 of
  \cite{Ginzburg1111} and in Fig.2 of  \cite{Ginzburg111111}.
  In other words, when ($-\lambda_{e}P_{c}$) increases,
  the effective "pumping" of soft laser photons into
  hard backscattered ones increases due to the
  Compton process.
  The analogous growth  of spectral luminosity
   $dL_{\gamma\gamma}/dW$ (W is the invariant mass of 
   $\gamma\gamma$ system) in the case  when  the 
   polarizations in both incoming 
   systems of beam electron and the laser photon
   satisfy the relation
   $2\lambda_{1e}P_{1c}=2\lambda_{2e}P_{2c}$ 
   is demonstrated in Figs.4 of 
   \cite{Ginzburg1111}  and \cite{Ginzburg111111}.
  As it was mentioned in \cite{Ginzburg1111},
  at $2\lambda_{e}P_{c} \approx -1$ the photons with the 
  maximal energy ($y_{i} \approx 0.7-0.85$) are circular 
  polarized and their helicity  is close to ($-P_{c}$).
  Thus, in the limit 
   $2\lambda_{1e}P_{1c}=2\lambda_{2e}P_{2c}=-1$,
   the produced pair of most energetic photons
   have total angular momentum
J=0 or J=2, depending on the signs of 
   $P_{1c}$ and $P_{2c}$.  This allows one to
   measure the cross sections $\sigma_{0}$ and
    $\sigma_{2}$ which correspond to collisions
    of  $\gamma  \gamma$-pairs having total angular momentum
    0 or 2, respectively.
   
   The other feature stems  from the fact that unlike
  the situation at an electron-positron collider, 
  the energy of the beams of backscattered photons  
  will vary  from  event to event. As already mentioned
  in the Introduction, we use the program CIRCE2 
  \cite{T.Ohl} for
 the energy  spectra of the 
  colliding backscattered photons, as well as the values
  of photon beam luminosities. CIRCE2 uses as input the
  data files  that were generated for TESLA using the 
  code 
  and the set of beam parameters described 
  in \cite{TelnovDatFils}, \cite{Telnov95} and \cite{TESLA}    
  \footnote{The spectra obtained by CIRCE2 
	    are in agreement \cite{Takahashi} 
	    with the code CAIN \cite{CAIN}.}.
   We use as a reasonable approximation
  the CIRCE2 output spectra obtained on the basis
  of the above mentioned data files that were originally
  generated  for  $E_{e-e-}^{tot}=800$ GeV 
  and  scale them (by 1000/800) to the higher beam  energy 
  $2E^{e}_{beam}=E_{e^{-}e^{-}}^{tot}=1000$ GeV.

  The photon energy spectrum obtained in this way 
  without of any cuts  with CIRCE2  for this total energy
   $E_{e^{-}e^{-}}^{tot}=1000$ GeV is shown in Fig.2 
              \footnote{Examples of energy, photon polarization 
	      and   $\gamma\gamma$ luminosity spectra,
	     obtained  for a set of  different values of total 
	      energy $E_{e-e-}^{tot}$, can be  seen in
	       \cite{Ginzburg111}-\cite{Telnov95}, 
              \cite{MoenigKlamke} and  \cite{TESLA}.}.
    Two peaks are clearly seen in this figure.
  The left one at a low photon energy is caused by multiple 
  Compton scattering  and beamstrahlung photons
  \cite{Telnov90},  \cite{Telnov95},  \cite{TelnovDatFils}
  and  \cite{TESLA}.   The second one, according to
   \cite{Ginzburg111}-\cite{Ginzburg111111}, 
  appears in the region  of  hard 
  photon production $y_{1,2} \approx 0.83$.
  It shows the degree of  monochromaticity of
  the produced backscattered high--energy photons.
  
  \begin{figure}[!ht]
     \begin{center}
\vskip -0.5cm   
    \begin{tabular}{ccc}
\mbox{\includegraphics[width=7.2cm, height=5.2cm]{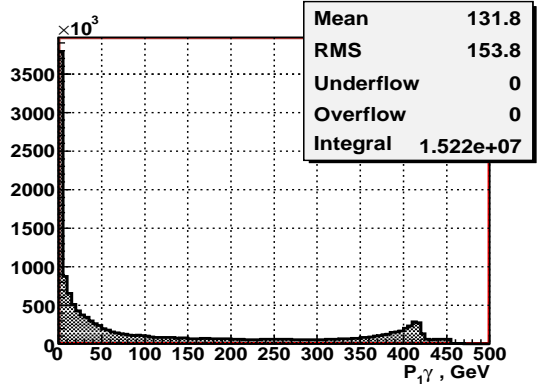}} 
    \end{tabular}
     \caption{\small \it The whole photon  momentum spectrum generated  by  CIRCE2.}
 \vskip -0.5cm     
     \end{center}        
     \end{figure} 
 
   The energy spectra  of backscattered 
  photons, as  provided  by CIRCE2, are used  as 
  input for PYTHIA for the  generation of 
  stop  pair production events. Due to the  stop 
  pair mass threshold $2M_{\widetilde{t}_1}$, only in
  about $0.3\%$  of the CIRCE2 events 
  the  energy of produced backscattered   $\gamma  \gamma$-system
  is high enough  for the  generation  of
  $\gamma \gamma \to \tilde t \tilde {\bar t}$ 
  signal events by PYTHIA. 

  \begin{figure}[!ht]
  \begin{center}
\vskip  -0.5cm     
\begin{tabular}{cc}
\mbox{{\bf a)}\includegraphics[width=7.2cm, height=5.2cm]{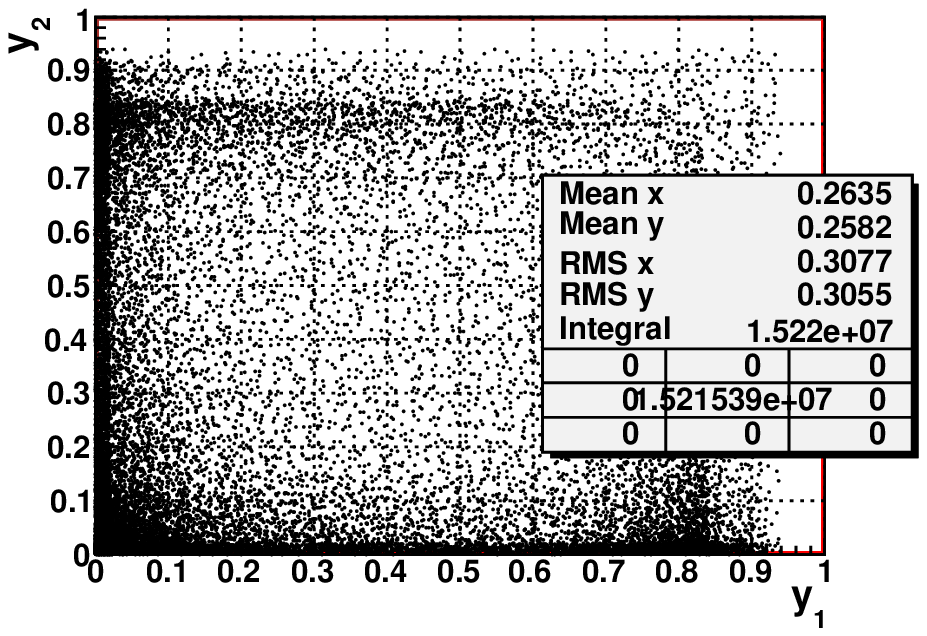}} 
\mbox{{\bf b)}\includegraphics[width=7.2cm, height=5.2cm]{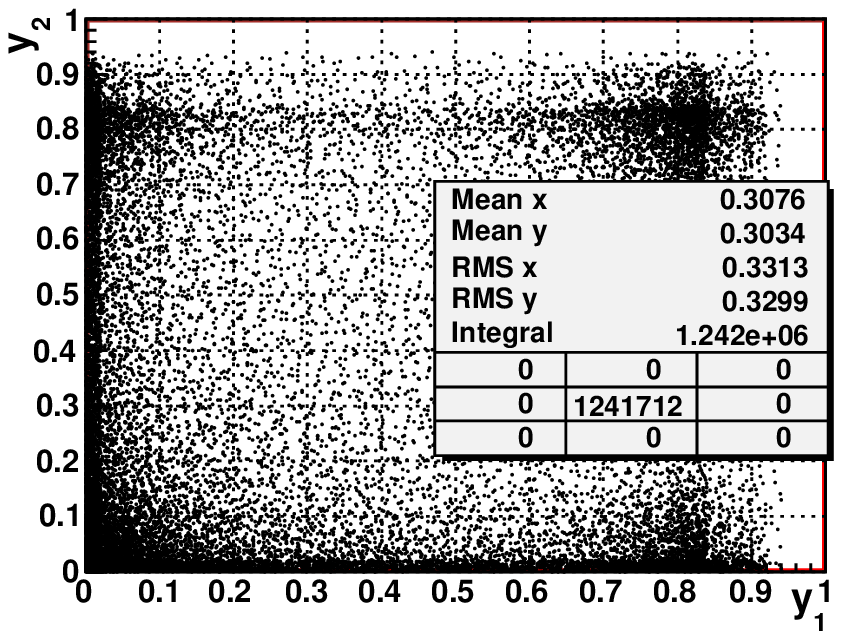}} \\
\mbox{{\bf c)}\includegraphics[width=7.2cm, height=5.2cm]{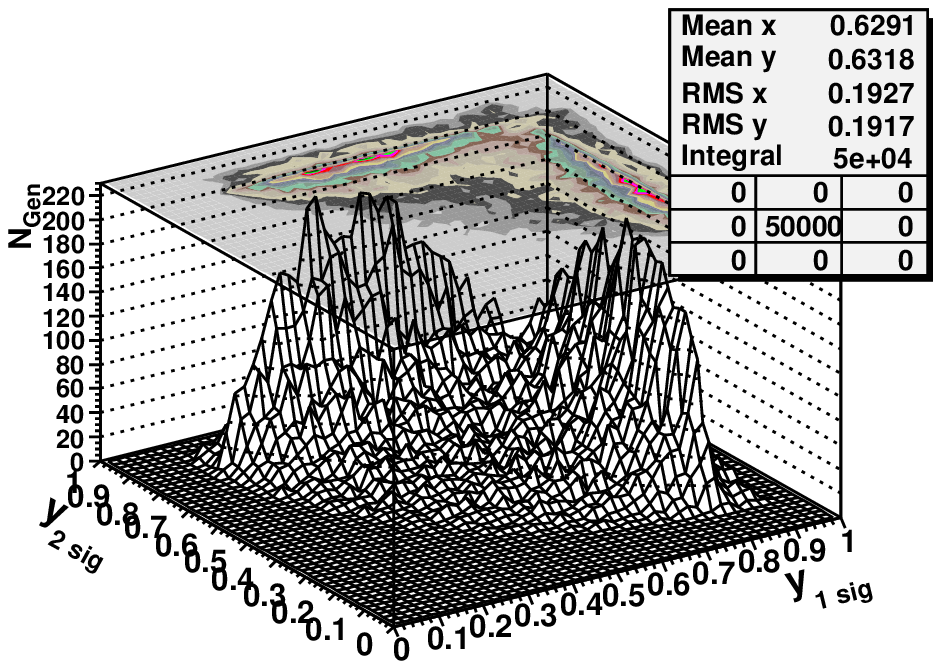}} 
\mbox{{\bf d)}\includegraphics[width=7.2cm, height=5.2cm]{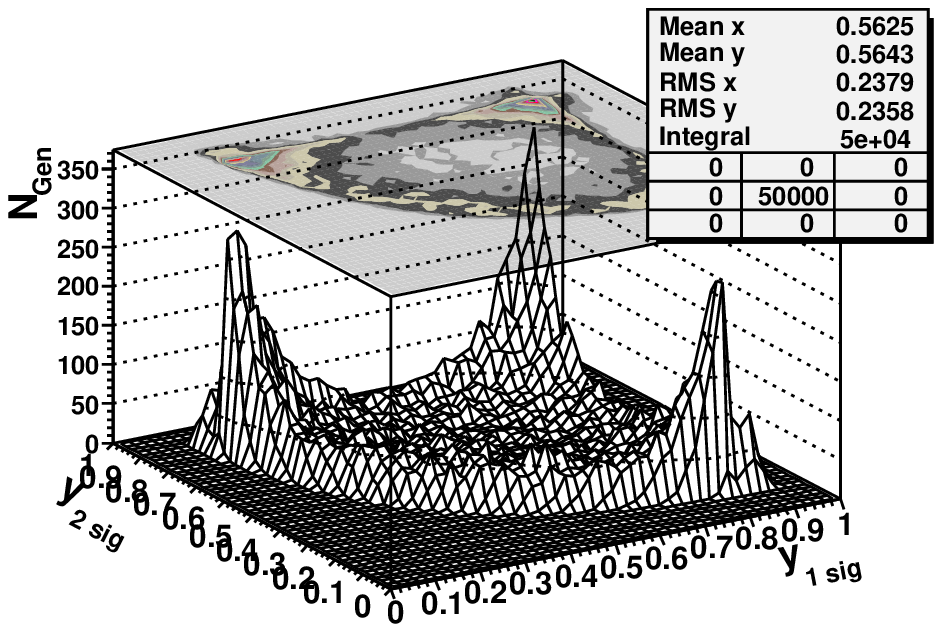}}\\   
\end{tabular}
   \caption{\small \it 
{\bf a)} correlation spectra of the energy fractions $y_1$ and $y_2$  for  events generated by
	    CIRCE2 for the case of the opposite sign
	    polarizations of backscattered photons  
	     ("J=2 case"). 
	    {\bf c)} correlation spectrum for that  part
	    of the events shown in plot {\bf a)} in which 
	    the energy of $\gamma  \gamma$-system
	    is above the threshold of stop pair  production. 
	   {\bf b)} and {\bf d)} are the same 
	    correlation spectra as {\bf a)} and {\bf c)} 
	    spectra but obtained for  the case of the same sign
	    polarizations of backscattered photons 
           ("J=0 case"). 
            $E_{e^{-}e^{-}}^{tot}=1000$ GeV. }    
\vskip -0.5cm    
  \end{center}        
  \end{figure}
   The correlations between the energies of two colliding
  photons given by CIRCE2 are shown in the plots 
  {\bf a)} and  {\bf b)} of Fig.3. 
  
   The two-dimensional plot {\bf a)} in Fig.3 
   shows  the  correlation between the energy
   fractions of produced photons
   $y_1$ and $y_2$ for the case
   that the two colliding backscattered photons have
   opposite  sign helicities,  i.e. when the total
   helicity of $\gamma  \gamma$-system  $J=2$
    \footnote{See Fig.2 of \cite{MoenigKlamke} as 
            an illustration.}.
   Plot {\bf b)} is 
  for $J=0$ i.e., it is for
the case that  the two  colliding
  backscattered photons  have the  same sign
  helicities  $^{5}$. 
  The  distribution in plot {\bf b)} shows maxima at
  $y_{1, 2} \approx 0.83$, which  corresponds to the 
  high-energy peak in Fig.2  and
   $2\lambda_{e}P_{c} \approx -1$. 
  The number of generated events in the cases {\bf a)} 
  and {\bf b)} are shown by the "Integral" 
  values in the statistic frames of both plots.
  They were chosen such to produce equal number
  (50000) of events at the PYTHIA level of simulation
  (see "Integral" values in the plots {\bf c)} and {\bf d)})
  of the two different  samples of signal stop production 
  events having  different polarization states of the
  incoming $\gamma\gamma$ pairs.  
   
  The lower two plots {\bf c)} and {\bf d)} of Fig.3 are 3-dimensional 
  plots with their  projections onto the 
  $y_1 - y_2$ plane. They also show the correlations  between the
  energy  fractions  $y_1$ and $y_2$ of the backscattered
  photons. In these plots we include only  those events that 
  lead to the production of a stop-antistop $ \tilde t \bar {\tilde t}$
  pair.  The left side of Fig.3 shows the plots for the  opposite sign
  polarization case  (i.e., $J=2$) and
  the right side  for the  same sign
  polarization case (i.e., $J=0$). Plots {\bf b)} and {\bf d)}
  show the enhancement of the  $J=0$ state
  contribution at $y_{1, 2} \approx 0.83$.
 
     It is worth mentioning that in a real photon-photon 
   collision experiment none of these cases would appear in 
   a pure form because  of  the unavoidable presence of 
   some admixture of other photon polarization states
   \footnote{Partly this is due to the fact that the source 
             electron beams are not $100\%$ polarized.}.
 
    \begin{figure}[!ht]
     \begin{center}
    \begin{tabular}{cc}
     \mbox{{\bf a)}\includegraphics[width=7.2cm, height=5.2cm]{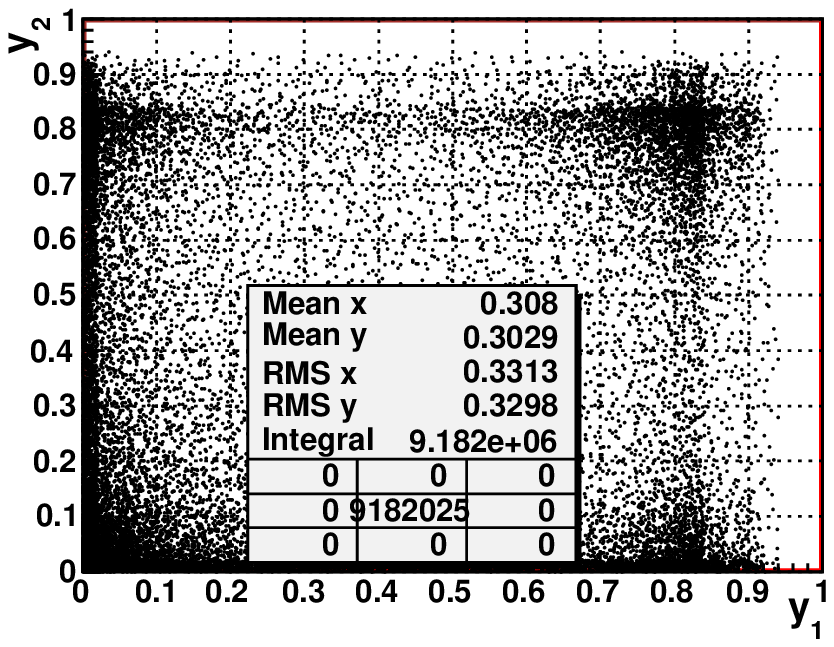}} 
     \mbox{{\bf b)}\includegraphics[width=7.2cm, height=5.2cm]{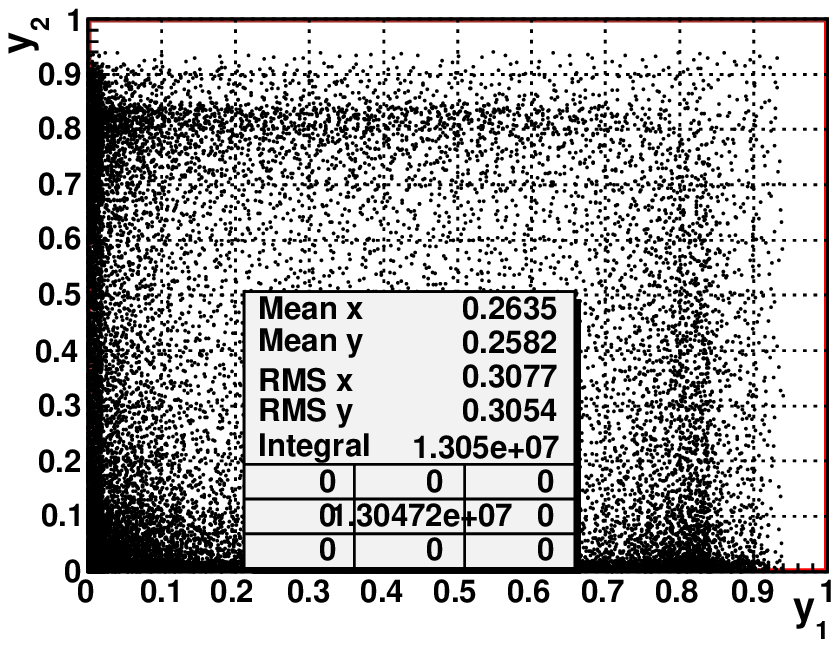}} \\
     \mbox{{\bf c)}\includegraphics[width=7.2cm, height=5.2cm]{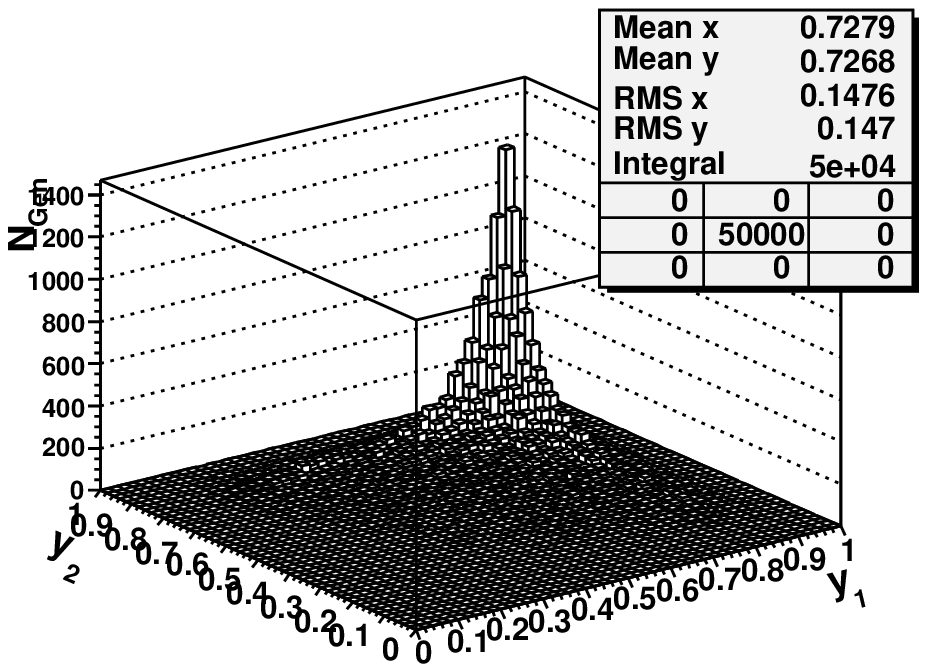}} 
     \mbox{{\bf d)}\includegraphics[width=7.2cm, height=5.2cm]{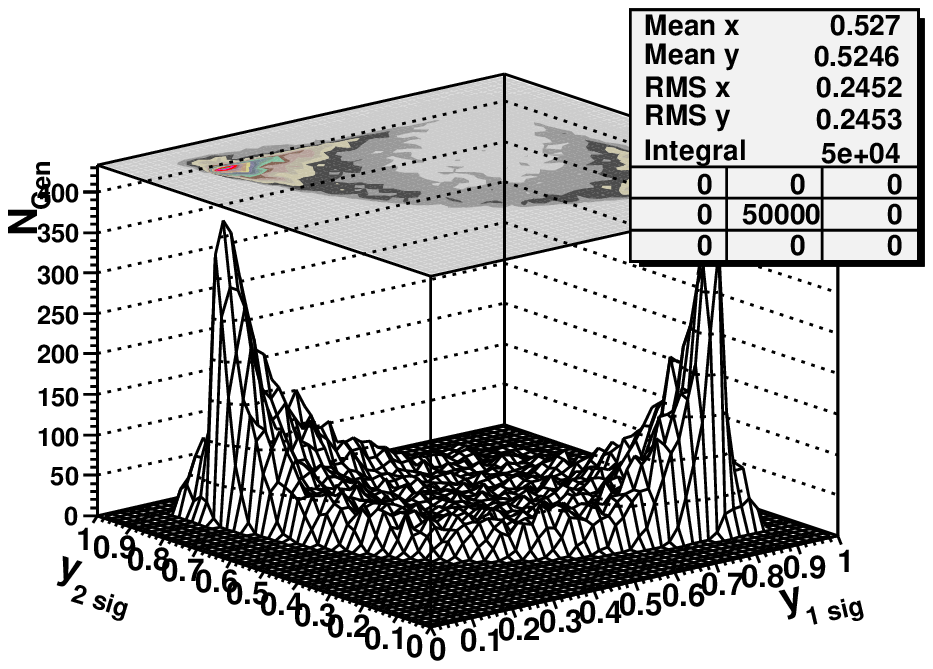}} \\   
    \end{tabular}
  \caption{\small \it The same as in Fig.3 but with the  $J=2$ state
   contribution enhanced.} 
      \end{center}        
     \end{figure}

   The simultaneous change of the signs of the laser 
   photon  and beam electron  helicities at only one 
   side  of the colliding  beams 
  does not change the equality $2\lambda_{1e}P_{1c}=2\lambda_{2e}P_{2c}$
   \cite{Ginzburg111}-\cite{Ginzburg111111}, 
   but leads  to a different  beam configuration,
   which may influence the shape of the luminosity spectrum.	    
       In Fig.4 we present the correlation plots that are
   analogous to those of Fig.3 but this time they  are
   for the case of the above mentioned  simultaneous 
   sign reversal  of the  laser photon and electron beam  
   polarizations at one side  ($i=2$, for
   example) of the colliding beams. 
   It is seen from the plots {\bf a)} and {\bf c)} that 
   this combination gives an increase of  the contribution 
   of the two-photon system of total angular momentum $J=2$.
 
  Finally we give the  values of 
  total photon-photon luminosities and the corresponding
  values of stop pair production cross sections (for the
  chosen value of the stop mass) obtained from  CIRCE2  
  and PYTHIA6 for $E_{e-e-}^{tot}=1000$ GeV
  for the opposite sign (${+- ~\& ~-+}$) and the 
  same sign (${++ ~\& ~--}$) backscattered photon helicities
  \footnote{ For simplicity in the following we shall use the 
                notation  $ "++"$   and $  "--"$ 
		for the same sign  photon helicities
		case and  $"+-"$  and $  "-+"$ for
		the opposite sign helicities case. }: \\

$\bullet$ for the plots shown in Fig.3 (i.e., enhanced $J=0$ state)
\begin{center}
  $L^{\gamma\gamma}_{+- ~\& ~-+}$ = 
      9.35 * $10^2$ $fb^{-1}year^{-1}$ ; 
~~~~~ $\sigma^{\tilde t_{1} \bar{\tilde t_{1}}} = 2.91 $ fb; \\
  $L^{\gamma\gamma}_{++ ~\& ~--}$ = 
      1.02 * $10^3$ $fb^{-1}year^{-1}$ ; 
~~~~~ $\sigma^{\tilde t_{1} \bar{\tilde t_{1}}} = 4.96 $ fb.\\
\end{center}

$\bullet$ for the plots shown in Fig.4 (i.e., enhanced $J=2$ state).
\begin{center}
 $ L^{\gamma\gamma}_{+- ~\& ~-+}$ = 
      1.02 * $10^3$ $fb^{-1}year^{-1}$ ;  
~~~~~~  $\sigma^{\tilde t_{1} \bar{\tilde t_{1}}} = 6.13 $ fb;\\ 
 $ L^{\gamma\gamma}_{++ ~\& ~--}$ = 
      9.35 * $10^2$ $fb^{-1}year^{-1}$ ;   
~~~~~~ $\sigma^{\tilde t_{1} \bar{\tilde t_{1}}} = 4.70 $ fb.

\end{center}

\section{Distributions of kinematical variables in 
         stop and top production.}

 ~~~~  In this Section we present various 
  plots for the kinematical distributions  of different  physical
  variables based on  two samples
  of $2.5 \cdot 10^{4}$ stop pair production events
  generated by CIRCE2 and PYTHIA6.4. They were weighted by the 
  photon-photon luminosity calculated with the help of 
  CIRCE2 and given above for the corresponding
  polarizations. Analogous plots are also given for
  $1.0 \cdot 10^{5}$  generated  background top events. 
   
 The generation of   all events, i.e. for the stop and top production,
  was done separately for both possible combinations
  of  photon polarizations,  i.e. for the same sign 
  ( $ "++"$   and $  "--"$) and for the opposite sign
   ($"+-"$  and $  "-+"$) helicities. 
   
  In the following we present only those plots which
  correspond to the case  where the relative 
  alignment of laser photon and beam electron
  helicities enhances the contribution of the colliding
  two-photon system with the total angular momentum 
  $J=0$ (i.e., corresponding to Fig.3). 
  \footnote{The case of $J=2$ is easier for background
            suppression due to spin $1/2$ of the top quark.}  
  
  To find the jets we  use  the subroutine PYCLUS of PYTHIA. 
    The parameters of this jet finder  are 
  chosen such that the number of jets is exactly    four.  
    

\subsection{ Distributions in stop events.}
~~~~~ Figures  5--8 show some  general  kinematical
  distributions characteristic of the  produced 
  stop pair system, i.e., the  distributions 
  of the energy of the stop  or antistop   $E_{\widetilde{t}_1}$, 
   the  transverse  momentum   $PT_{\widetilde{t}_1}$,  the  polar angle
  $\theta_{\widetilde{t}_1} $  (all in $e^{-}e^{-}$ c.m.s.)
  and the invariant mass of the produced stop pair
  $M_{inv}(\widetilde{t}_1 +\bar{\widetilde{t}}_1) $. 
  In these plots we do not distinguish between 
  stop and antistop. By comparing the left  hand
  side of these figures with the right side 
  one sees the effects of the different chosen 
  polarizations  (and corresponding
  luminosities).

   In Fig.5 one can see that the stop energy 
   $E_{\widetilde{t}_1}$  spectra start close to the value of the stop mass
   $M_{\widetilde{t}_1}=167.9$ GeV. In the case of opposite sign photon
   polarizations (plot {\bf a)}) the spectrum   has a peak at 
    $E_{\widetilde{t}_1}  \approx 320$  GeV
   and it is characterized by a high mean value  
   $E_{\widetilde{t}_{1}}^{mean} =  311$  GeV.
   It means that   the  produced stops are  rather energetic. 
   In the case of  the same sign polarizations
   (plot {\bf b)})  the energy spectrum is softer, 
   having the main peak at  
   $E_{\widetilde{t}_1}  \approx 200$  GeV
   and the mean value about 274 GeV. 
   So, one may expect  that the  stops  
   produced in the  same sign case 
   are on the average less energetic  
   than in the  opposite sign case. 
   One can also see a  second smaller peak at 
    $E_{\widetilde{t}_1}  \approx 400$.
   This is due to   the effect of the luminosity and cross 
   section enhancement in the $J=0$ case at  
   $y_{1} \approx y_{2} \approx 0.83$ (see 
   the right-hand plots of Fig.3).
   
    \begin{figure}[!ht]
    \begin{center}
    \begin{tabular}{cc}       
     \mbox{{\bf a)}\includegraphics[width=7.2cm, height=4.4cm]{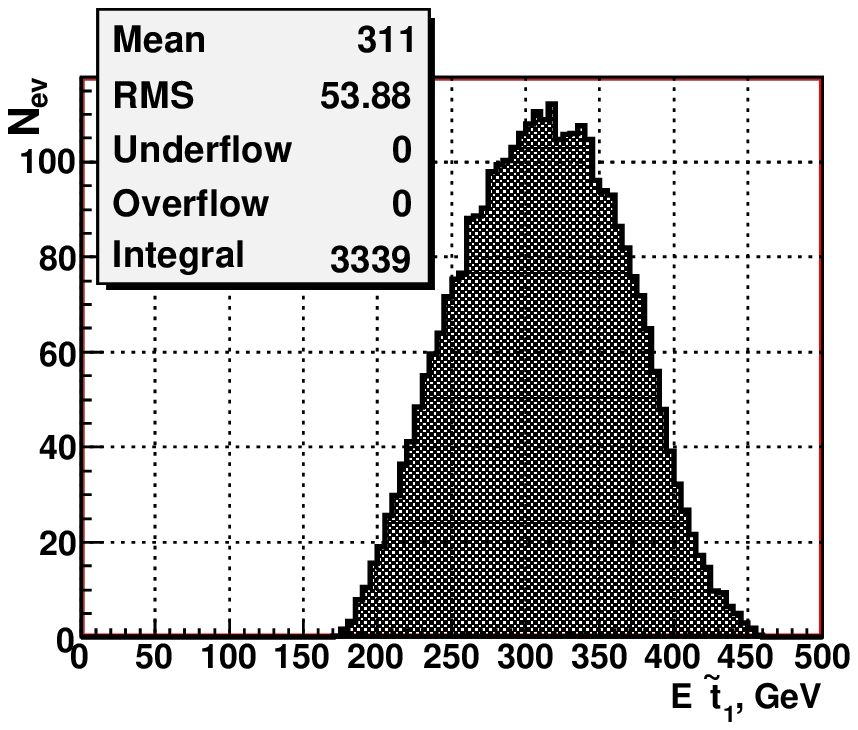}}      
     \mbox{{\bf b)}\includegraphics[width=7.2cm, height=4.4cm]{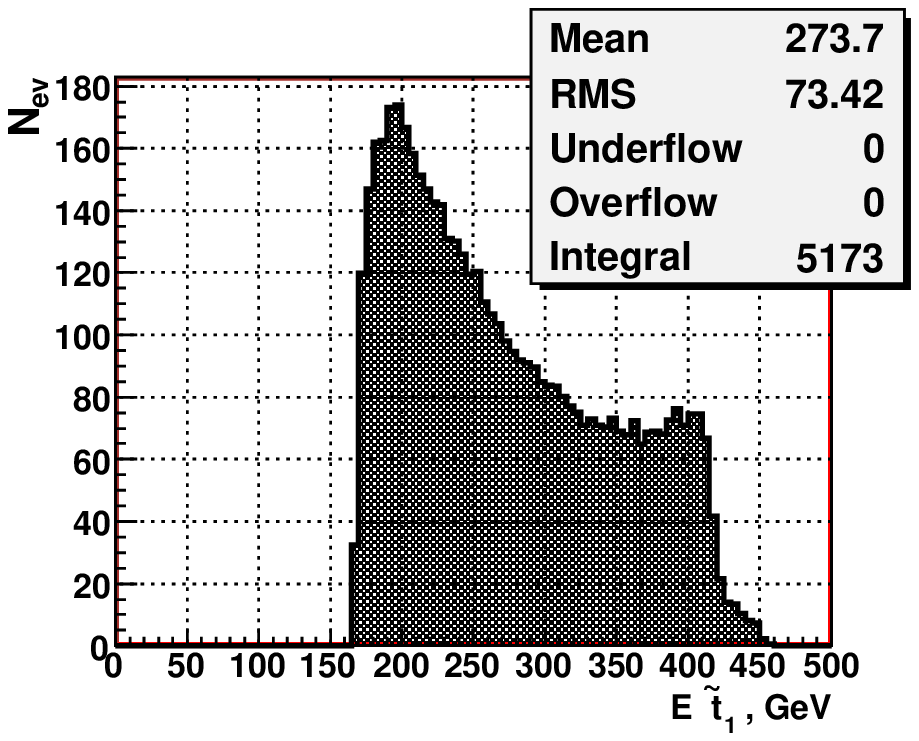}} \\
   \end{tabular}
 \caption{\small \it  Stop energy 
       $E_{\widetilde{t}_1}$ spectra. 
              {\bf a)}  $"+-"$  and $"-+"$ polarizations,
              {\bf b)}  $"++"$ and $"--"$ polarizations.}     
       \end{center} 
\vskip -0.5cm                
     \end{figure}  
     
     Figure 6 shows analogous distributions 
   for the stop transverse momentum    $PT_{\widetilde{t}_1} $.
   The $PT_{\widetilde{t}_1}$ spectrum for 
    the same sign polarizations (plot {\bf b)}) is much
   softer than for  the opposite sign polarizations
   (plot {\bf a)}), with mean values of 111 GeV
   and 214 GeV, respectively.
   
   \begin{figure}[!ht]
 \begin{center}
    \begin{tabular}{cc}  
     \mbox{{\bf a)}\includegraphics[width=7.2cm, height=4.4cm]{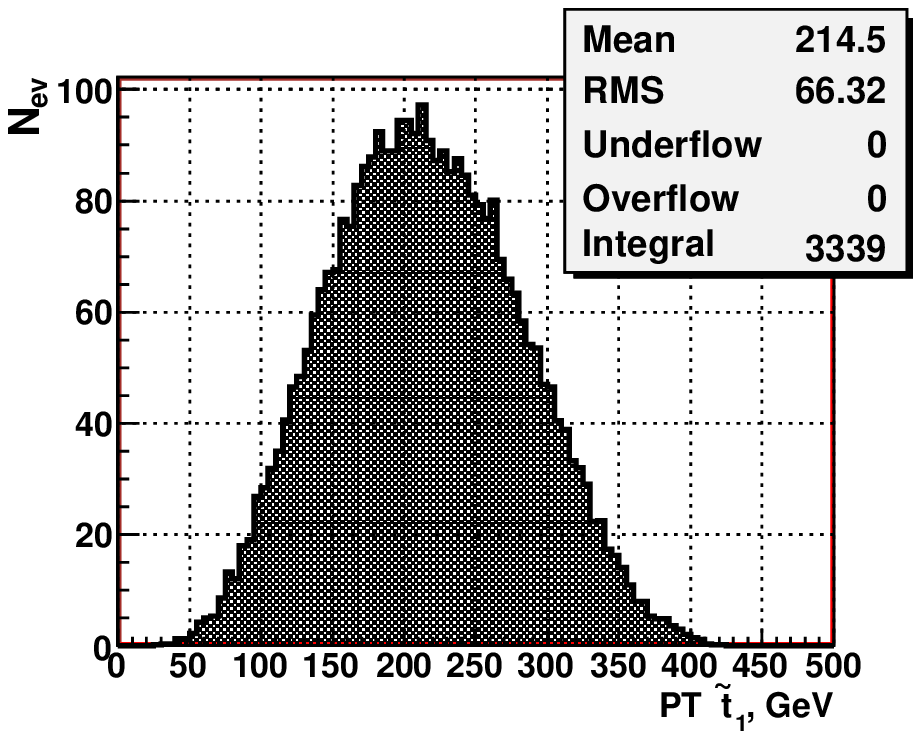}}      
     \mbox{{\bf b)}\includegraphics[width=7.2cm, height=4.4cm]{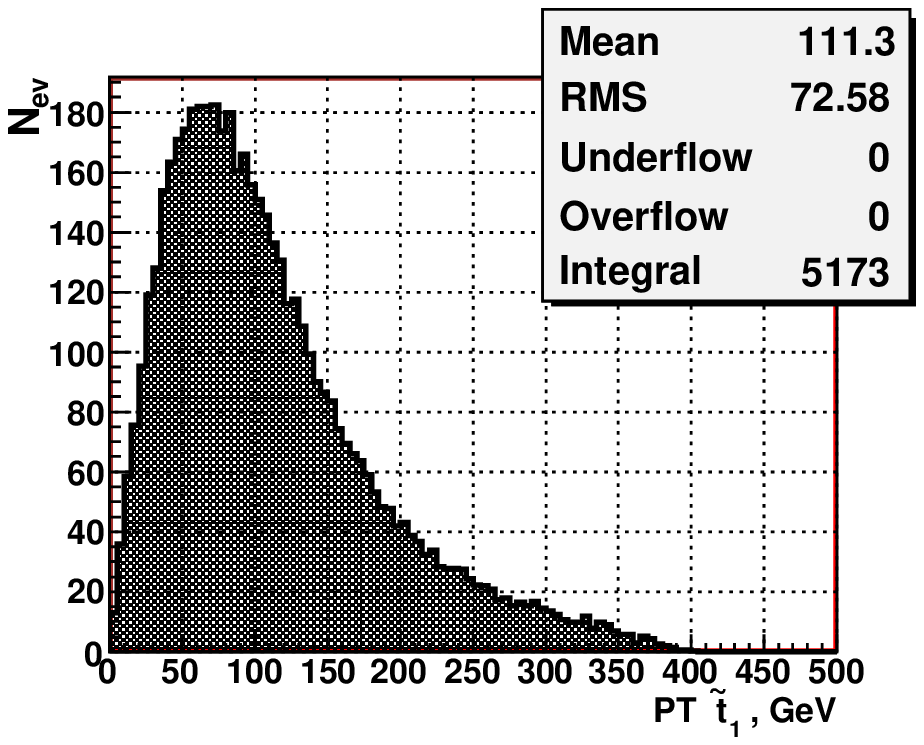}} \\
    \end{tabular}
     \caption{\small \it Stop transverse momentum 
      $PT_{\widetilde{t}_1} $ spectra. 
              {\bf a)} $"+-"$  and $"-+"$ polarizations,
              {\bf b)} $"++"$ and $"--"$  polarizations.}
     \end{center} 
\vskip -0.5cm             
     \end{figure}

   The  polar angle $\theta_{\widetilde{t}_1}$  distributions are 
   shown in  Fig.7.  One can  see that the 
   distribution for   $"+-"$  and $"-+"$ 
   polarizations (plot {\bf a)}) is very different
   from that for   $"++"$  and $"--"$ 
   polarizations (plot {\bf b)}).
        
    \begin{figure}[!ht]
     \begin{center}
\vskip -0.5cm     
    \begin{tabular}{cc}
     \mbox{{\bf a)}\includegraphics[width=7.2cm, height=4.4cm]{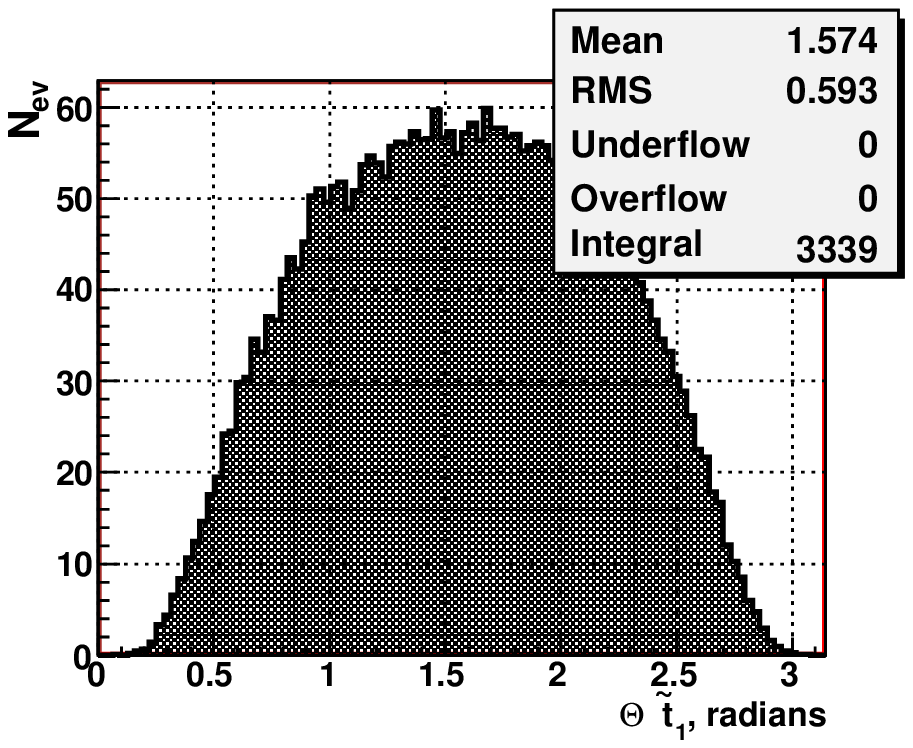}} 
     \mbox{{\bf b)}\includegraphics[width=7.2cm, height=4.4cm]{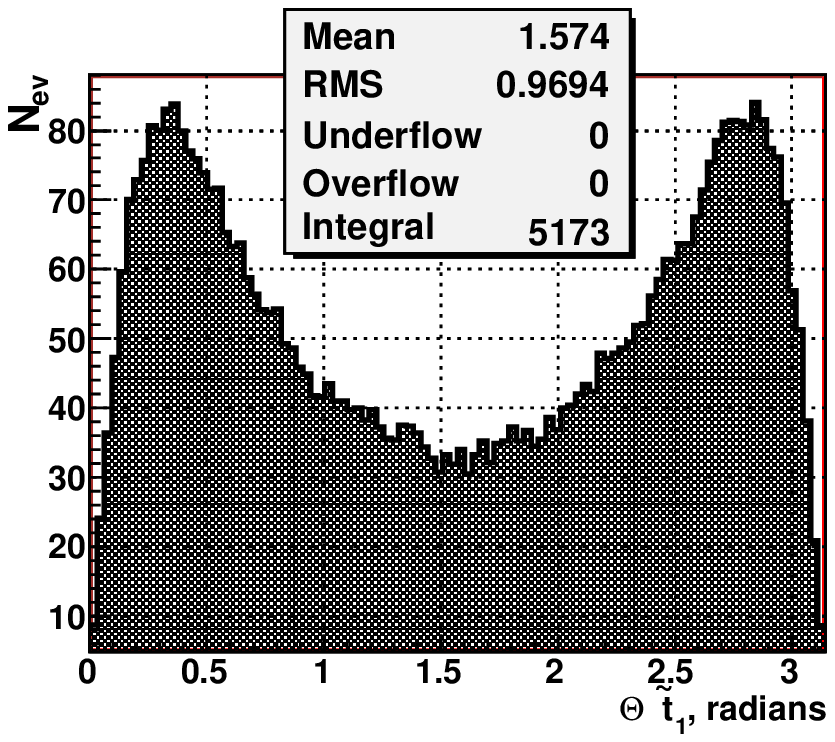}} \\
    \end{tabular}
     \caption{\small \it  Stop polar angle   
       $\theta_{\widetilde{t}_1}$ spectra.
              {\bf a)} $"+-"$  and $"-+"$ polarizations,
              {\bf b)} $"++"$ and $"--"$  polarizations.}
      \end{center}   
\vskip -0.5cm            
     \end{figure}  
        
   The  invariant mass
    $M_{inv}(\widetilde{t}_1 +\bar{\widetilde{t}}_1)$
   spectra of the produced 
   stop-antistop system are shown in Fig.8.
   For   $"+-"$  and $"-+"$ polarizations (plot {\bf a)}) 
   it has a peak around 550 GeV, which is about 
   170 GeV higher than  the analogous peak at 380 
   GeV for $"++"$  and $"--"$ polarizations (plot {\bf b)}). 
   Note that the shapes of the distributions of 
   the invariant mass of the stop pairs  shown
   in Fig.8  follow the energy spectra given 
   in Fig.5. Thus, the second peak in plot {\bf b)} of    Fig.8 at 
    $M_{inv}(\widetilde{t}_1 +\bar{\widetilde{t}}_1) 
   \approx 800$ GeV has the same origin as the peak 
   in the plot {\bf b)} of Fig.5 at
    $E_{\widetilde{t}_1} \approx 400$ GeV.   
         
     \begin{figure}[!ht]
 \begin{center}
    \begin{tabular}{cc}
    \mbox{{\bf a)}\includegraphics[width=7.2cm, height=4.4cm]{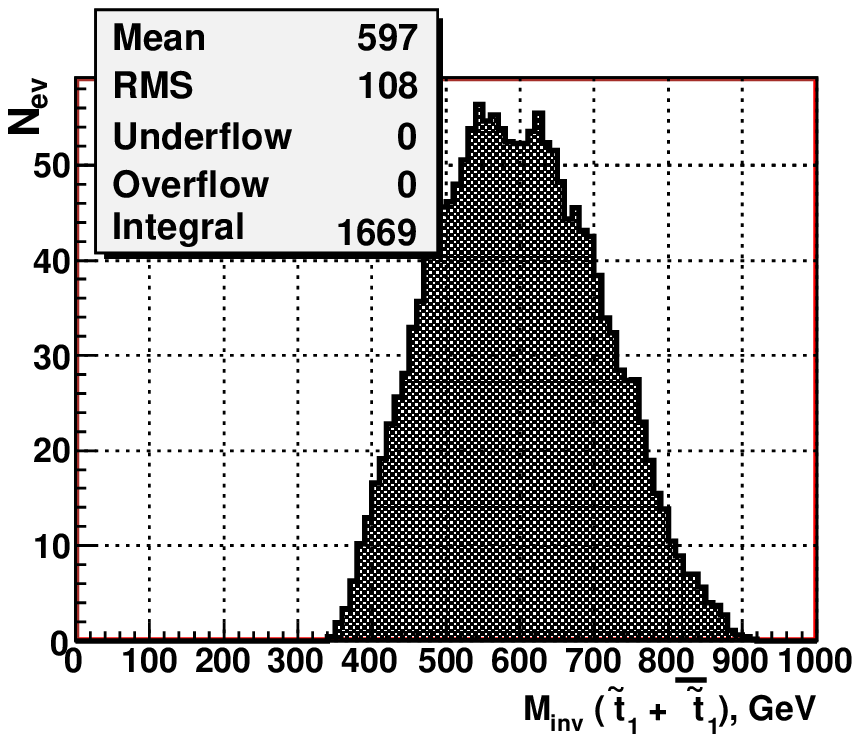}}
    \mbox{{\bf b)}\includegraphics[width=7.2cm, height=4.4cm]{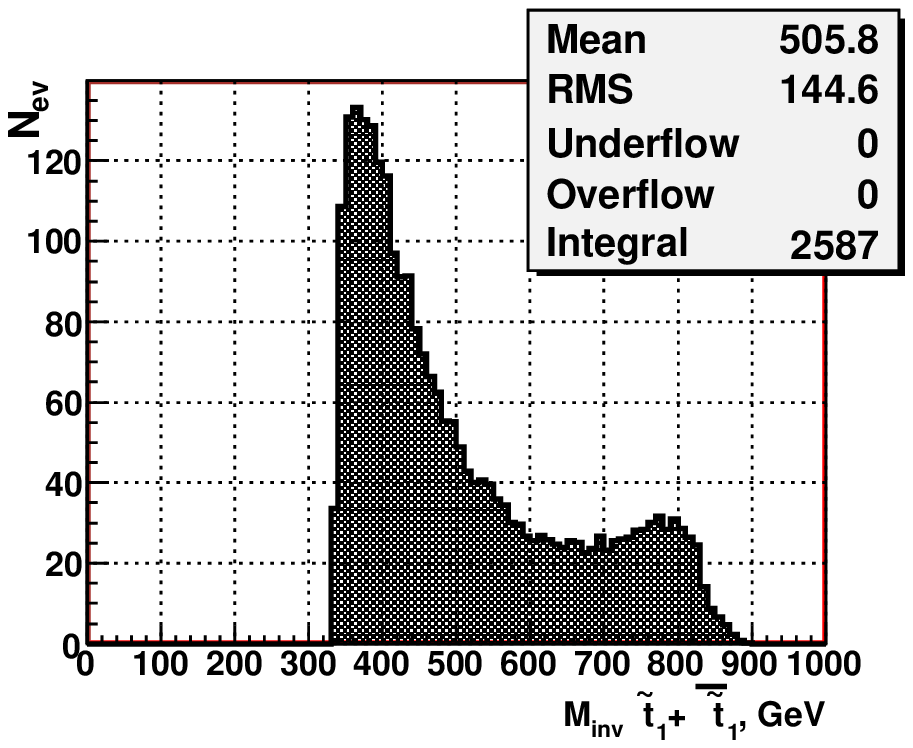}} \\
    \end{tabular}
     \caption{\small \it Stop pair invariant mass  
     $M_{inv}(\widetilde{t}_1 +\bar{\widetilde{t}}_1$) spectra.            
              {\bf a)} $"+-"$  and $"-+"$ polarizations,
              {\bf b)} $"++"$ and $"--"$  polarizations.}
     \end{center}  
 \vskip -0.5cm           
     \end{figure}
   
   The vertical axis in the plots shows the number
   of stops and antistops produced per year ($=10^{7}sec$)
   in each  bin. Taking the integral of the distributions 
   and dividing its value by two (there is one stop-antistop
   pair in each event) one can get  the total  number
   of events expected per year for the applied cuts.
   These numbers   are shown as  "Integral" values 
   within the statistical frames in the  upper
   corners of the plots. They are calculated by taking 
   into account the ratio of the photon-photon 
   luminosity in the energy region above the stop pair  
   threshold   over the total photon-photon luminosity. 
   In the case of  $"+-"$  and $"-+"$ polarizations
   this ratio is approximately 0.419. From Fig.8 it 
   is seen that the number of events per year for the 
   $"++"$ and $"--"$ backscattered photon polarizations
   is equal to $N_{++/--}=2587$.
   It  is appreciably larger
   than the corresponding number of events  per year
   $N_{+-/-+}=1669$
   for  $"+-"$  and  $"-+"$  polarizations. 
    
%
\subsection{ Jet distributions from W decay.} 
%

 ~~~~    According to the decay chain (2), the final state
   has to contain  two jets due to the decay of
   one W boson  into two quarks  $W \to q_{i} + \bar q_{j}$ (see Fig.1).

    Fig.9 shows the distributions    of the energy
   $E_{W-quark}$ of the quarks  produced in the 
    W boson decay (which we  call  "$W$-quarks") 
    for  stop (plots {\bf a)} and {\bf b)}) and top  
    (plots {\bf c)} and {\bf d)}) production. 
    Plots {\bf a)} and {\bf c)}  present  $"+-"$  and 
    $"-+"$ polarizations, while plots {\bf b)} and 
    {\bf d)}  present  $"++"$ and $"--"$  polarizations.   
   The stop-quark spectra begin  at zero  and 
   go up to 220 GeV, with a mean values of 
   $\approx 65$  GeV,   while the top-quark  spectra  
   go  up to approximately 300-350 GeV, 
   with the mean values of 85-97 GeV.  
  
    \begin{figure}[!ht]
     \begin{center}
    \begin{tabular}{cc}
  \mbox{a)\includegraphics[width=7.2cm, height=4.4cm]{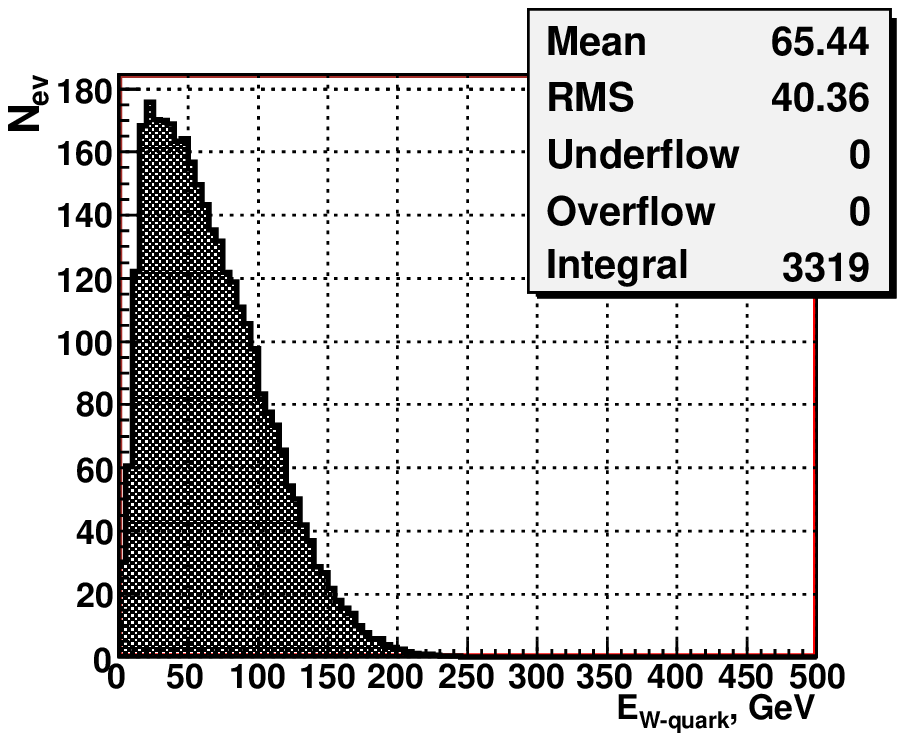}}
  \mbox{b)\includegraphics[width=7.2cm, height=4.4cm]{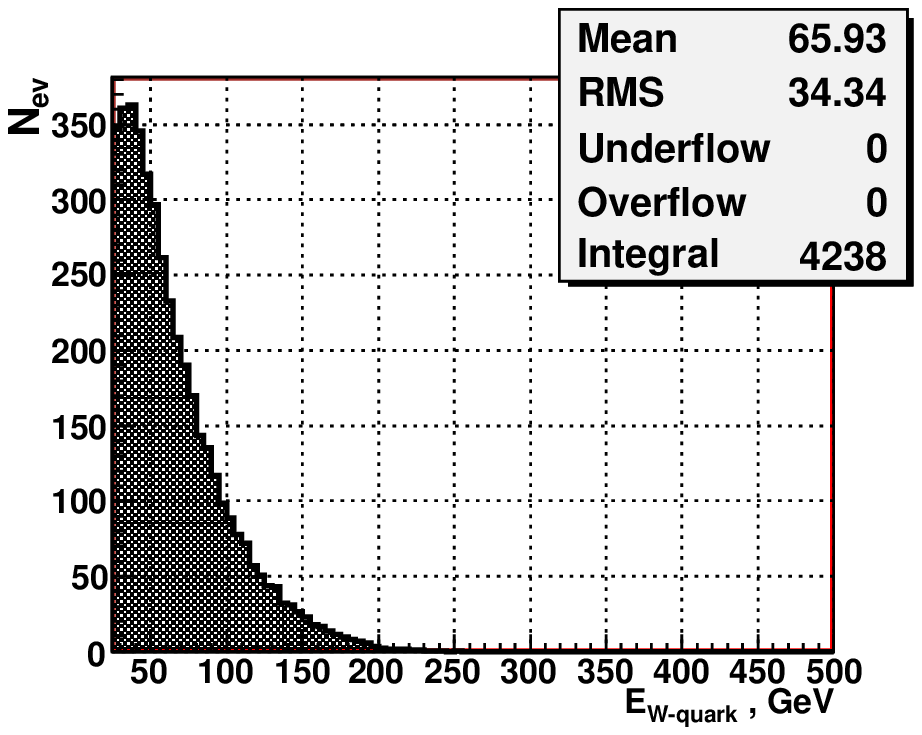}} \\
  \mbox{c)\includegraphics[width=7.2cm, height=4.4cm]{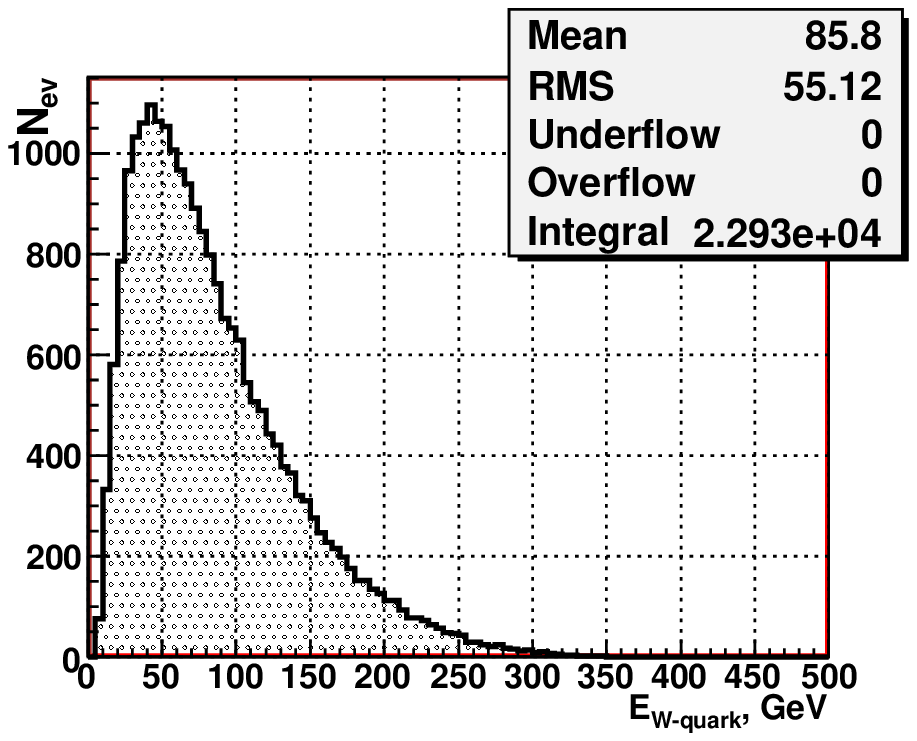}}
  \mbox{d)\includegraphics[width=7.2cm, height=4.4cm]{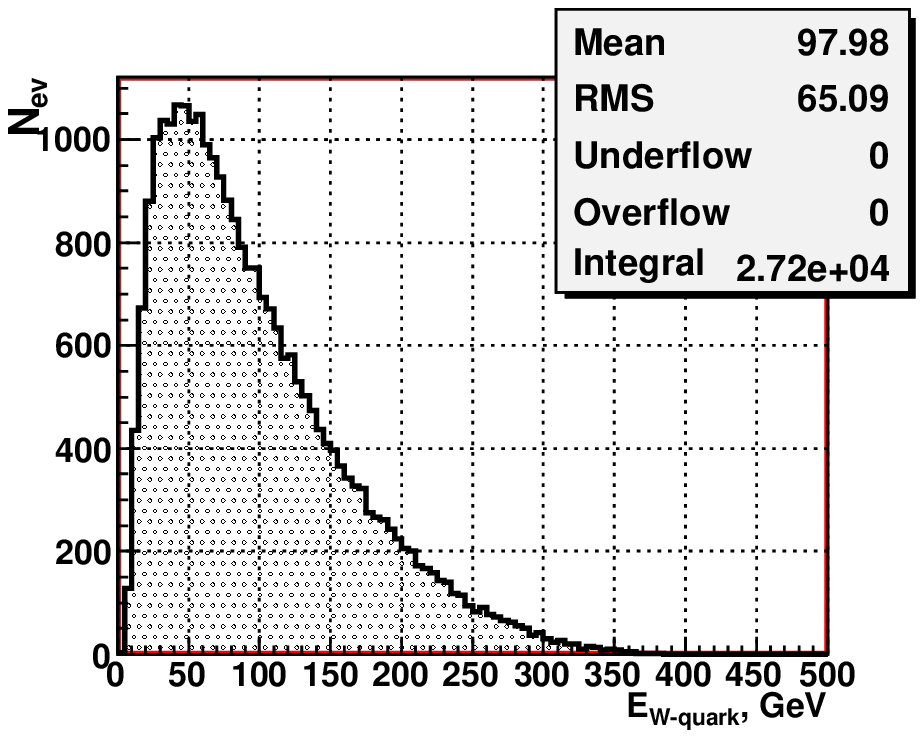}}		   
    \end{tabular}
     \caption{\small \it Energy spectra of 
                the quarks from W boson decay.  
	       {\bf a)} and {\bf b)} are for the stop pair production;
	       {\bf c)} and {\bf d)} are for the top pair production.
	  {\bf a)} and {\bf c)} $"+-"$  and $"-+"$ polarizations,
          {\bf b)} and {\bf d)} $"++"$ and $"--"$  polarizations.}       
     \end{center} 
\vskip -0.5cm            
     \end{figure}

  Fig.10  shows the transverse momentum   $PT_{W-quark}$ spectra 
  of the quarks produced  in the W boson decay 
  for stop (plots {\bf a)} and {\bf b)}) and top 
  (plots {\bf c)} and {\bf d)}) production. Plots
   {\bf a)} and {\bf c)}  are for  $"+-"$  and 
  $"-+"$ polarizations,  plots {\bf b)} and {\bf d)}
   are for  $"++"$ and $"--"$  polarizations.   
   The shapes of the  $PT_{W-quark}$ spectra of 
   these "$W$-quarks" are rather    similar  to the
   $E_{W-quark}$ spectra. In the case of top 
   production the "$W$-quarks" are slightly 
   more energetic and have a larger transverse 
   momentum than those from  stop pair production.

    \begin{figure}[!ht]
     \begin{center}
    \begin{tabular}{cc}
        \mbox{a)\includegraphics[width=7.2cm, height=4.4cm]{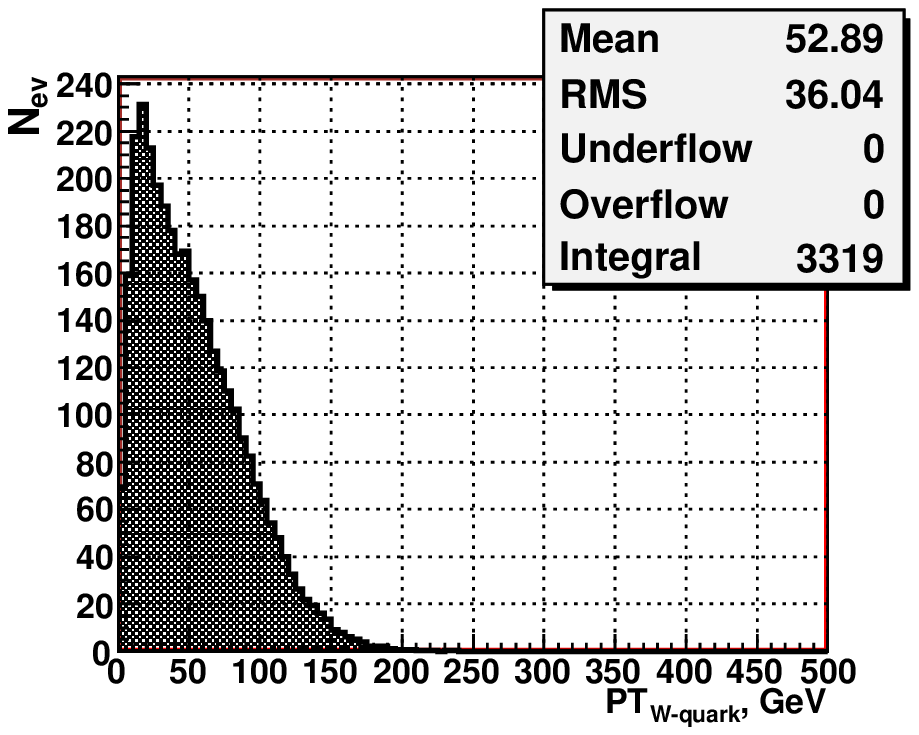}}      
        \mbox{b)\includegraphics[width=7.2cm, height=4.4cm]{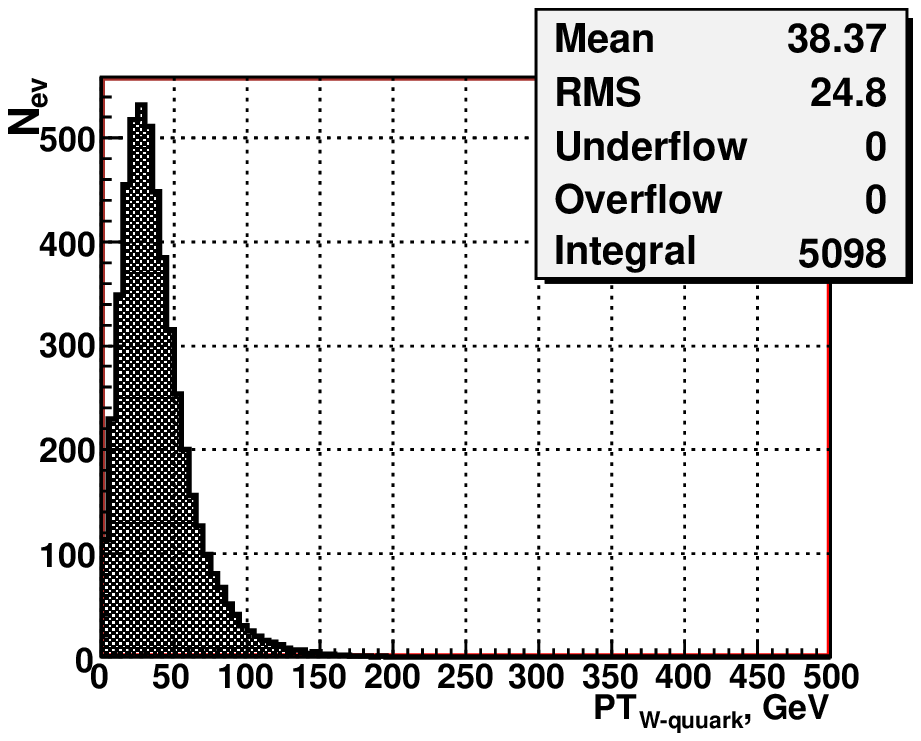}} \\
       \mbox{c)\includegraphics[width=7.2cm, height=4.4cm]{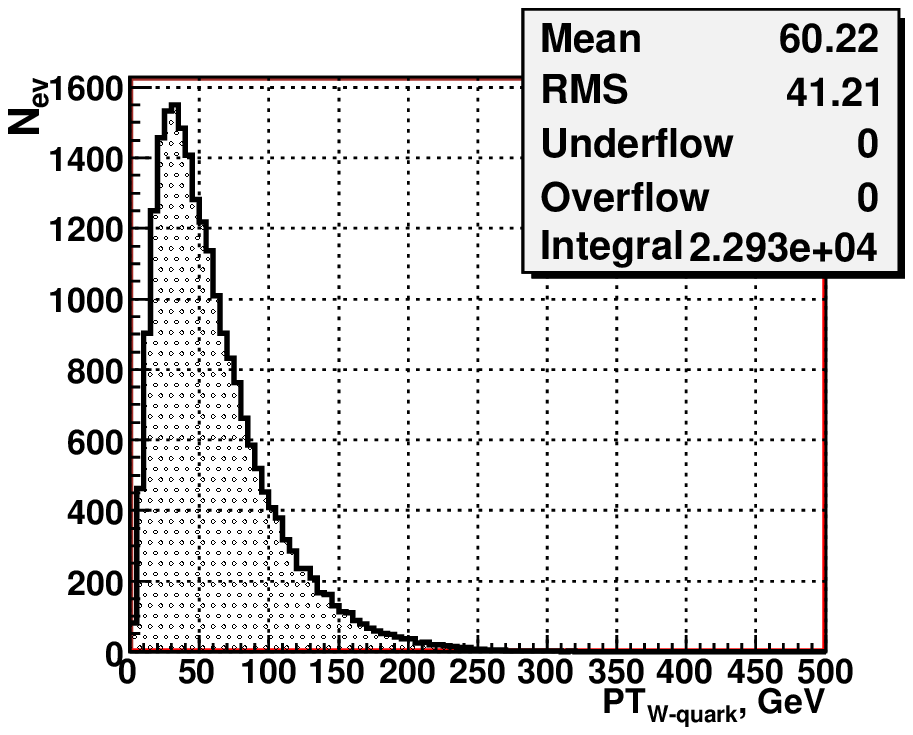}}      
        \mbox{d)\includegraphics[width=7.2cm, height=4.4cm]{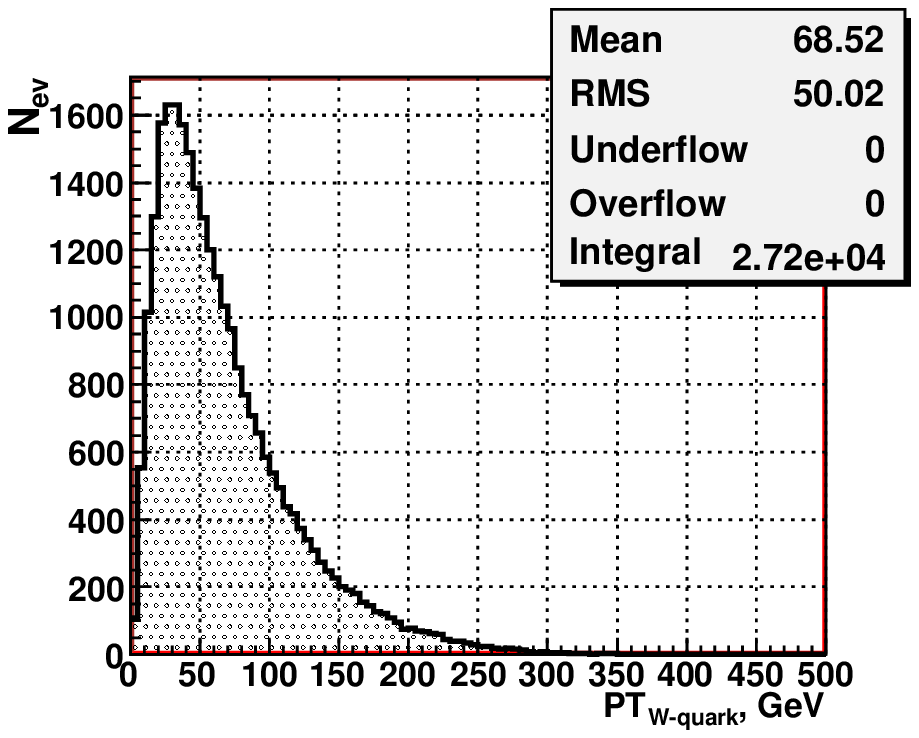}} 		  
     \end{tabular}
     \caption{\small \it PT spectra of the quarks   
                     produced in the W boson decay.
	       {\bf a)} and {\bf b)} are for the stop pair production;
	       {\bf c)} and {\bf d)} are for the top pair production.
	  {\bf a)} and {\bf c)} $"+-"$  and $"-+"$ polarizations,
          {\bf b)} and {\bf d)} $"++"$ and $"--"$  polarizations.}           
     \end{center}  
\vskip -0.5cm             
     \end{figure} 
        
   As the next step we take into account 
    the  hadronization of the "$W$-quark"
   into a jet which we call  "$jet_W$".
    Figs.11 and 12 show the energy
   $E_{jet_W}$  and  transverse momentum  $PT_{jet_W}$
   distributions  of the corresponding "$W$-jets".
  Plots   {\bf a)} and {\bf b)} are for  stop, 
  plots {\bf c)} and {\bf d)} are for top  production. 
  Plots {\bf a)} and {\bf c)} present  $"+-"$  and 
  $"-+"$ polarizations, while plots {\bf b)} and {\bf d)} 
  present  $"++"$ and $"--"$  polarizations.   
   According to our   choice  of PYCLUS jet finder 
   parameters there are   two   "$jet_W$"  in     the  event.
   
     \begin{figure}[!ht]
     \begin{center}
    \begin{tabular}{cc}
 \mbox{a)\includegraphics[width=7.2cm, height=4.4cm]{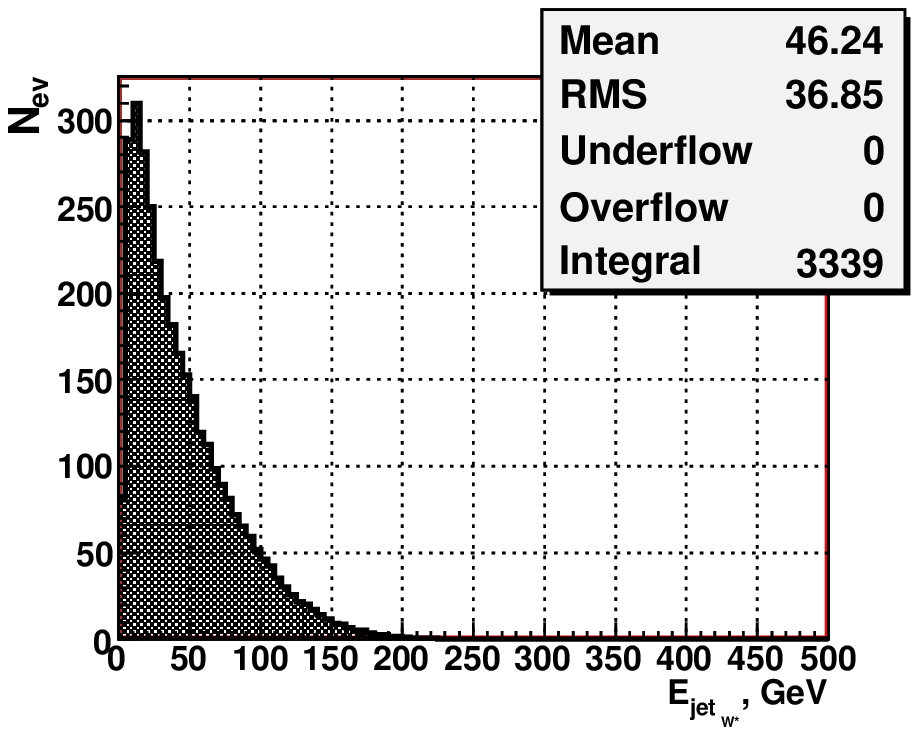}}
 \mbox{b)\includegraphics[width=7.2cm, height=4.4cm]{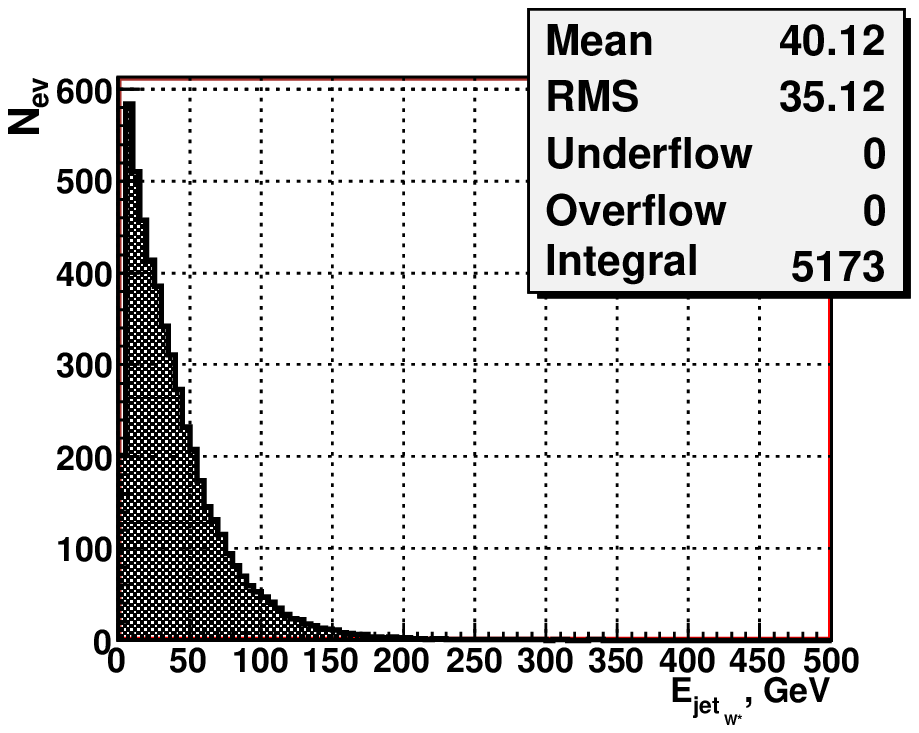}} \\
  \mbox{c)\includegraphics[width=7.2cm, height=4.4cm]{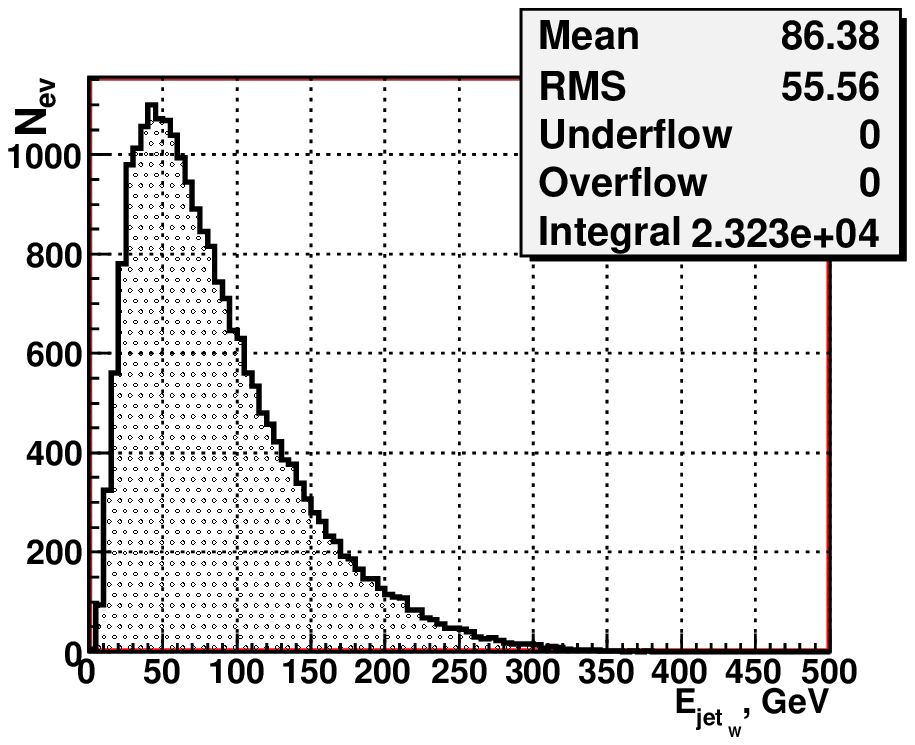}}
  \mbox{d)\includegraphics[width=7.2cm, height=4.4cm]{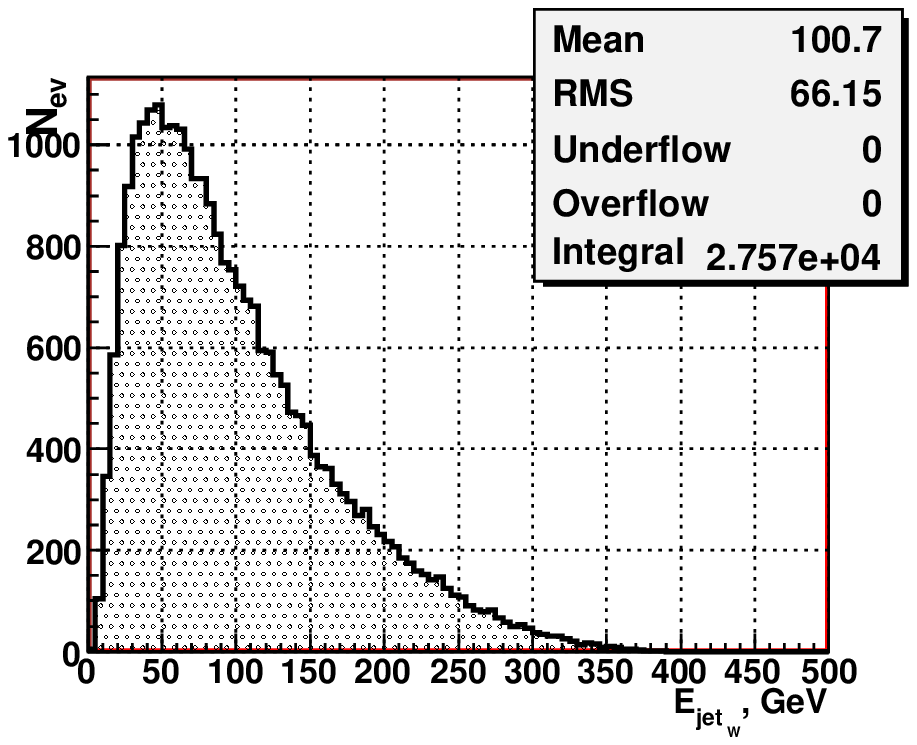}}		   
     \end{tabular}
     \caption{\small \it $E_{jet_W}$ energy spectra.
	       {\bf a)} and {\bf b)} are for the stop pair production;
	       {\bf c)} and {\bf d)} are for the top pair production.
	  {\bf a)} and {\bf c)} $"+-"$  and $"-+"$ polarizations,
          {\bf b)} and {\bf d)} $"++"$ and $"--"$  polarizations.}
                 \end{center} 
\vskip -0.5cm              
     \end{figure}
    
   Comparing plots {\bf a)}  and {\bf b)} of Fig.11 
   for the  energy  distribution  of "W-jets" 
  in  stop production with   the corresponding plots of Fig.9, 
   one observes that    the corresponding mean
   values of the   "W-jets" energy  $E_{jet_W{*}}$ 
   (we use  the notation $W^{*}$ for the virtual W)
in  Fig.11   is about 19-25 GeV lower than  the   mean  energy 
   $E_{W-quark}$  of "$W$-quarks".
   It is also seen (Fig.9)  that the peak positions of 
   "$W$-quark" energy distribution in the stop case 
   ($E^{peak}_{W-quark} \approx 25$ GeV) 
    is shifted  to the left by about 15 GeV
   ($E^{peak}_{jet_W{*}}\approx 10$ GeV)
   when  passing to the jet level
   (see plots {\bf a)} and {\bf b)} of Fig.11).
   The end point of the  $E_{jet_W{*}}$ 
   distribution in the stop case is somewhat lower
   than that for the corresponding  quarks. 
     
    Analogously, the mean values and the peak 
   positions of  the distribution of the 
   transverse momentum  the "$W^{*}$-quarks" 
    $PT_{W^{*}-quark}$, shown in Fig.10 {\bf a)} and {\bf b)},
   decrease by about 12-20 GeV when passing 
   to the jet level (see Fig.12 {\bf a)}  and {\bf b)}),
   while the  end point of  the  $PT_{jet_{W}{*}}$  distribution 
   is a bit lower than the  end point of 
   $PT_{W^{*}-quark}$ distribution.     
     
     \begin{figure}[!ht]
     \begin{center}
    \begin{tabular}{cc}
 \mbox{a)\includegraphics[width=7.2cm, height=4.4cm]{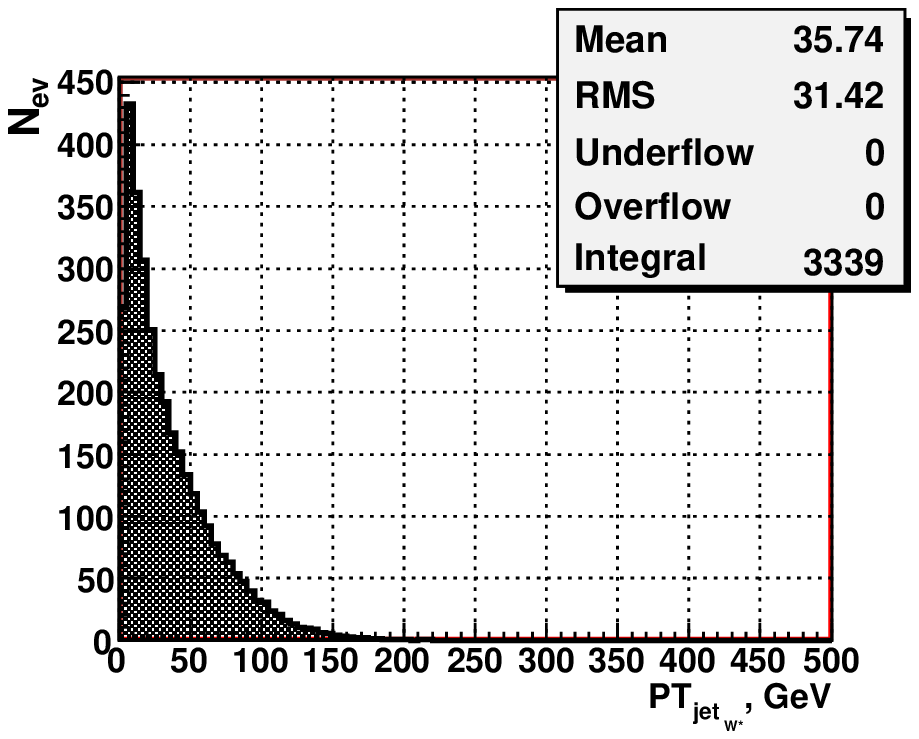}}
 \mbox{b)\includegraphics[width=7.2cm, height=4.4cm]{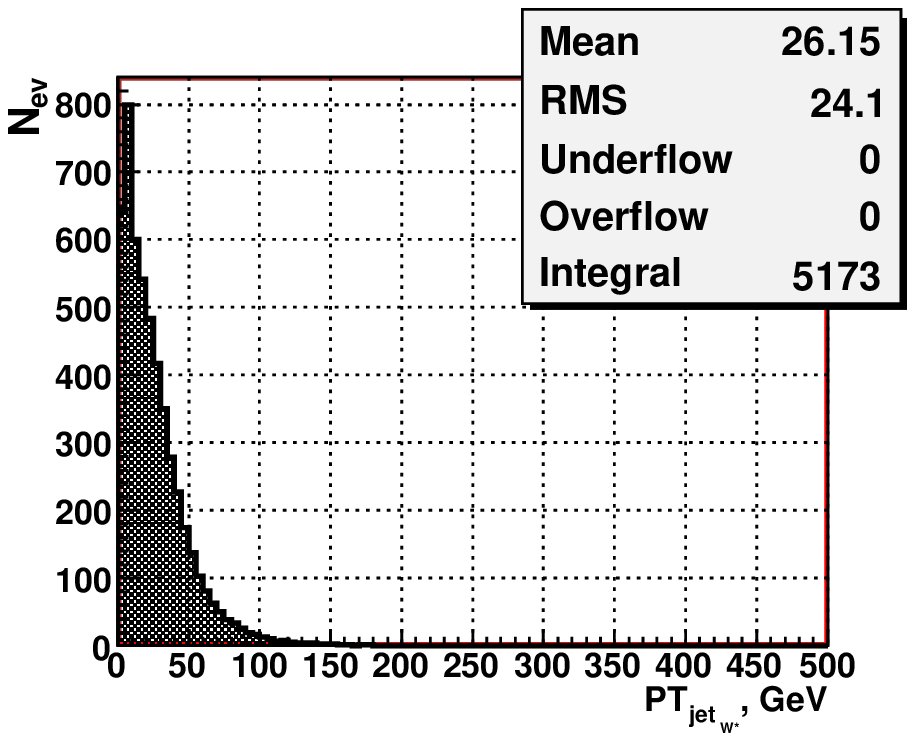}} \\
  \mbox{c)\includegraphics[width=7.2cm, height=4.4cm]{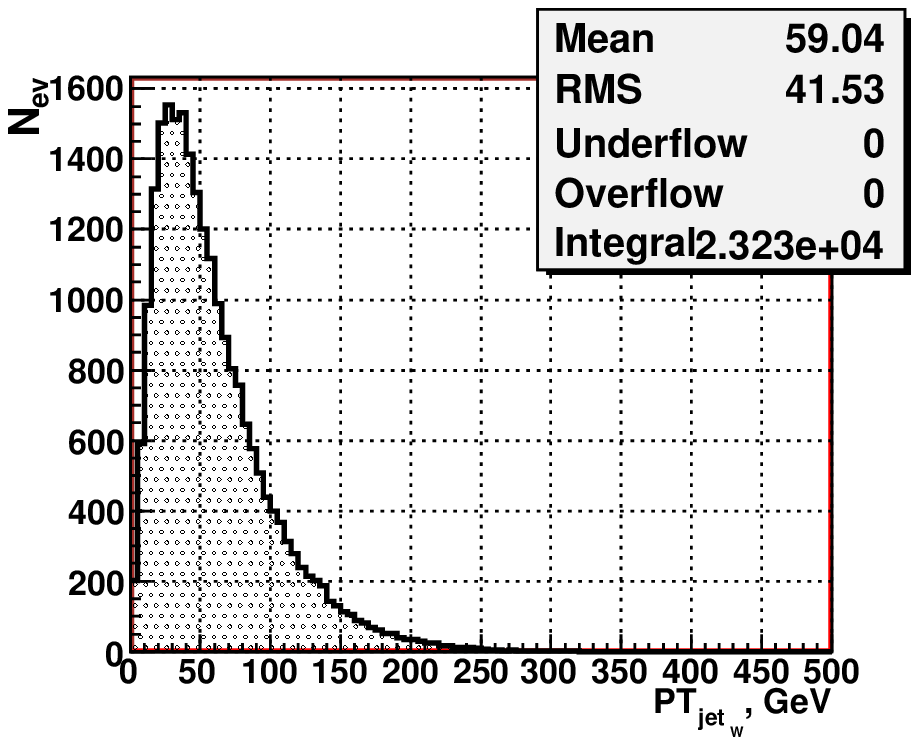}}
  \mbox{d)\includegraphics[width=7.2cm, height=4.4cm]{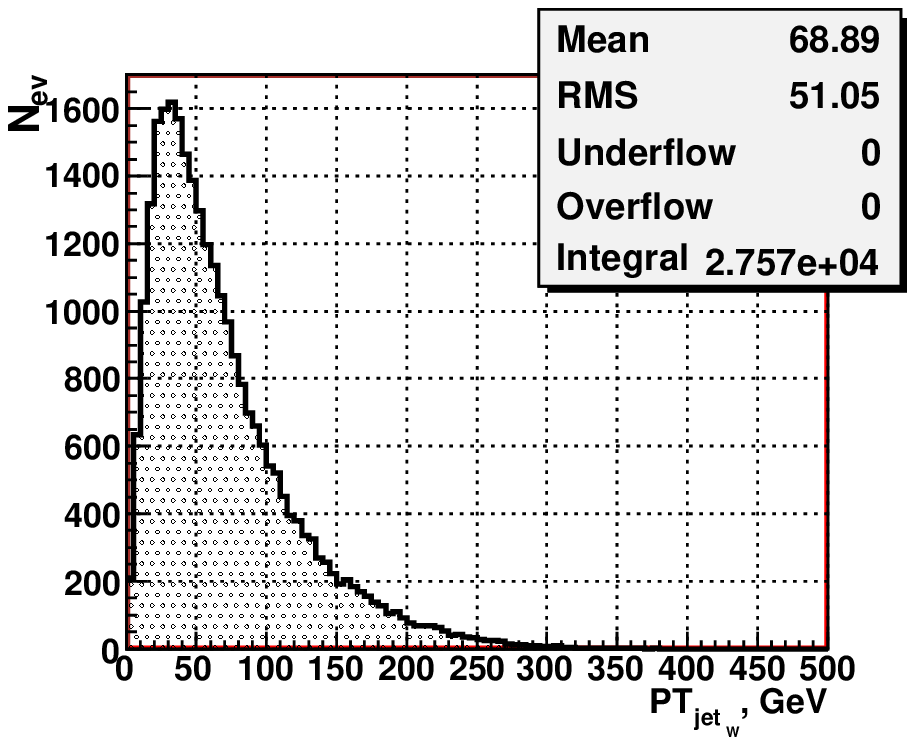}}		   
     \end{tabular}
     \caption{\small \it "$jet_W$"  PT- spectra. 
	  {\bf a)} and {\bf b)} are for the stop pair production;
	  {\bf c)} and {\bf d)} are for the top pair production.
	  {\bf a)} and {\bf c)} $"+-"$  and $"-+"$ polarizations,
         {\bf b)} and {\bf d)} $"++"$ and $"--"$  polarizations.}
                 \end{center} 
\vskip -0.5cm              
     \end{figure} 
   
   Due to the  different kinematics in  top production
    mentioned above, the spectrum of the
   the energy  $E_{jet_W}$,   its peak position 
   and the mean value of the   "$jet_W$"
   energy distribution in the top case 
   are practically equivalent  to the  $E_{W-quark}$  spectrum, 
   peak position and the  mean value of the 
   corresponding  "$W$-quark" energy
   distribution (see Fig.9 {\bf c)},  {\bf d)}
   and Fig.11 {\bf c)}, {\bf d)}). 
Analogously, by comparing plots {\bf c)} and {\bf d)} of Figs.10 and 12  for $PT_{W-quark}$ and $PT_{jet_W}$, one can see that the transverse momentum distribution in top production is stable under hadronization.    
    
      Figure 13 shows  the spectrum   of the invariant mass
   $M_{W}=M_{inv}(quark1 + quark2)$ reconstructed from the 
   vectorial sum of 4-momenta of  the two "W-quarks".
 The main features of these plots practically   do not differ for
   $"+-"$  and  $ "-+"$   and the $"++"$  and  $"--"$ 
   polarization cases. Therefore we do not show them separately. 
   Plot {\bf a)} is for  stop pair production,  plot
   {\bf b)} is for  top production. 
 In  plot {\bf a)} of Fig.13 one clearly sees the virtual nature 
    of the W boson  in the stop pair  production case. 
    Hence, in the stop case the invariant mass of two quarks
    produced in the decay of the virtual W boson ($W^{*}$) is 
    smaller than the mass of a real W boson. In  top production 
    (see plot {\bf b)} of Fig.13) there is a peak in the 
    invariant mass  distribution at the mass value of the
    real  W boson.
    
  \begin{figure}[!ht]
     \begin{center}
    \begin{tabular}{cc}
     \mbox{{\bf a)}\includegraphics[width=7.2cm, height=4.4cm]{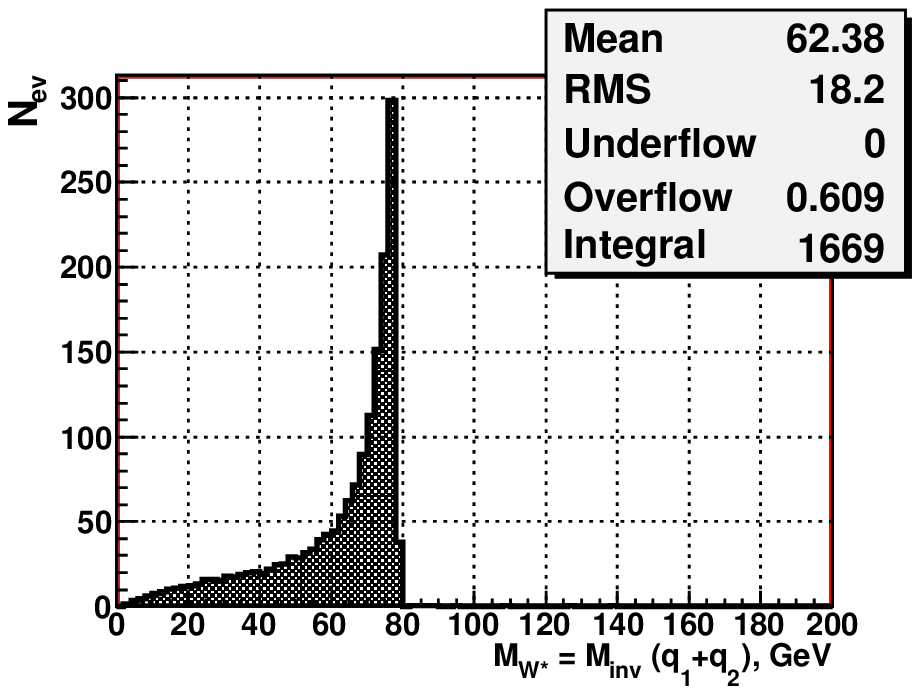}} 
    \mbox{{\bf b)}\includegraphics[width=7.2cm, height=4.4cm]{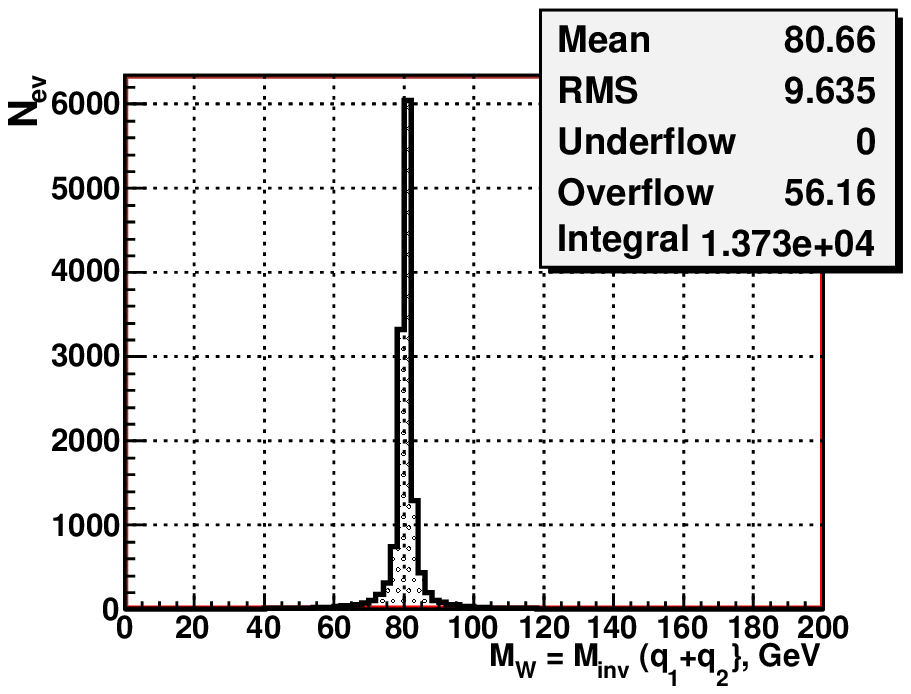}}        
    \end{tabular}
     \caption{\small \it The invariant mass of two 
              quarks     $M_{W}=M_{inv}(quark1 + quark2)$, 
	      reconstructed from the  vectorial
	      sum of 4-momenta of two quarks that
              are produced in  $W \to q_{i} + \bar q_{j}$ 
	      decay. {\bf a)} stop pair production;
	       {\bf b)} top pair production.}     
     \end{center}  
\vskip -0.5cm             
     \end{figure}     
     
       Figure 14 shows  the corresponding  plots  at the jet
    level. The invariant mass  is built 
        \begin{figure}[!ht]
     \begin{center}
    \begin{tabular}{cc}
     \mbox{{\bf a)}\includegraphics[width=7.2cm, height=4.4cm]{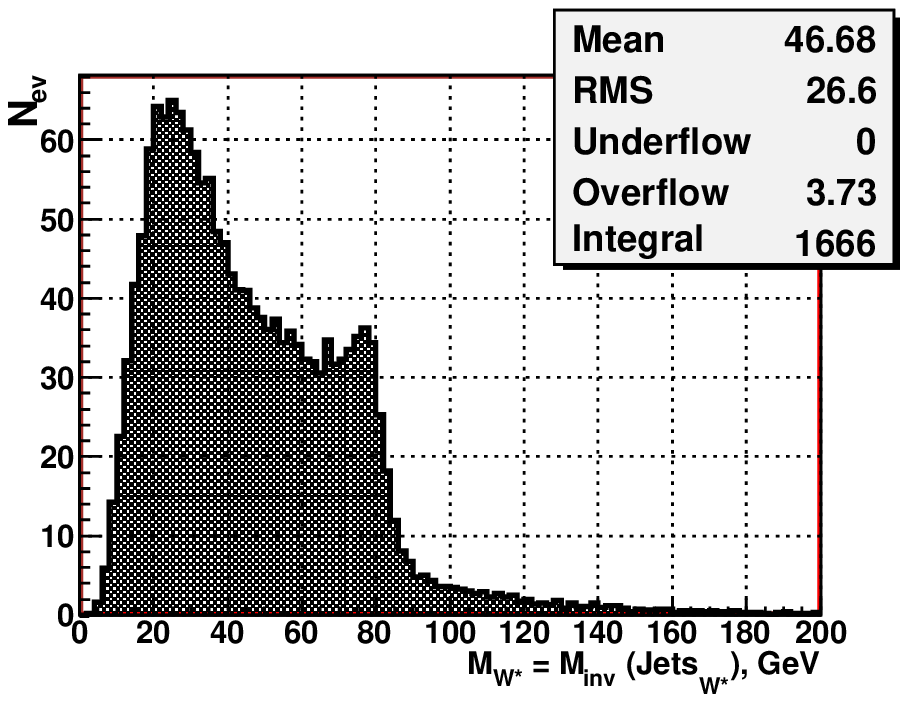}}      
     \mbox{{\bf b)}\includegraphics[width=7.2cm, height=4.4cm]{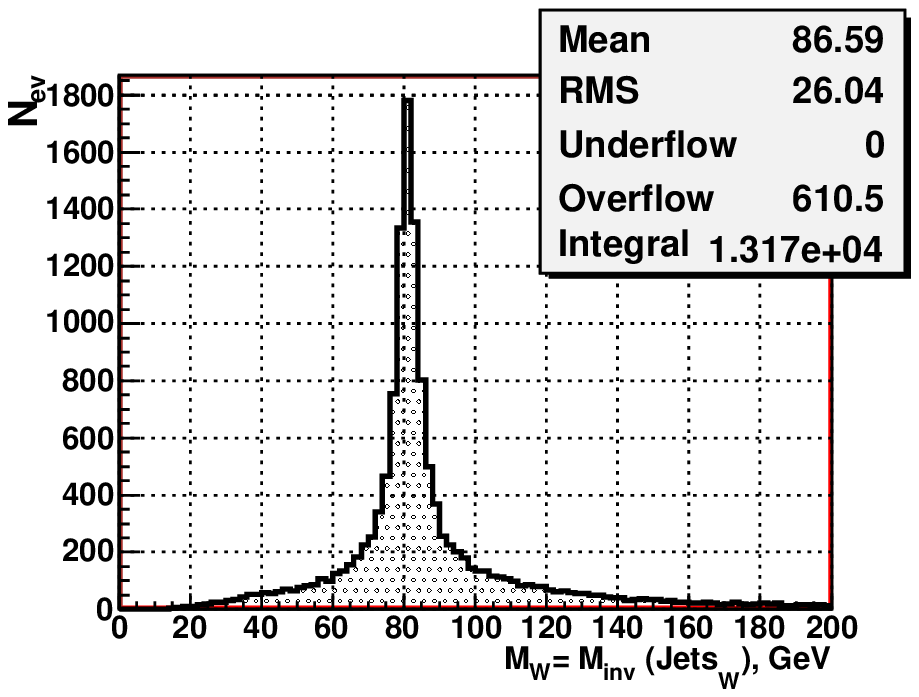}} \\
    \end{tabular}
     \caption{\small \it Number of generated events
              versus the reconstructed invariant
	      mass of "all-non-$b$-jets". 
	      {\bf a)} stop pair production; 
	      {\bf b)} top pair production.}     
     \end{center}  
\vskip -0.5cm             
     \end{figure}  
    of  "all-non-$b$-jets" (or, shortly,  ``$jets_{W^{*}}$''). 
    One can see from plot {\bf a)} that in the stop case 
    the peak position of   $M_{inv}(jets_{W^{*}})$ 
     is shifted to the left  and a 
    long tail for higher invariant masses  appears.     
    As seen from plot {\bf b)}, in the top case at the jet level
    the position of the W-peak remains at the 
    same value  of $M_{W}$ (with  a high precision)
    as in plot {\bf b)} of Fig.13,
    except  some shifting of the mean value.   
    From comparison of plots  {\bf a)} and 
    {\bf b)} of Fig.14 we conclude  that the cut 
     $M_{inv}(jets_{W^{*}}) \leq 70$ GeV
     may allow us to eliminate  this tail and a big amount 
     of the top background. 
   
%
\subsection{ $b$-quark and $b$-jet distributions in  stop 
             and top production.} 
%
   ~~~~  In the case of stop  decay into a  
   $b$-quark and a chargino,
   $\tilde t_{1} \to b \tilde \chi_{1}^{\pm}$, the 
   jets produced in $b$-quark hadronization are
   observable objects. Their features are interesting 
   from the viewpoint of experimentally distinguishing 
   the stop signal events from the top background.
    
   In Fig.15 we show in plots {\bf a)} and {\bf b)}  
   the distributions of the energies 
    $E_b$ of the  $b$- and $\bar b$-quarks
   (which we do not distinguish in
   the  following) produced in the decay 
   $\tilde t_{1} \to b \tilde \chi_{1}^{\pm}$  for 
   the $"+-"$, $"-+"$  and   
   $"++"$, $"--"$ polarizations, respectively.
  Both   spectra begin  at $E_{b} \approx 4$ GeV, 
  corresponding to the b-quark mass,   and 
   go up to  $E_{b} \approx 34$ GeV. The mean
   values are about 14 GeV and 13 GeV, respectively.    
   Plots {\bf c)} and {\bf d)} of Fig.15  are two  analogous 
   plots for  top pair production. The corresponding 
   spectrum in top production is much harder
   and its main part is concentrated within
   the interval $45 < E_b  < 150$ GeV. The mean values are  
   $E_b \approx 113$ GeV   and $E_b\approx 130$ GeV,
     respectively, which is almost four times higher  than
     the end point of the   $b$-jet  energy  spectra 
     in the stop events.   It means that in the stop case  
   the b-quark takes a smaller part of the 
stop energy  than the b-quark  gets in the  background top case.  
    \begin{figure}[!ht]
     \begin{center}
    \begin{tabular}{cc}
     \mbox{{\bf a)}\includegraphics[width=7.2cm, height=4.4cm]{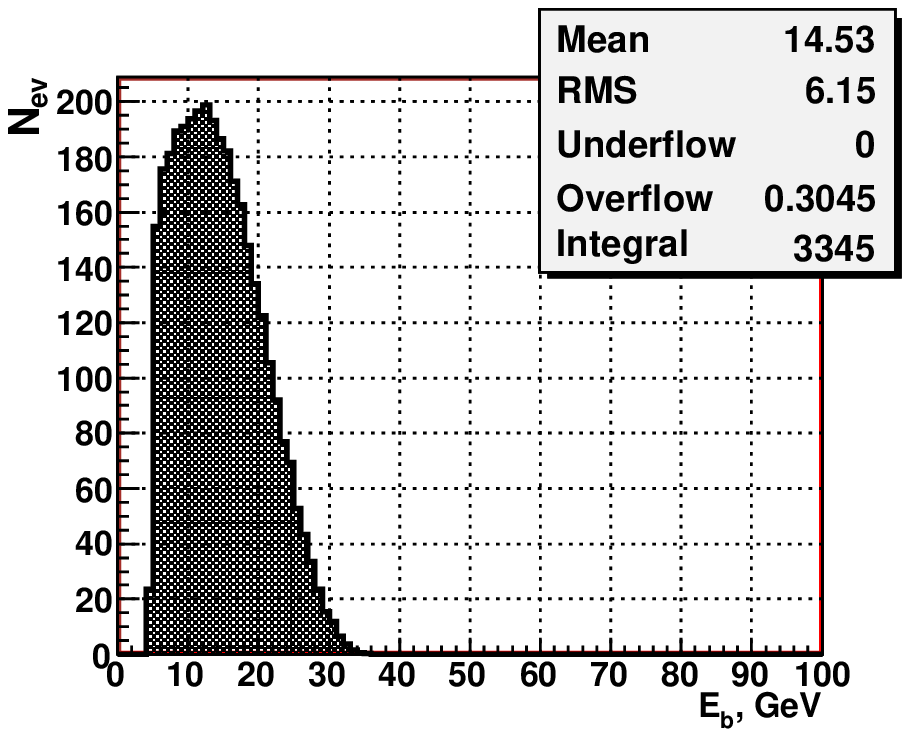}}      
     \mbox{{\bf b)}\includegraphics[width=7.2cm, height=4.4cm]{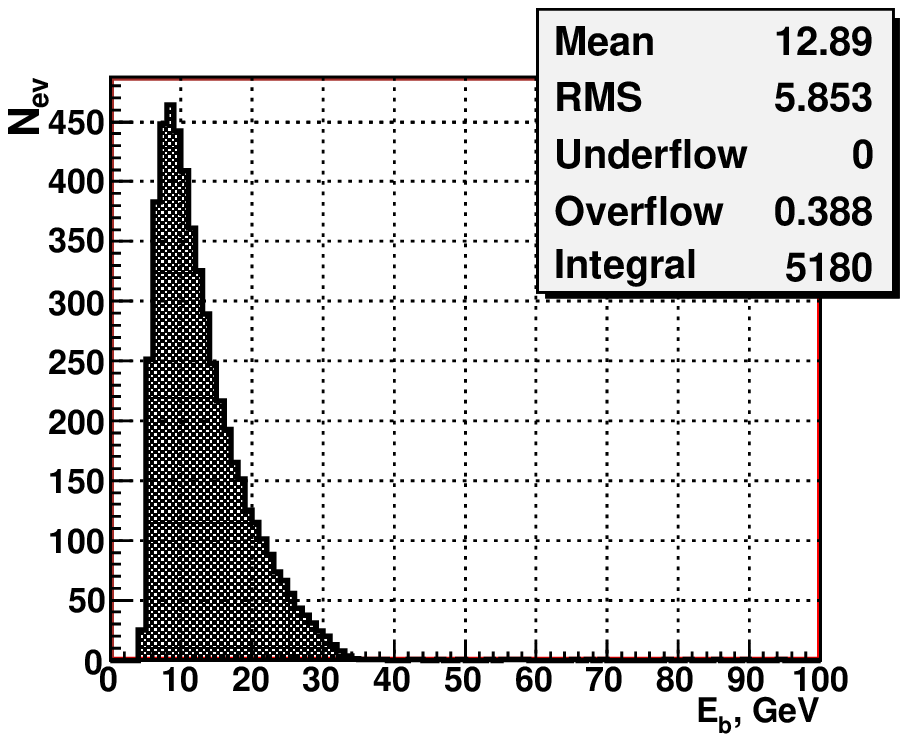}} \\
     \mbox{{\bf c)}\includegraphics[width=7.2cm, height=4.4cm]{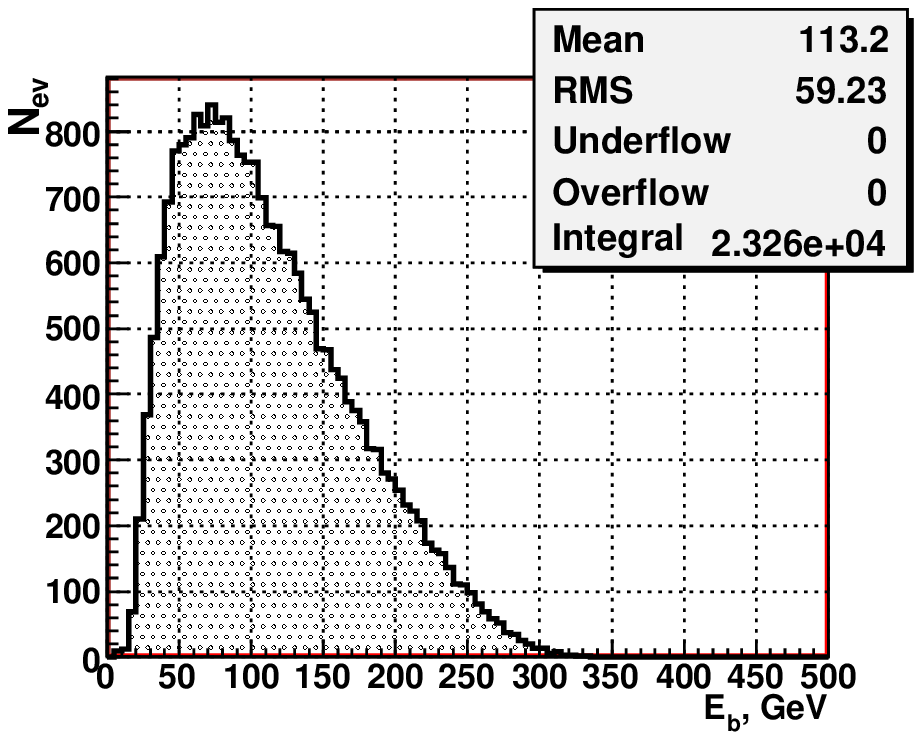}}      
     \mbox{{\bf d)}\includegraphics[width=7.2cm, height=4.4cm]{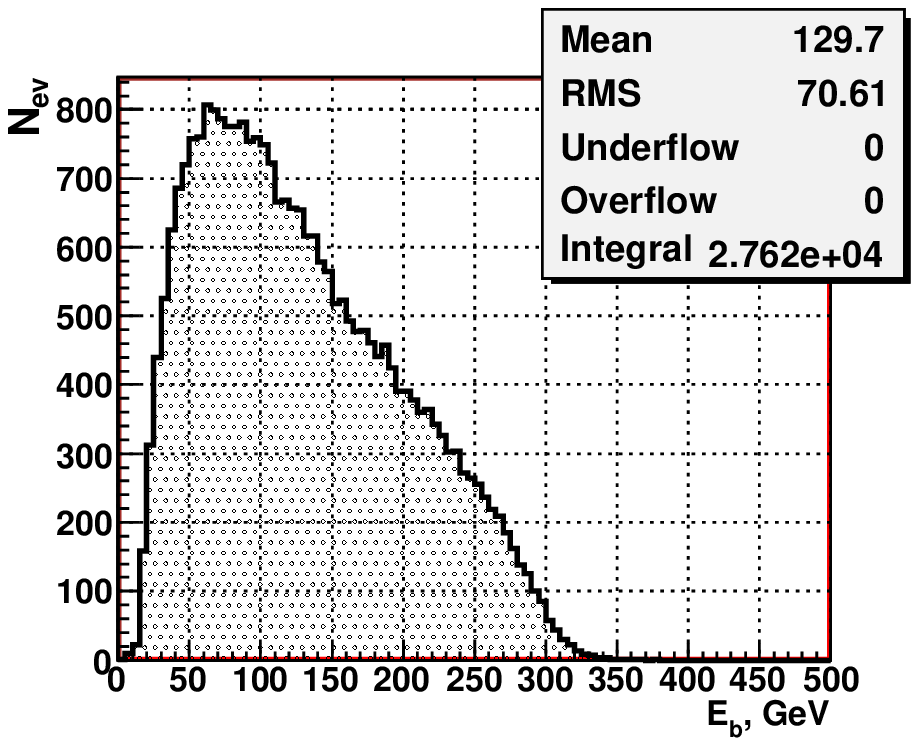}} \\
    \end{tabular}
     \caption{\small \it $b$- and $\bar b$-quark 
              energy spectra.
	       {\bf a)} and {\bf b)} are for the stop pair production;
	       {\bf c)} and {\bf d)} are for the top pair production.
	  {\bf a)} and {\bf c)} $"+-"$  and $"-+"$ polarizations,
          {\bf b)} and {\bf d)} $"++"$ and $"--"$  polarizations.}
     \end{center}  
 \vskip -0.5cm          
     \end{figure}

 Figure 16  shows the transverse   momentum   $PT_{b}$ spectra
   of $b$-quarks for  stop (plots {\bf a)} and {\bf b)}) and top 
   (plots {\bf c)} and {\bf d)}) production. Plots 
   {\bf a)} and {\bf c)} are for    $"+-"$  and $"-+"$ 
   polarizations, plots {\bf b)} and {\bf d)}  
   are for  $"++"$ and $"--"$  polarizations.
    Comparing the distributions in Figs.15 and 16  with the
    corresponding ones in  Figs.5 and 6, one can conclude
   that in  stop pair production  the  $b$-quarks  have only a
   small fraction of the energy and transverse 
   momentum of the parent stops. The shape 
   of the  $PT_{b}$ spectra  of $b$-quarks 
in  the  stop case  is  similar to the shape of  the $E_{b}$ spectra.   This means that in the
   stop decay the  transverse component of 
the $b$-quark momentum is larger than the longitudinal component. 
    \begin{figure}[!ht]
     \begin{center}
    \begin{tabular}{cc}
     \mbox{{\bf a)}\includegraphics[width=7.2cm, height=4.4cm]{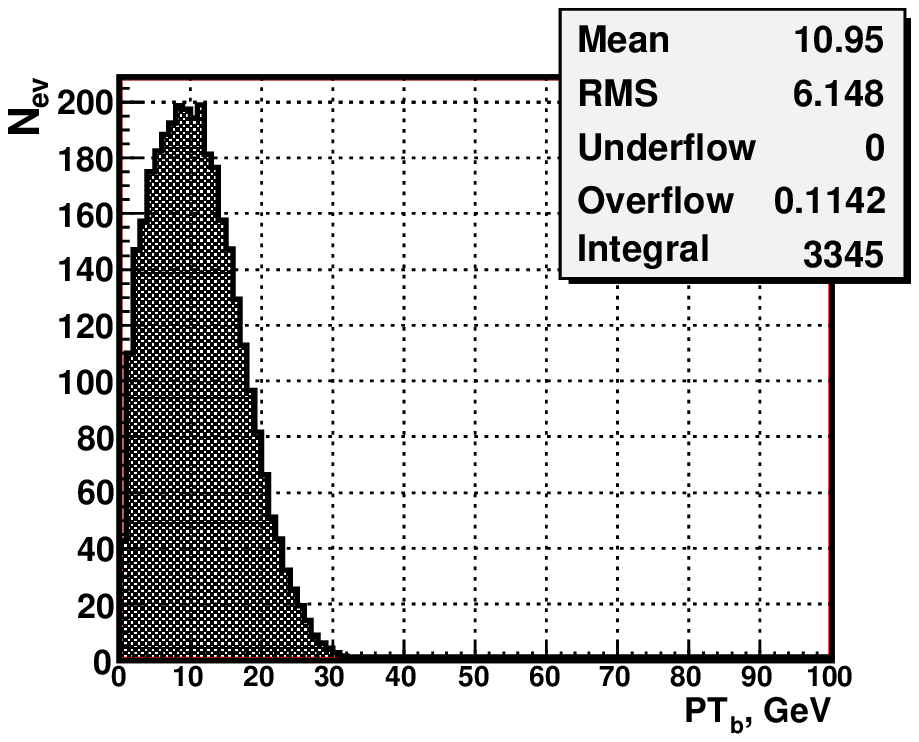}}      
     \mbox{{\bf b)}\includegraphics[width=7.2cm, height=4.4cm]{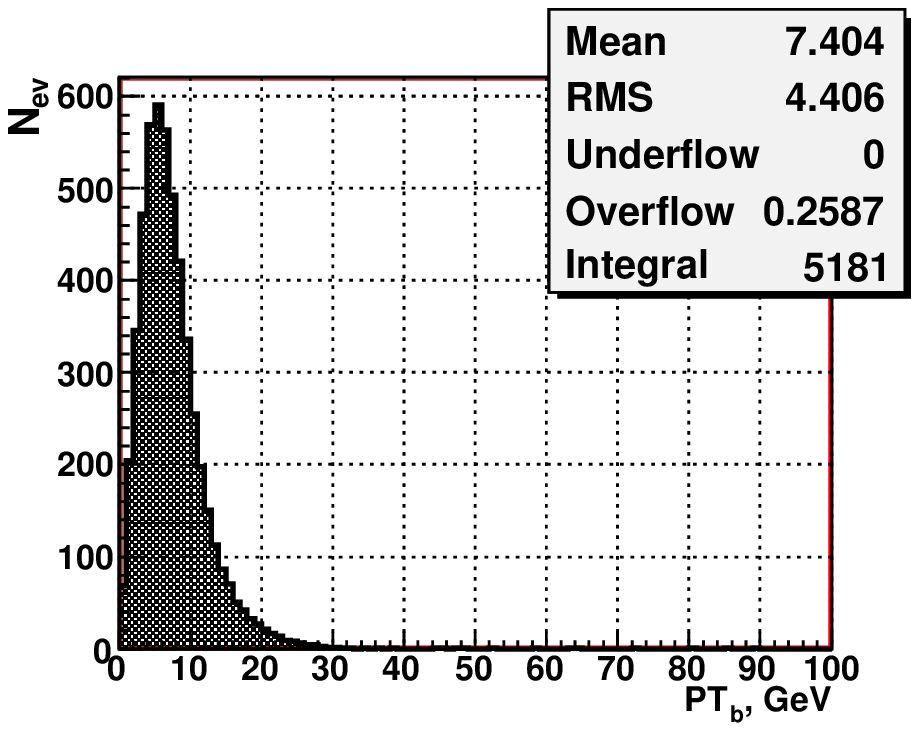}} \\
     \mbox{{\bf c)}\includegraphics[width=7.2cm, height=4.4cm]{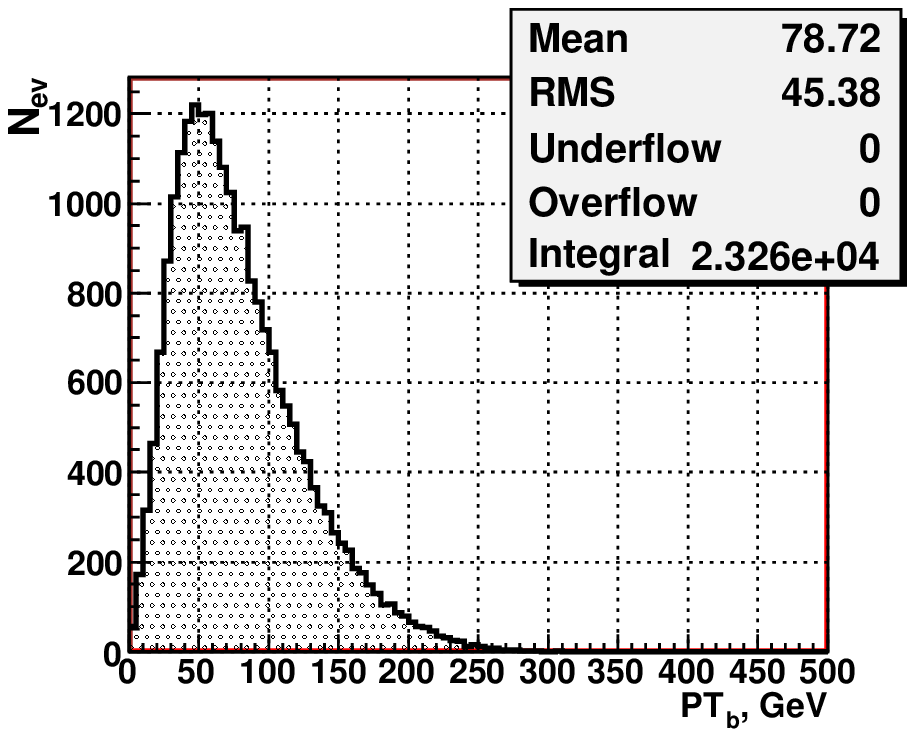}}      
     \mbox{{\bf d)}\includegraphics[width=7.2cm, height=4.4cm]{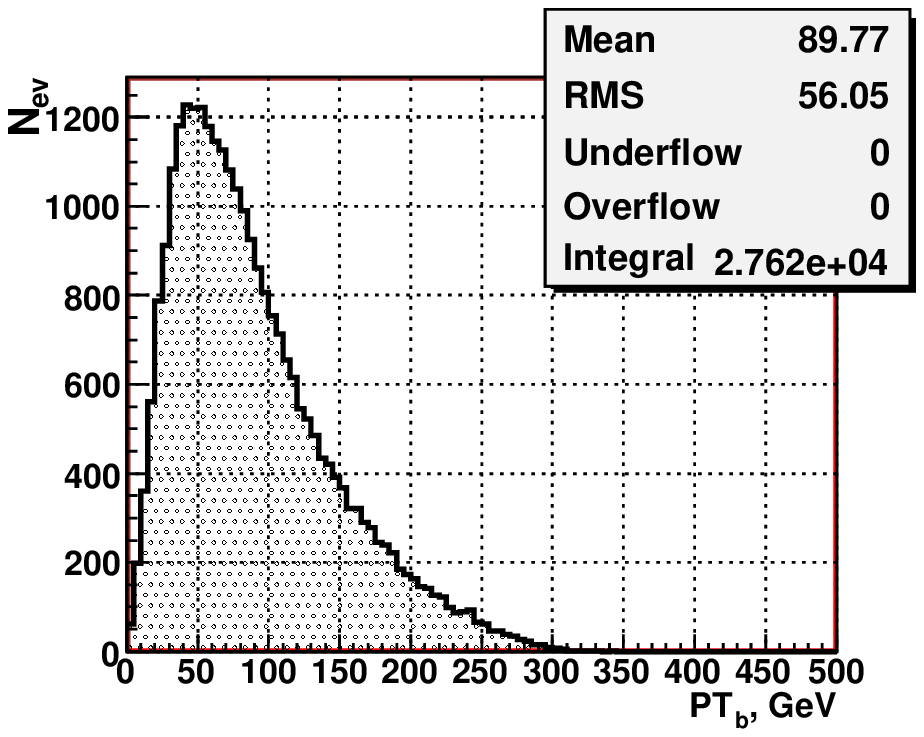}} \\
     
    \end{tabular}
     \caption{\small \it  b- and $\bar b$-quark PT spectra.
            {\bf a)} and {\bf b)} are for the  stop production;	   {\bf c)} and {\bf d)} are for the top case.
	  {\bf a)} and {\bf c)} $"+-"$  and $"-+"$ polarizations,
          {\bf b)} and {\bf d)} $"++"$ and $"--"$  polarizations.}
     \end{center} 
 \vskip -0.5cm          
     \end{figure}

   The kinematical distributions of the
   $b$-quarks in  top decay  are quite 
   different. The $b$-quarks produced in top decays
   are very energetic. Most of the top events have 
   $E_{b} \geq 25$  GeV and $PT_{b} \geq 20$ GeV.
   The difference  to    stop decay is easily understandable. 
   The stop decays into a heavy chargino,
     whereas the top decays  into a real
   W boson whose mass is only half of the mass
   of the chargino  $M_{\chi_{1}^\pm}$. Therefore, the 
   $b$-quarks  in top decays have a larger
   phase space than the $b$-quarks  in stop decays.
 
  The  distribution  of the  polar angle  $\Theta_{b}$ 
    of the $b$-quarks in 
   stop  production are  presented in Fig.17. Plot {\bf a)} is
   for  $"+-"$  and $"-+"$ polarizations, plot {\bf b)} is
   for  $"++"$  and  $ "--"$ polarizations.
   The difference between these distributions due to
   the polarization effects is clearly seen in this figure. 
   
       \begin{figure}[!ht]
     \begin{center}
    \begin{tabular}{cc}
\mbox{{\bf a)} \includegraphics[width=7.2cm, height=4.4cm] {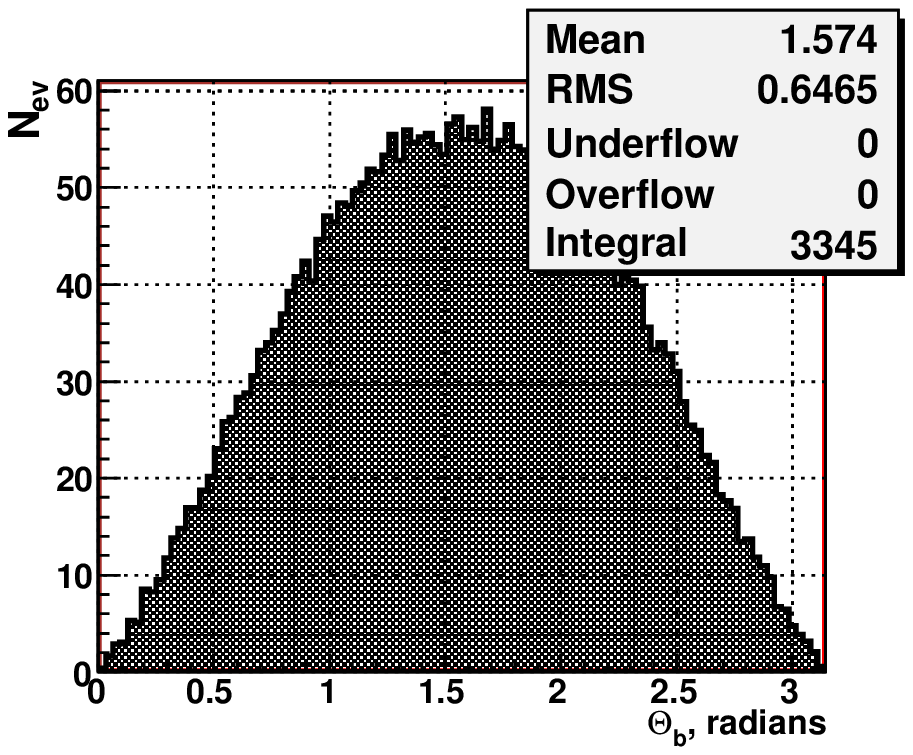}}      
\mbox{{\bf b)} \includegraphics[width=7.2cm, height=4.4cm] {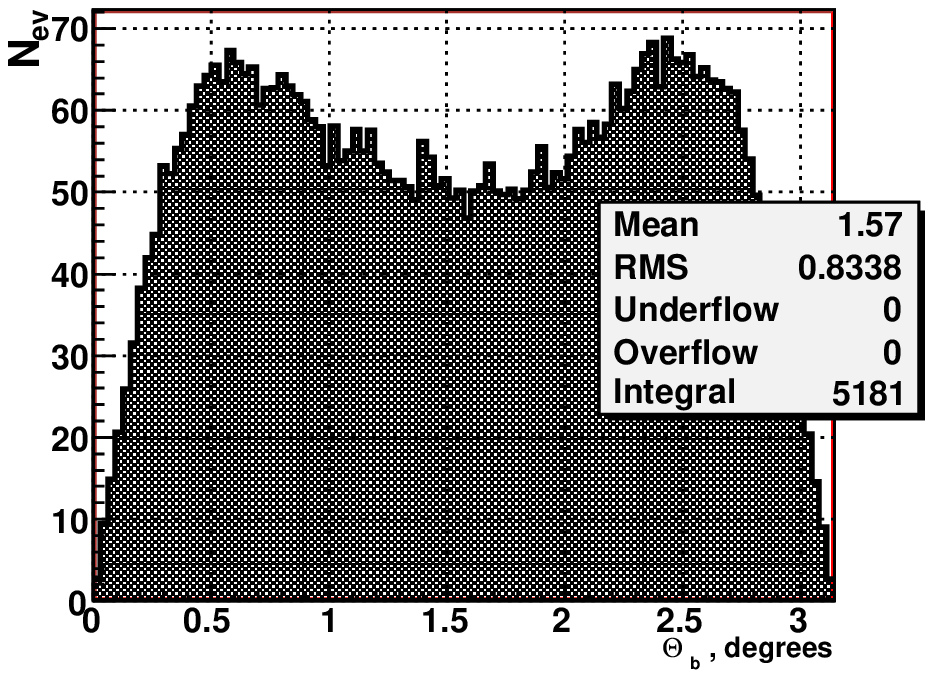}} \\
    \end{tabular}
\caption{\small \it The spectra of the b- and $\bar b$-quark polar    angle  $\theta_{b} $  for stop production events.
           {\bf a)}  $"+-"$  and $"-+"$ polarizations,
           {\bf b)}  $"++"$ and $"--"$  polarizations.}  
     \end{center}  
\vskip -0.5cm            
     \end{figure}
      
   In Fig.18 the  $cos(b,\bar b)$ distribution is shown, 
   where  $cos(b,\bar b)$  is the cosine  of the 
   opening angle between the 3-momenta  of the 
   $b$- and $\bar b$-quarks produced in 
    the same stop event.   Plot {\bf a)} is
   for  $"+-"$  and $"-+"$ polarizations, plot {\bf b)} is
   for $"++"$  and  $ "--"$ polarizations.
   It demonstrates that  most of the 
   $b$- and $\bar b$-quarks  move   in
   approximately opposite directions, but some 
   are in the same  hemisphere. Thus, in the 
   experiment we may expect a similar  angular
    distribution of the corresponding $b$- and   $\bar b- jets$. 

    \begin{figure}[!ht]
     \begin{center}
    \begin{tabular}{cc}
     \mbox{{\bf a)} \includegraphics[width=7.2cm, height=4.4cm]  {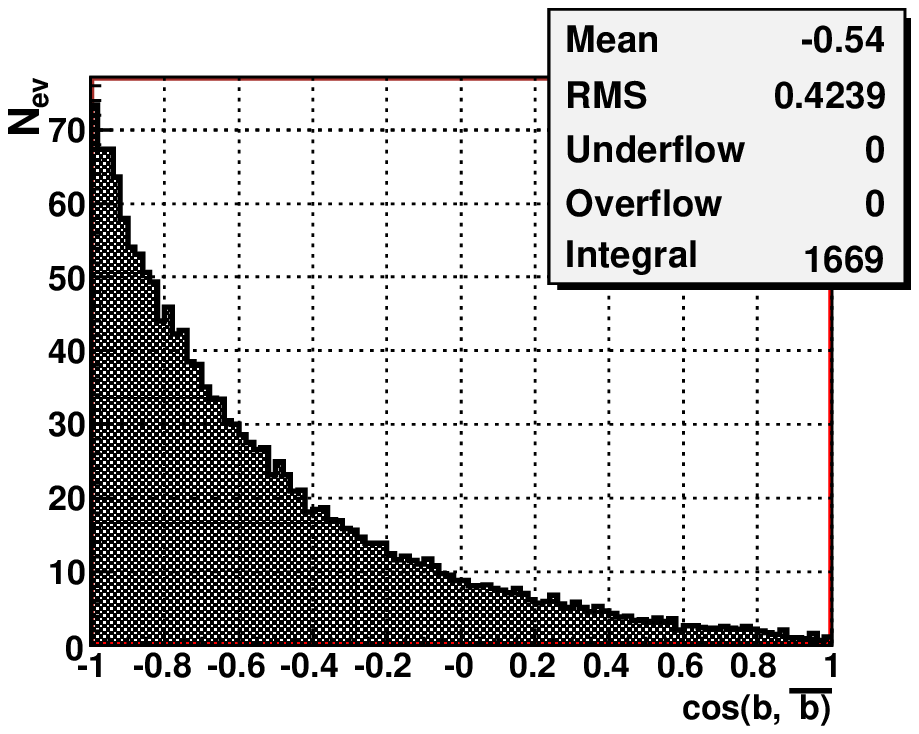}}      
     \mbox{{\bf b)} \includegraphics[width=7.2cm, height=4.4cm]{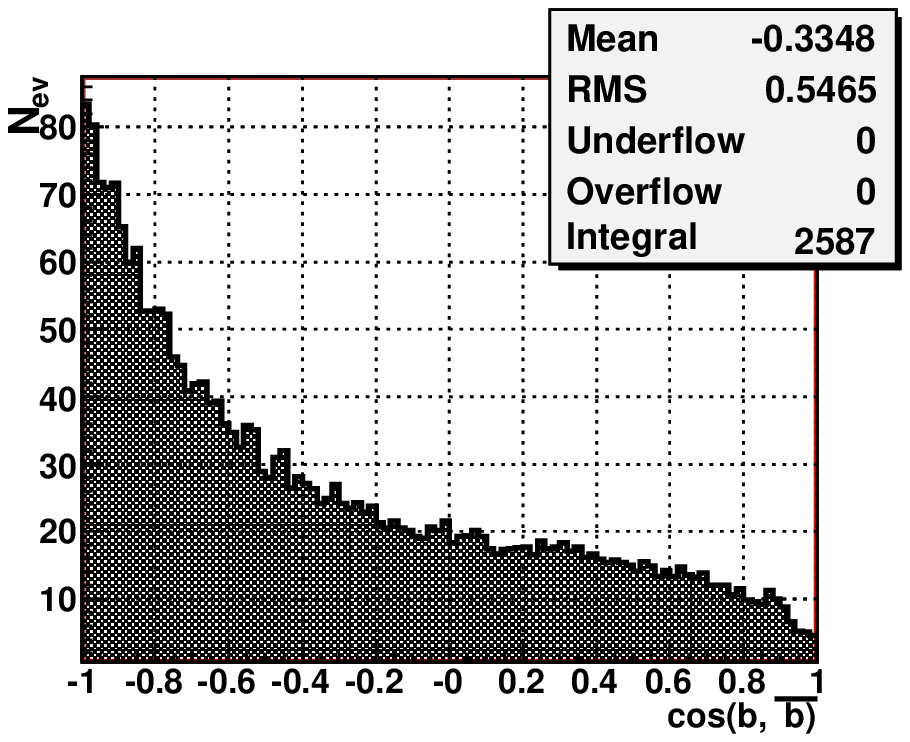}} \\
    \end{tabular}
     \caption{\small \it The spectra of $cos(b,\bar b)$ for stop 
                         production events.
	   {\bf a)} $"+-"$  and $"-+"$ polarizations,
           {\bf b)} $"++"$ and $"--"$  polarizations.}	 
     \end{center}  
\vskip -0.5cm              
     \end{figure}       
       As the next step,  we take into account  $b$-quark
   hadronization into a $b$-jet.   Fig.19 
  \begin{figure}[!ht]
     \begin{center}
    \begin{tabular}{cc}
\mbox{{\bf a)}\includegraphics[width=7.2cm, height=4.4cm]{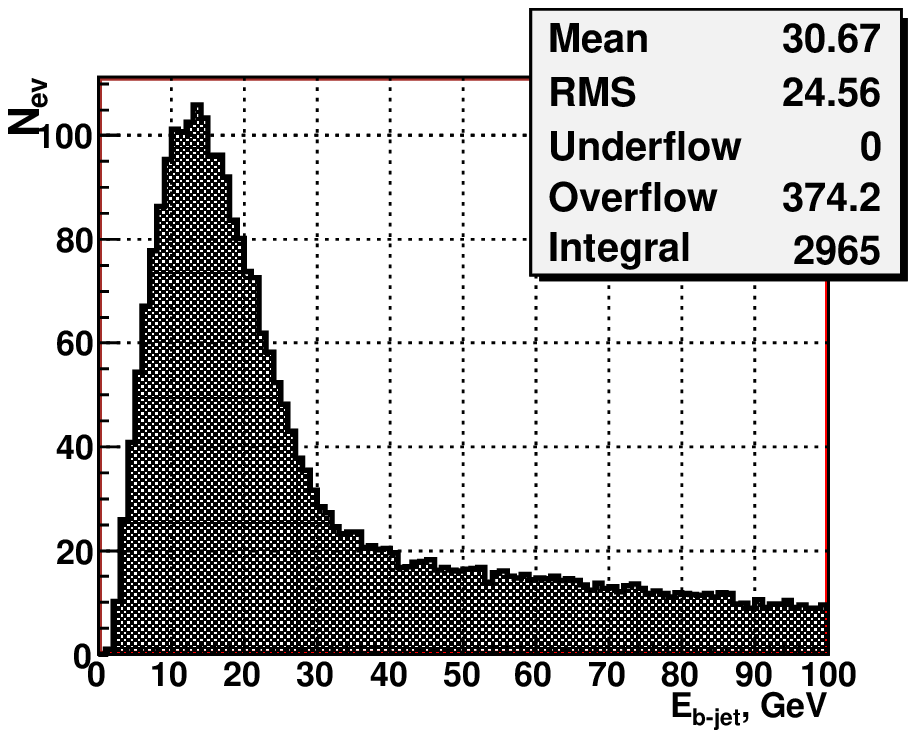}} 
\mbox{{\bf b)}\includegraphics[width=7.2cm, height=4.4cm]{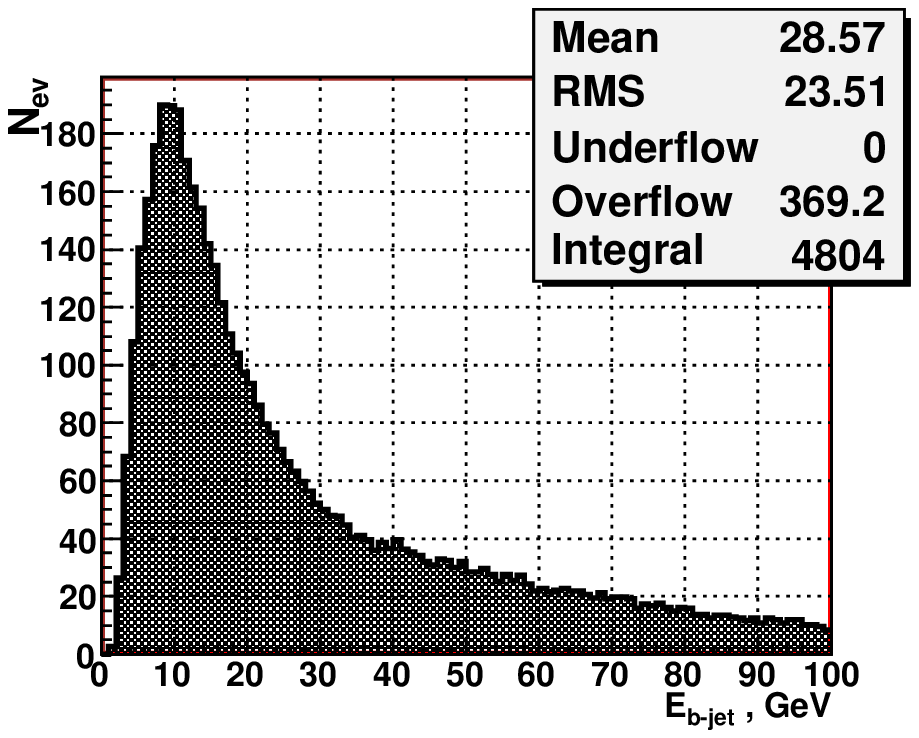}} \\
\mbox{{\bf c)}\includegraphics[width=7.2cm, height=4.4cm]{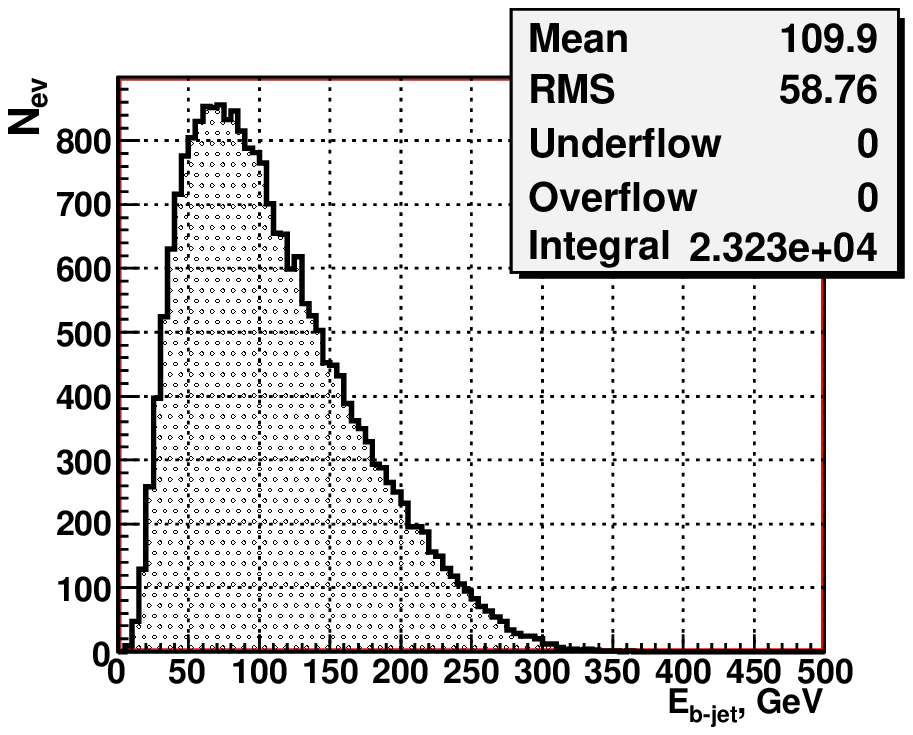}} 
\mbox{{\bf d)}\includegraphics[width=7.2cm, height=4.4cm]{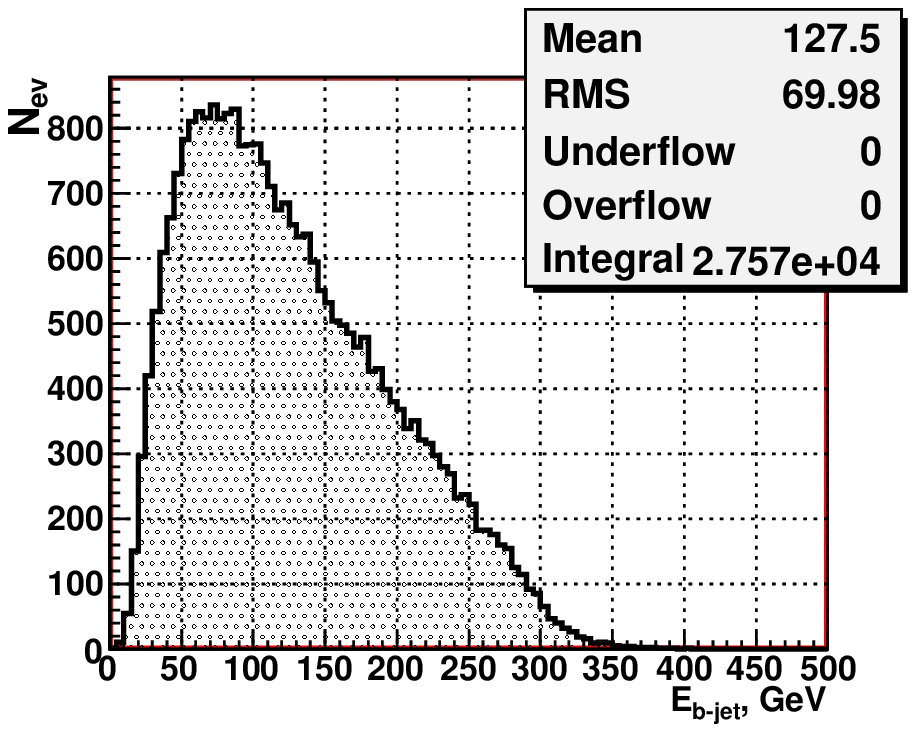}} \\            
    \end{tabular}
     \caption{\small \it  b-jet energy spectra. 
              {\bf a)} and {\bf b)} are for stop par production;
	      {\bf c)} and {\bf d)} are for top par production.
	  {\bf a)} and {\bf c)} $"+-"$  and $"-+"$ polarizations,
     {\bf b)} and {\bf d)} $"++"$ and $"--"$  polarizations.}    
   \end{center} 
\vskip -0.5cm                
     \end{figure}   
      shows the energy
   $E_{b-jet}$ distributions  of the corresponding $b$-jets.
  Plots  {\bf a)} and {\bf b)} of Fig.19 are for  stop production
   and plots {\bf c)} and {\bf d)} are for top production. 
These and the following plots for jets are obtained using the distance measure of the "Durham algorithm" implemented in the PYCLUS jet finder of PYTHIA.  
 Technically, $b$-jets are defined as jets
  that contain at least  one B-hadron. Their decay 
may be identified by the presence of a secondary vertex \cite{Haw}. 
   
Comparing the upper plots {\bf a)} and {\bf b)}  of Fig.19 for the   $b$- and  $\bar b$- jet energy distributions in  stop 
production with  the upper plots {\bf a)} and {\bf b)} of Fig.15 
   for the $b$-quark and $\bar b$-quark energy distributions,
   one observes  the appearance of long tails at higher energies. One sees that the end points of the energy   distributions
   for the $b$-jets and  $\bar b$-jets  are  higher than  those 
   for the corresponding  quarks. Furthermore, 
   the corresponding mean values of the jet 
   energies  $E_{b-jet}$  in  Fig.19 are 
   about 15 GeV  higher than those  in Fig.15.
It is interesting to note that the  peak positions of the energy distributions   $E_{b-jet}$ in the stop  case
  shown in the plots {\bf a)} and {\bf b)} of Fig.19 for the 
   $b$- and  $\bar b$- jets  practically coincide 
   with those shown in the plots {\bf a)} and {\bf b)} of
   Fig.16 for $b$-quarks.
  
   At the same time, due to the different 
   kinematics in  top production,
   the mean values of the $b$-jet and $\bar b$-jet  
   energy distributions  $E_{b-jet}$ in the  top case 
   are  only by about 2 GeV smaller  than the mean 
   values of the  corresponding $b$-quark
   and $\bar b$-quark energy
   distributions (see plots {\bf c)} and {\bf d)} in Fig.15 
   and Fig.19).
 
   Figure 20 shows the transverse momentum  $PT_{b-{jet}}$
    distributions  of the $b$-jets and $\bar b$-jets in 
    stop production (plots {\bf a)} and {\bf b)}) and  top
    production (plots {\bf c)} and {\bf d)}). By comparing with 
    Fig.16 for the corresponding PT distributions at the 
    $b$-quark level we can see, in analogy with 
    the  energy distributions,  long tails at
    high PT  which increase  the mean 
     PT values for stop production  by about 15 GeV 
    for  $ "+-"$ and $"-+"$   polarizations  and  by
    about 13 GeV for   $ "++" $ and $"--"$ polarizations. 
    Note that the peak positions of the PT distributions
shown in the plots {\bf a)} and {\bf b)} at the  jet level practically  do not differ from the positions of the peaks at the quark level 
(see plots {\bf a)} and {\bf b)} of  Fig.16).
      
     \begin{figure}[!ht]
    \begin{center}
    \begin{tabular}{cc}
     \mbox{{\bf a)}\includegraphics[width=7.2cm, height=4.4cm]{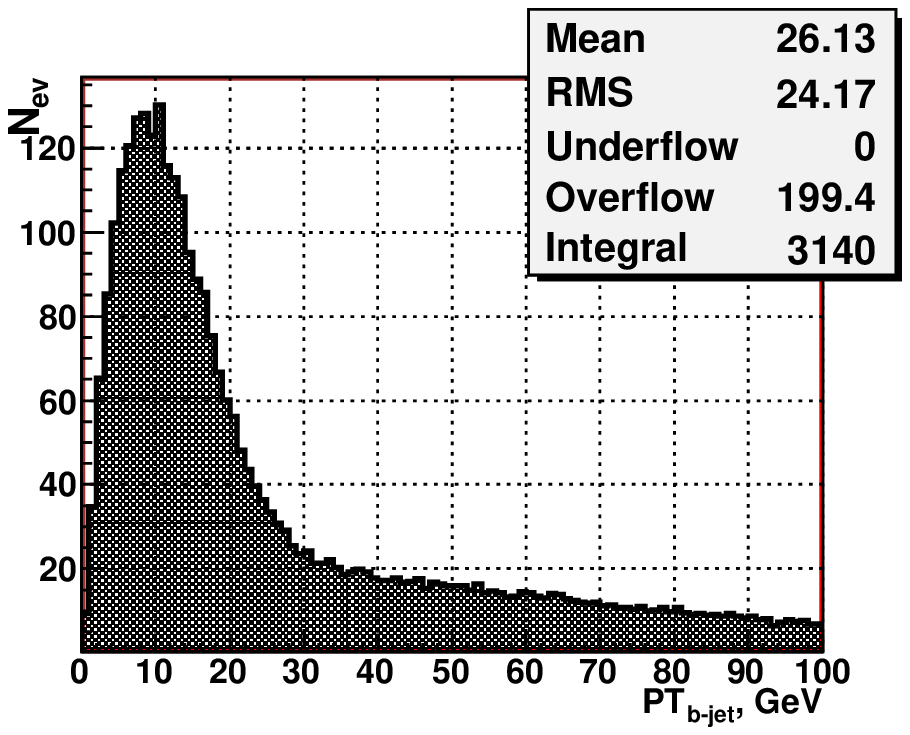}}           
     \mbox{{\bf b)}\includegraphics[width=7.2cm, height=4.4cm]{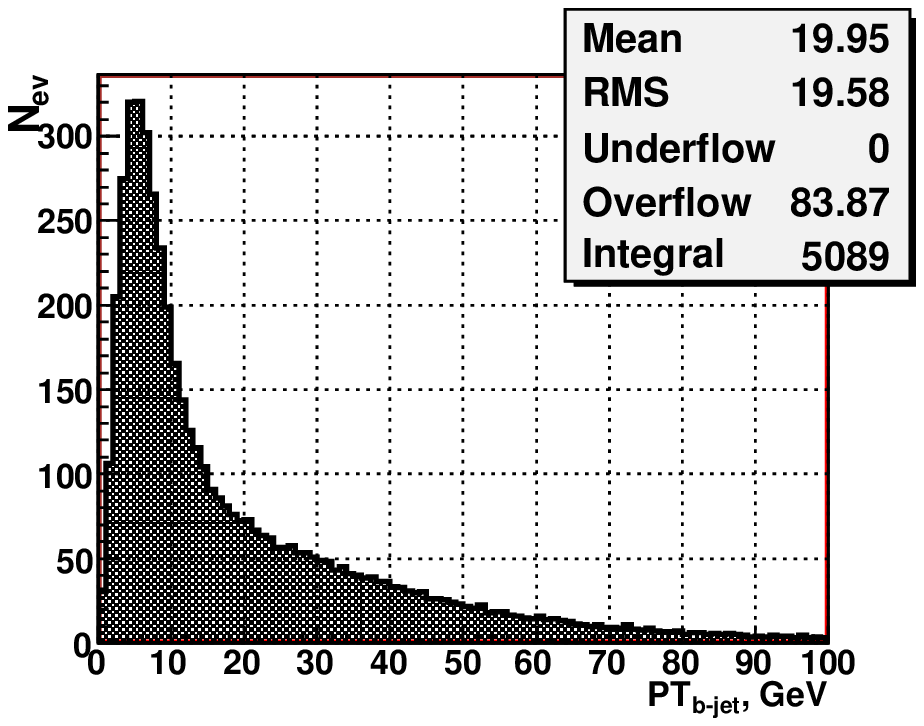}} \\
     \mbox{{\bf c)}\includegraphics[width=7.2cm, height=4.4cm]{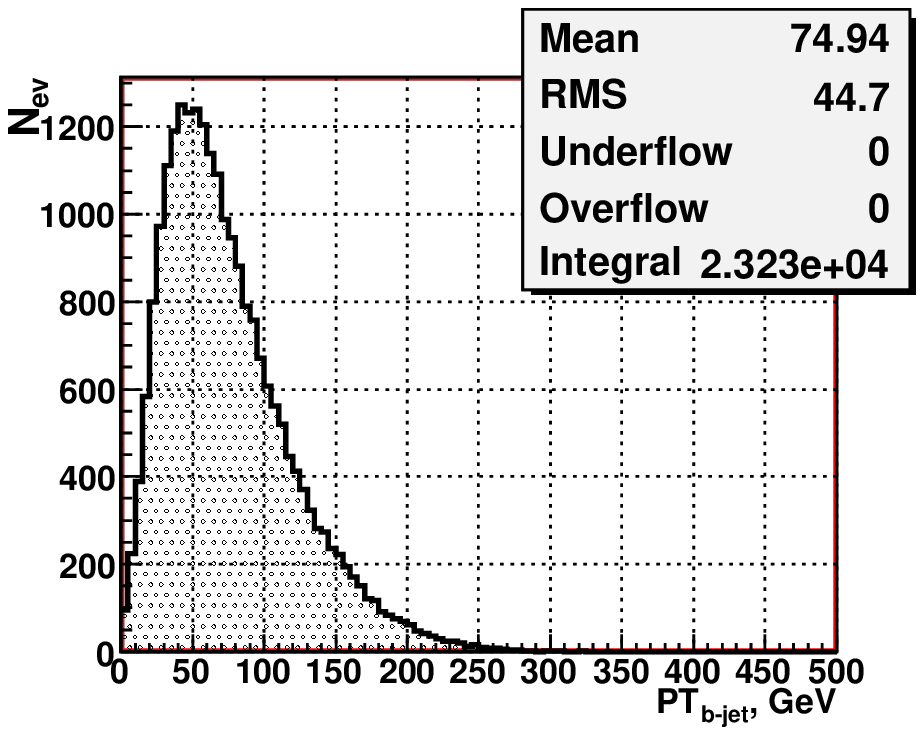}}      
     \mbox{{\bf d)}\includegraphics[width=7.2cm, height=4.4cm]{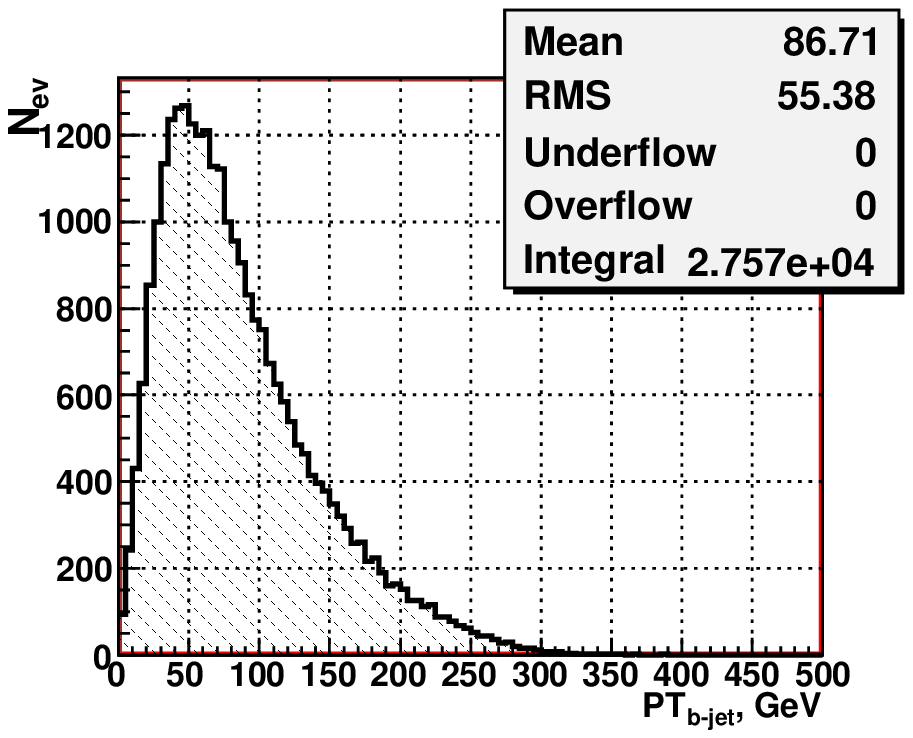}} \\          
     \end{tabular}
     \caption{\small \it   b-jet  PT spectra.  
             {\bf a)} and {\bf b)} are for stop pair production; 
	      {\bf c)} and {\bf d)} are for top pair production. 
	  {\bf a)} and {\bf c)} $"+-"$  and $"-+"$ polarizations,
{\bf b)} and {\bf d)} $"++"$ and $"--"$  polarizations.}         
     \end{center} 
\vskip -0.5cm                
     \end{figure}
      
It is also seen that  in the case of  top production the mean values of the  PT distributions of the $b$-jets  practically do
not differ from the analogous ones shown in the plots {\bf c)} and {\bf d)}  of Fig.16  for  $b$-quarks.

  Let us summarize the results which were
   obtained  in subsections  4.2 and 4.3
   by the use of PYCLUS jet finder. First, it was 
   found that in  the  case of top background
   production the characteristic parameters of
   energy and transverse momentum distributions
   of jets stemming from  W decay and of $b$-jets,
   produced in $b$-quark hadronization, 
   practically do not differ from the
   parameters  of their parent quarks   distributions.

   This picture changes quite noticeably
   when we consider  the  case of stop production 
   with its further decay through the channel
   $\tilde t_{1} \to b \tilde \chi_{1}^{\pm}$. 
   In this case the $b$-quarks
   are essentially less energetic than the
   b-quarks produced in top decay   $t \to bW^{\pm}$. 
   The simulation has shown that the
   usage of the same PYCLUS jet finder 
   in stop case leads to a noticeable
   redistribution of the jet energies 
    between "W-jets" and  $b$-jets
   and, correspondingly, of jet transverse 
   momenta. Namely, in the stop case the mean values of the jet
   energy $E_{jet_W{*}}$ and jet transverse momentum 
   $PT_{jet_{W^{*}}}$ are  about 12-25 GeV 
  smaller than the  energy $E_{W-quark}$ and transverse momentum
   $PT_{W-quark}$  of parent  "$W$-quarks"
   stemming from W boson decay.  On the contrary, the mean
   values of the $b$-jet energy $E_{b-jet}$ and the jet 
   transverse momentum $PT_{b-{jet}}$ are 
   about 5-15 GeV higher than the energy
   $E_{b}$ and $PT_{b}$  of the parent $b$-quarks.
   
It is worth emphasizing that  the positions of the  peak of the energy and transverse  momentum distributions are stable when going from the quark to the jet level. 
   In the following we shall return to this
   subject and consider the set of physical
   variables which will take into account
    the  effect of energy redistribution  in the 
   case of stop production.
              
    \subsection{ Distributions of the signal muons.}

  ~~~~ To select the signal stop pair production events,  see
    the left plot of Fig.1,  one has to identify the
   muon from the W decay. The corresponding energy $E_{sig-mu}$ and transverse  momentum $PT_{sig-mu}$
  distributions of the signal muons are shown  in 
   Fig.21 for both polarization combinations.
     
  \begin{figure}[!ht]
     \begin{center}
\vskip -0.5cm         
    \begin{tabular}{cc}
    \mbox{{\bf a)}\includegraphics[width=7.2cm, height=4.4cm]{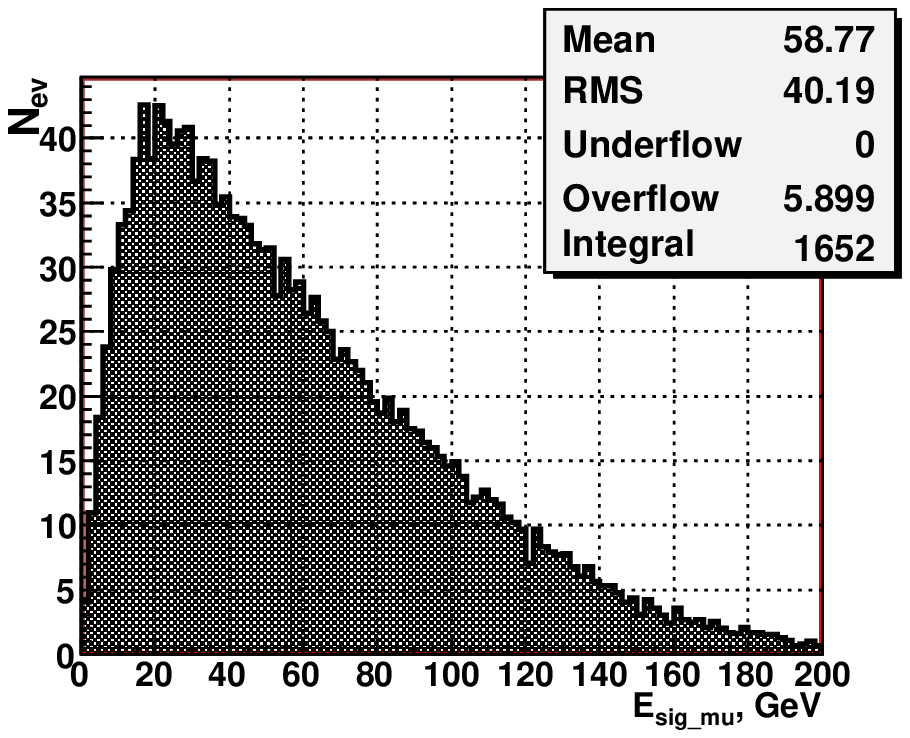}}      
    \mbox{{\bf b)}\includegraphics[width=7.2cm, height=4.4cm]{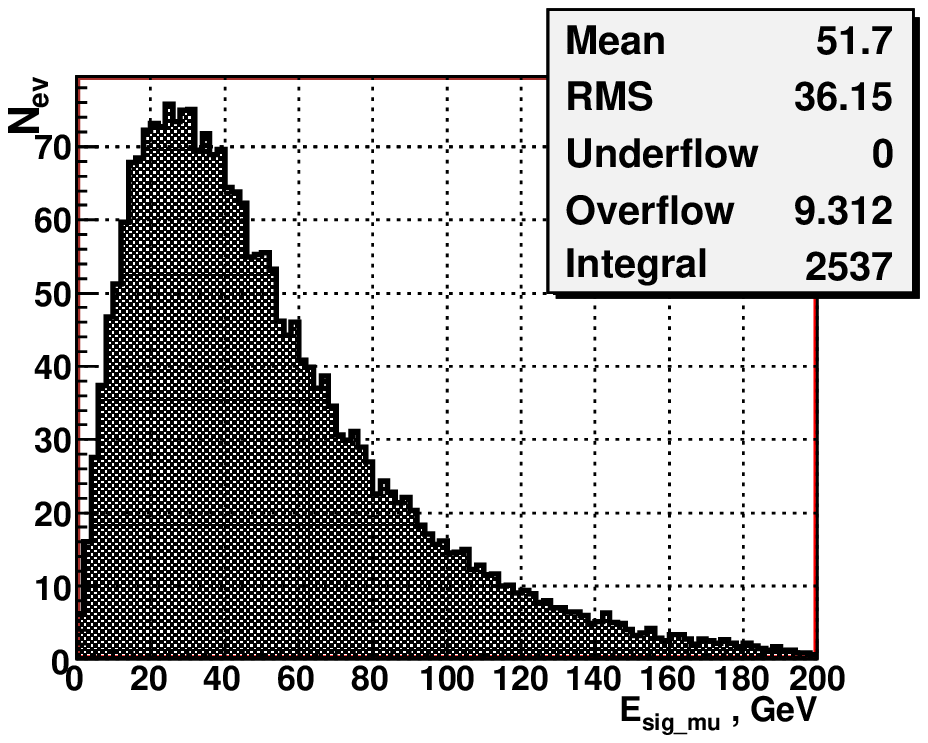}} \\
    \mbox{{\bf c)}\includegraphics[width=7.2cm, height=4.4cm]{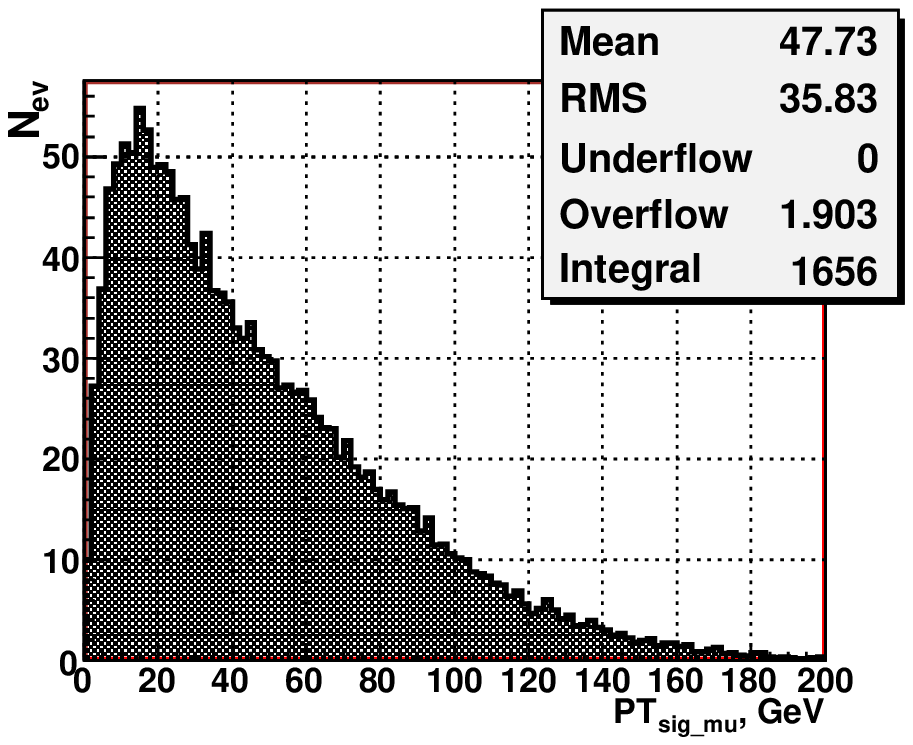}}      
    \mbox{{\bf d)}\includegraphics[width=7.2cm, height=4.4cm]{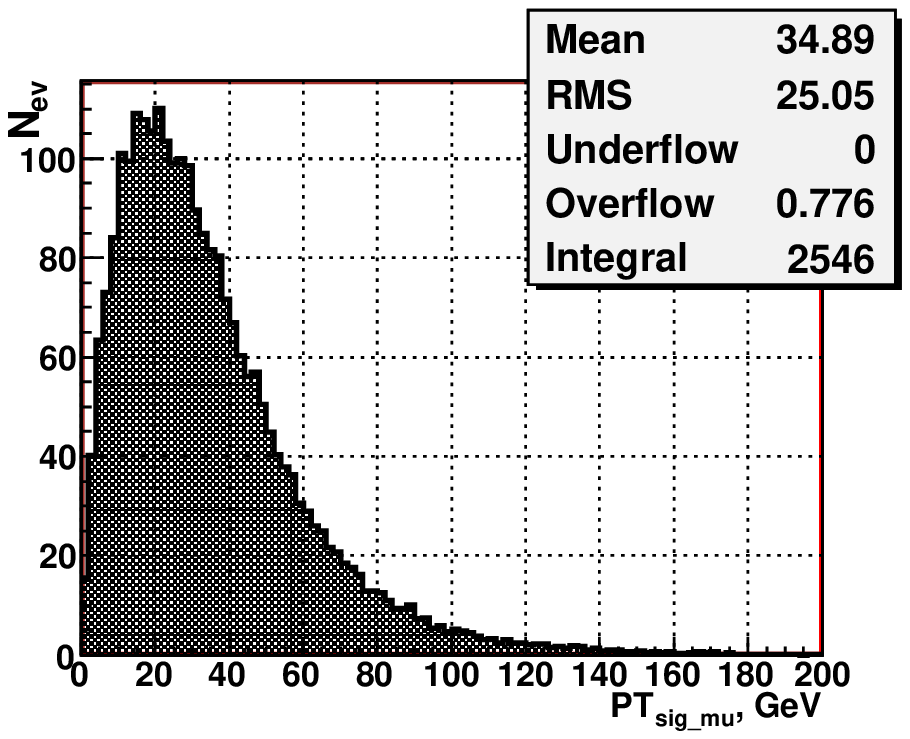}} \\    
    \end{tabular}
     \caption{\small \it Energy and PT distributions 
     of signal muons. 
           {\bf a)} and {\bf c)} $"+-"$  and $"-+"$ polarizations,
           {\bf b)} and {\bf d)} $"++"$  and $"--"$  polarizations.}     
     \end{center} 
\vskip -0.5cm                
     \end{figure}

    There are, however, also muons in the event coming from
   leptonic and semileptonic decays of hadrons. 
   The left and right plots of  Fig.22
   show the energy $E_{dec-mu}$ and the transverse momentum
   $PT_{dec-mu}$ spectra of these muons  stemming from hadron 
   decays within the detector volume, for which we took the size 
    parameters from  \cite{ILCRDR1}, \cite{ILCRDR2}.  
 It can be  seen that the decay muons have a rather small energy 
   $E_{dec-mu}$ and transverse momentum $PT_{dec-mu}$.
     Their mean values are about $0.85$ and $0.59$ GeV,
   respectively. The analogous spectra for the signal muons 
   in Fig.21 show that the  signal muons 
   have a much higher energy   $E_{sig-mu}$ and 
   transverse momentum  $PT_{sig-mu}$.  The 
   mean value of the signal muons  
    energy $E^{mean}_{sig-mu} = 47.6$ GeV 
   is about  60  times higher than the 
    mean value of the energy
   of the decay muons. An analogous
   difference can be seen between the 
   mean values of transverse momenta
   PT of signal and decay muons.
   Therefore one can cut off
   most low--energy decay muons rejecting those
   with  $E_{mu} \le 6$ GeV. Such
   a cut leads to a loss of about  $2\%$ of
   signal events as seen from the  Fig.21
   (the bin width in this plot is 2 GeV).

  \begin{figure}[!ht]
     \begin{center}
    \begin{tabular}{ccc}
     \mbox{{\bf a)}\includegraphics[width=7.2cm, height=4.4cm]{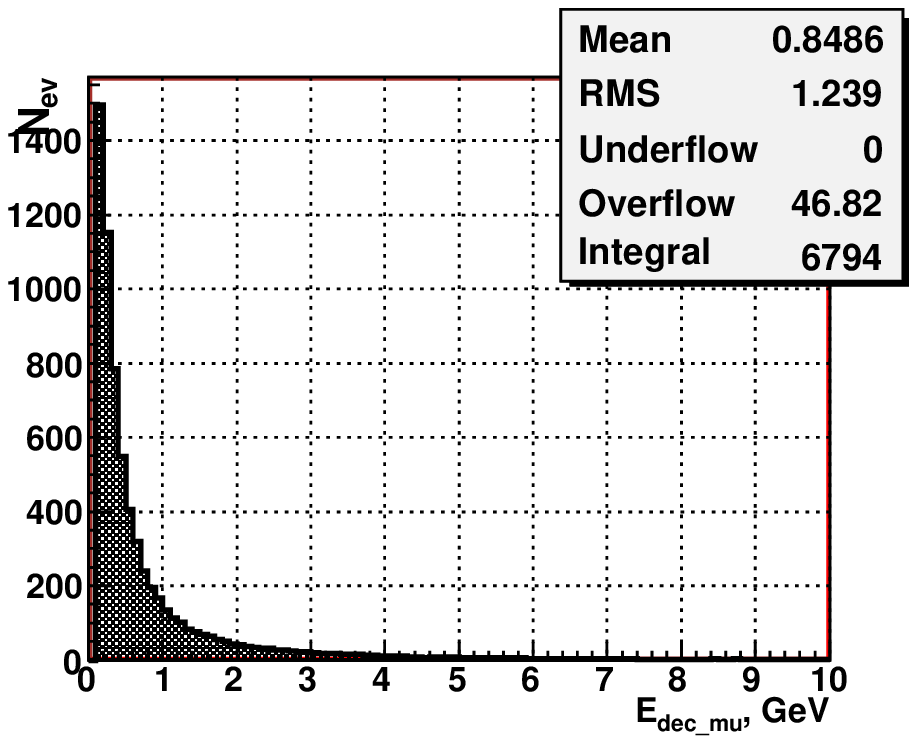}}      
     \mbox{{\bf b)}\includegraphics[width=7.2cm, height=4.4cm]{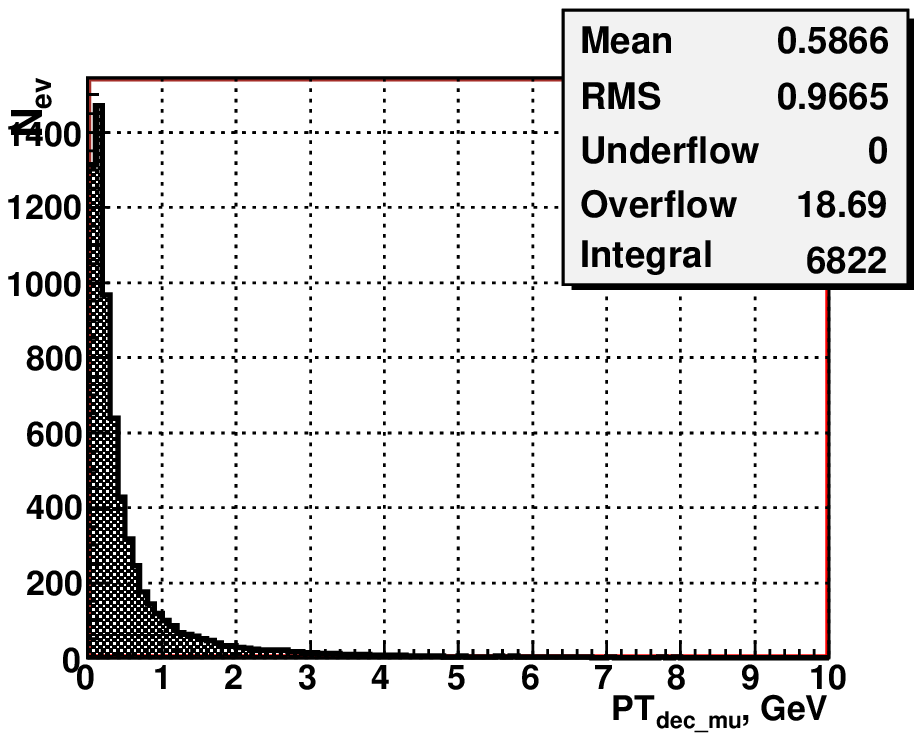}} \\
    \end{tabular}
     \caption{\small \it 
     Distributions  muons from leptonic and semileptonic decays. 
     {\bf a)} Energy distribution; {\bf b)} PT distribution.}    
     \end{center}  
\vskip -0.5cm               
     \end{figure}
      
  We have also studied another way to select the signal muon from W decay.
   If the axes of all four jets in the event are known,
  then in general the signal muon has the largest transverse momentum
  with respect to any of these jet axes.

%
\section{ Some global variables.}
%

~~~  In stop pair production the  two  neutralinos and 
  the energetic neutrino from the W boson decay escape detection. 
  The simulation with PYTHIA6 allows us to estimate the 
  missing energy and the missing transverse momenta that are
  carried away by these
  particles. We also take into account the non-instrumented
  region around the beam pipe given by the polar angle 
  intervals  $\Theta < 7^o$ and  $\Theta > 173^o$.
   
     \begin{figure}[!ht]
     \begin{center}
    \begin{tabular}{cc}   
     \mbox{{\bf a)}\includegraphics[width=7.2cm, height=4.4cm]{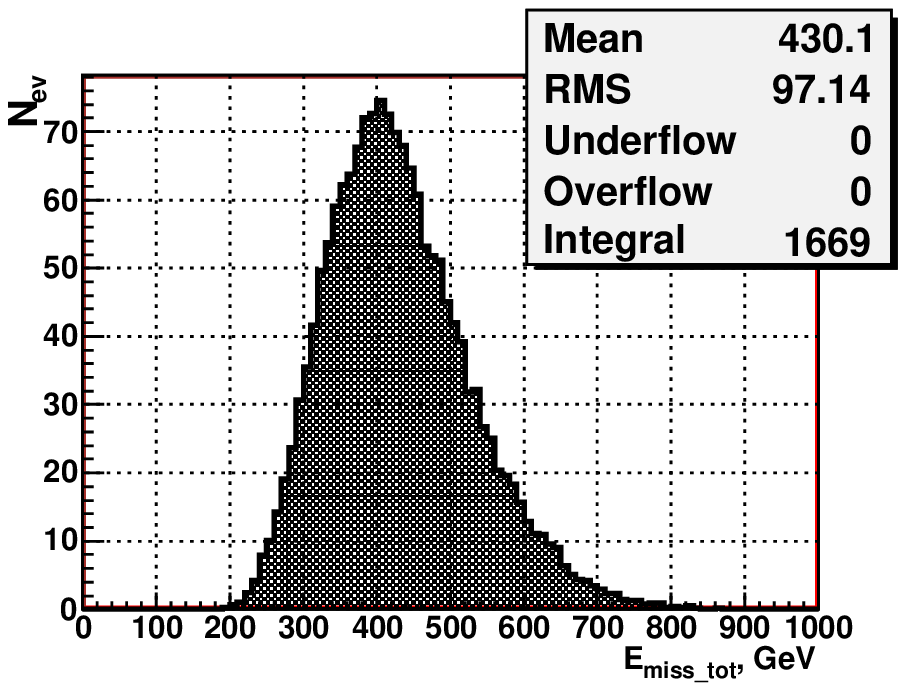}}      
     \mbox{{\bf b)}\includegraphics[width=7.2cm,  height=4.4cm]{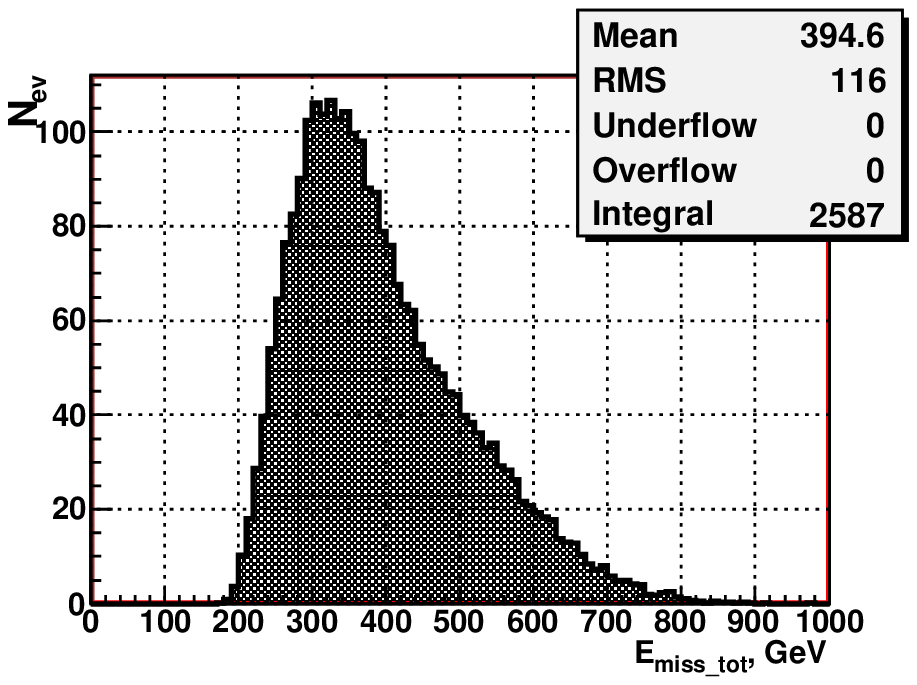}} \\
     \mbox{{\bf c)}\includegraphics[width=7.2cm,  height=4.4cm]{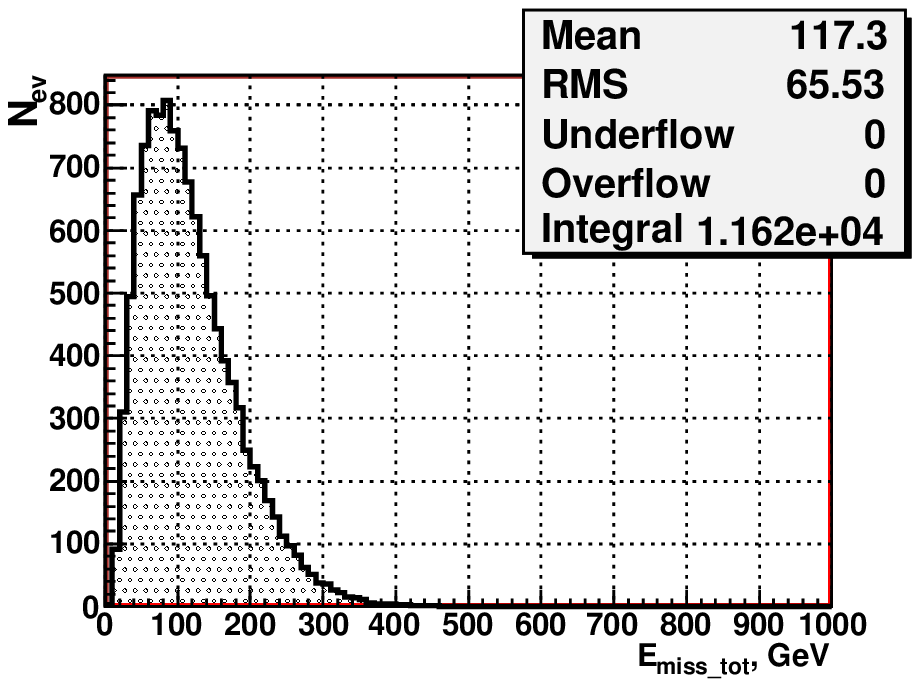}}      
     \mbox{{\bf d)}\includegraphics[width=7.2cm,  height=4.4cm]{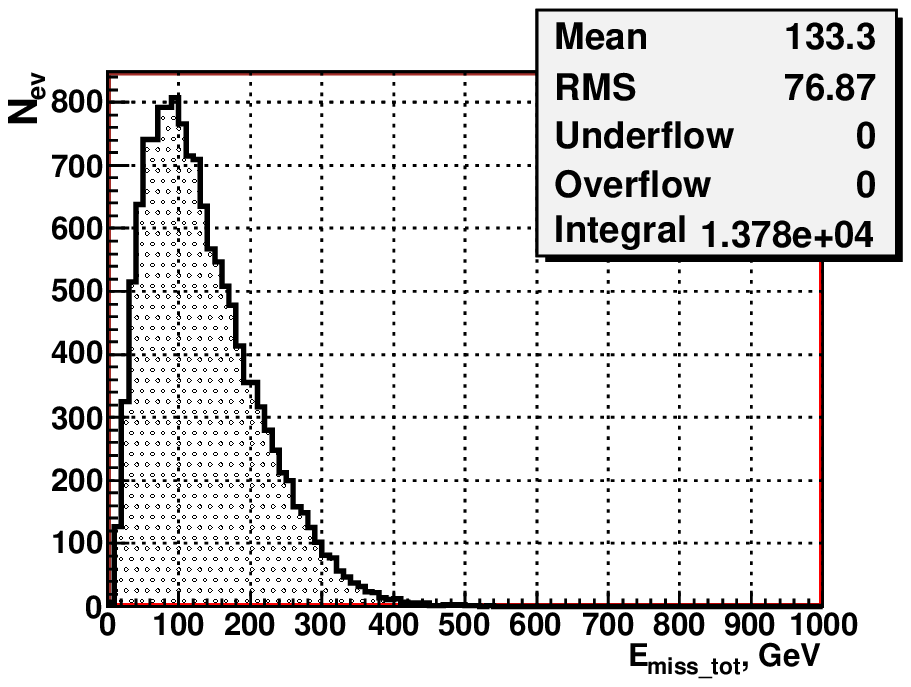}} \\         
    \end{tabular}
     \caption{\small \it Missing energy $E_{miss-tot}$   distribution. 
	      {\bf a)} and {\bf b)} are for stop pair production;
	      {\bf c)} and {\bf d)} are for top pair production. 
	  {\bf a)} and {\bf c)} $"+-"$  and $"-+"$ polarizations,
          {\bf b)} and {\bf d)} $"++"$ and $"--"$  polarizations.}             
    \end{center} 
   \vskip -0.5cm               
    \end{figure}
    
 The distributions of the  total missing energy for 
  stop production and top production are presented in
  the upper and lower plots of Fig.23, respectively. In 
  stop pair production,  see plots {\bf a)} and {\bf b)},
  the $E_{miss-tot}$ spectrum starts at 190-200 GeV and  at ends 800 GeV.
  In  top pair production (plots {\bf c)} and {\bf d)}), where
  the two neutralinos are not present, the missing energy
  $E_{miss-tot}$ is much smaller.   It starts from 
  $ \approx 10$ GeV and finishes at   $ \approx 380-420$ GeV. 
    
 Figure 24 shows the distributions of the total visible 
 energy in  event  $E_{vis-tot}$ in stop production 
 (plots {\bf a)} and {\bf b)}) and  in top production 
 (plots {\bf c)} and {\bf d)}). The large missing 
  energy in stop production (Fig.23) is related to
  the low visible energy (Fig.24), while in  top production 
  the low missing energy correlates with the large 
  visible energy. A cut on the total visible energy of 
  approximately  $E_{vis-tot} < 250 $ GeV
  \footnote{that is equivalent to setting a lower 
            limit for the missing energy.}
  would eliminate most of the top background while 
  approximately 10$\%$ of the signal events are lost.
  
    \begin{figure}[!ht]
     \begin{center}
    \begin{tabular}{cc}
     \mbox{{\bf a)}\includegraphics[width=7.2cm,  height=4.4cm]{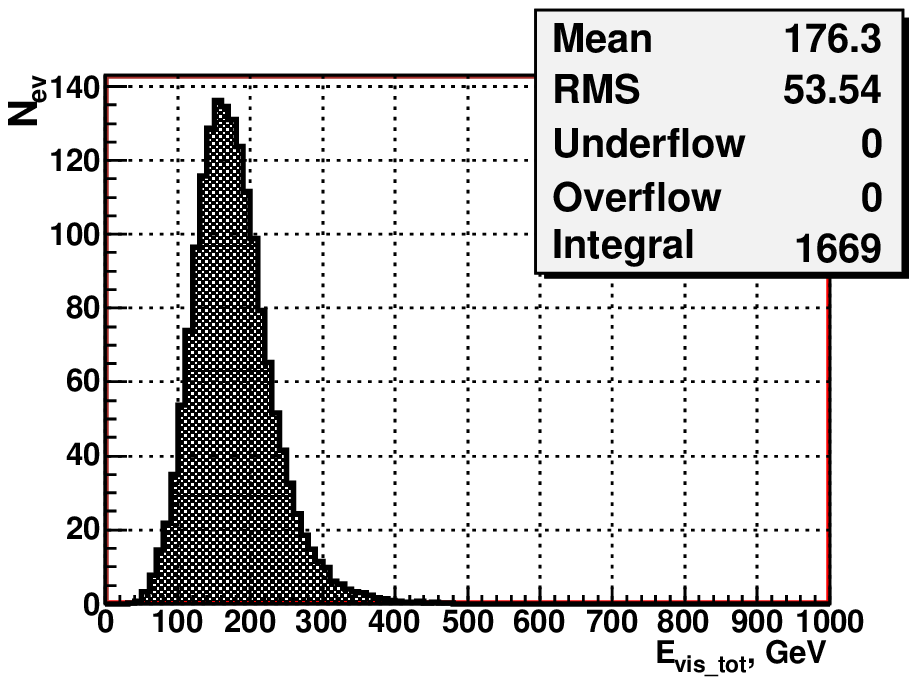}}      
     \mbox{{\bf b)}\includegraphics[width=7.2cm,  height=4.4cm]{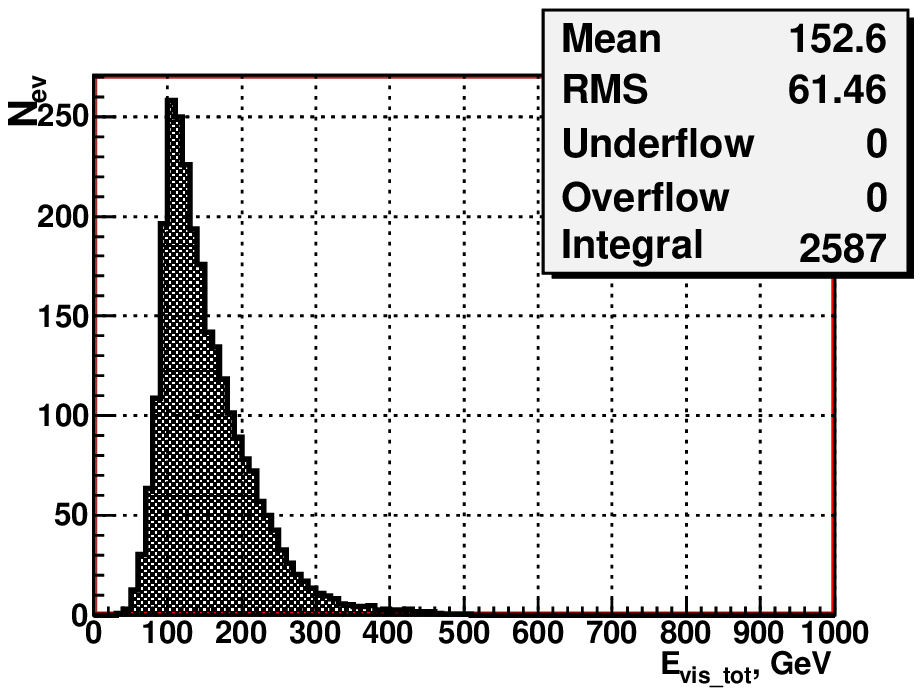}} \\
     \mbox{{\bf c)}\includegraphics[width=7.2cm,  height=4.4cm]{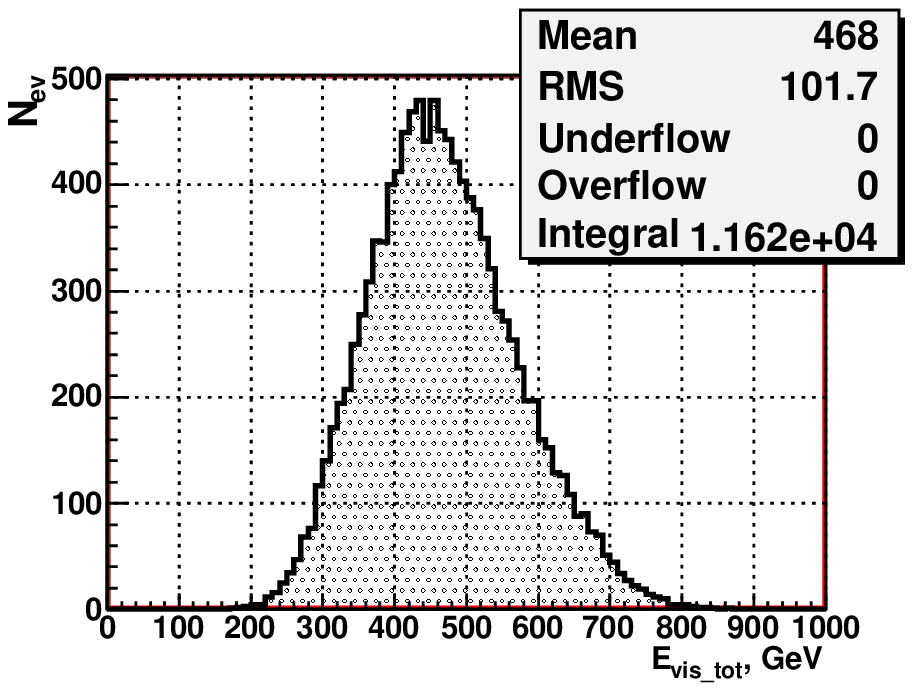}}      
     \mbox{{\bf d)}\includegraphics[width=7.2cm,  height=4.4cm]{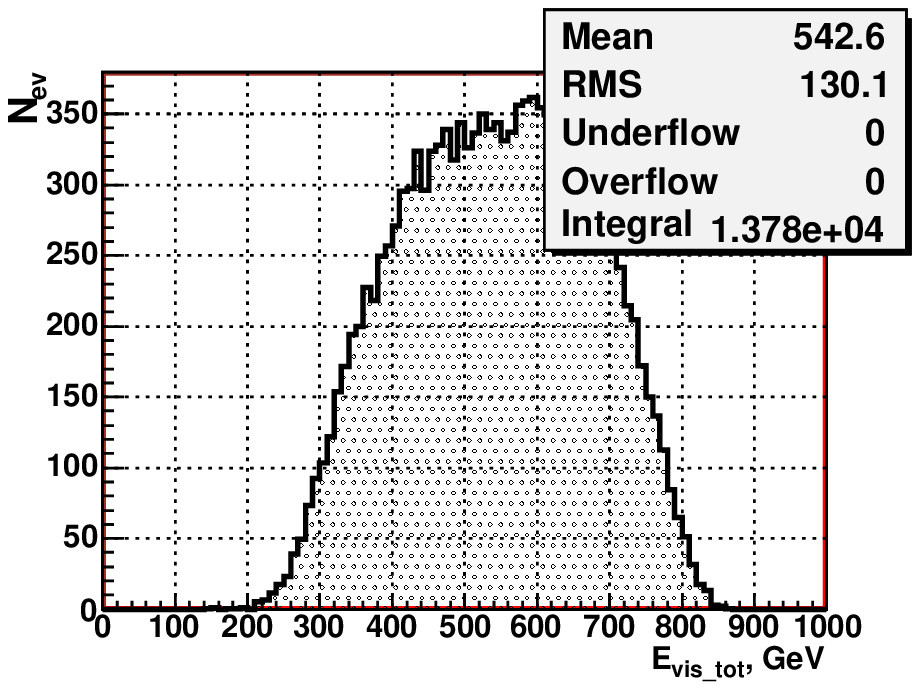}} \\          
    \end{tabular}
     \caption{\small \it Total energy $E_{vis-tot}$  distribution.
              {\bf a)} and {\bf b)} are for stop pair production;
	      {\bf c)} and {\bf d)} are for top pair production. 
	  {\bf a)} and {\bf c)} $"+-"$  and $"-+"$ polarizations,
          {\bf b)} and {\bf d)} $"++"$ and $"--"$  polarizations.} 
     \end{center}    
\vskip -0.5cm             
     \end{figure}
     
   Another useful observable is the scalar sum  of the 
   moduli of the transverse   momenta in an event 
  $PT_{scalsum} =\sum\nolimits_{i=1}^{N_{part_i}} |PT_{i}|$,
   where the sum goes over all  ($N^{part}$) detectable
   particles  (i)  in the 
   event.  Fig.25 shows the distributions of the scalar sum 
   of the  transverse momenta for
    stop production (plots {\bf a)} and {\bf b)})
   and for  top production (plots {\bf c)} and {\bf d)}). 
   It is seen that the restriction
   $PT_{scalsum} \leq 180$ GeV  would lead to a good
   separation of the stop signal events from the top background.

     \begin{figure}[!ht]
     \begin{center}
    \begin{tabular}{cc}
     \mbox{{\bf a)}\includegraphics[width=7.2cm,  height=4.4cm]{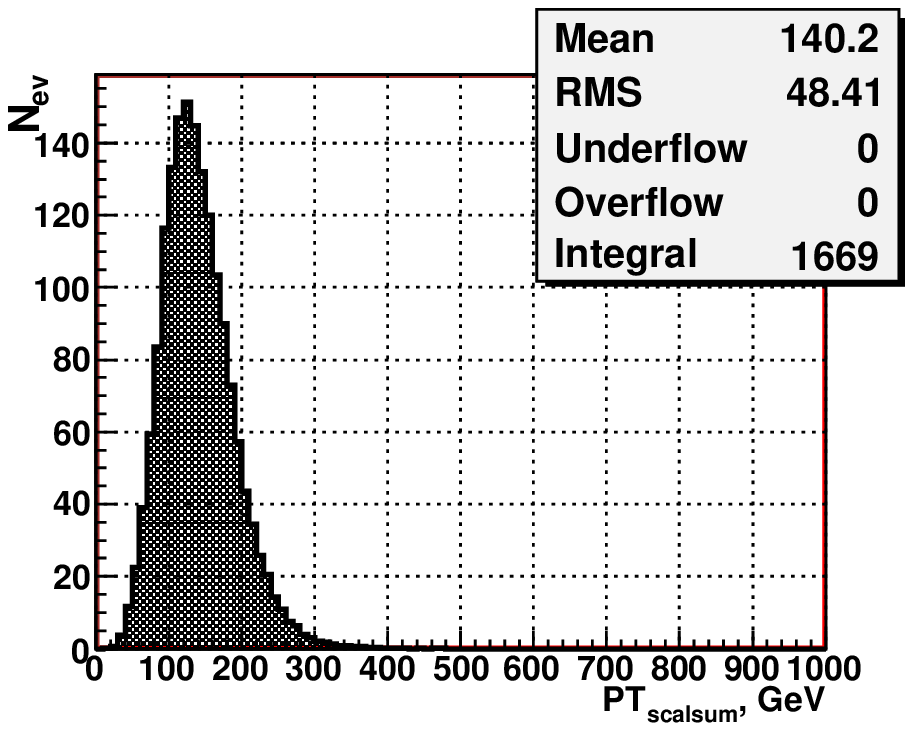}}      
     \mbox{{\bf b)}\includegraphics[width=7.2cm,  height=4.4cm]{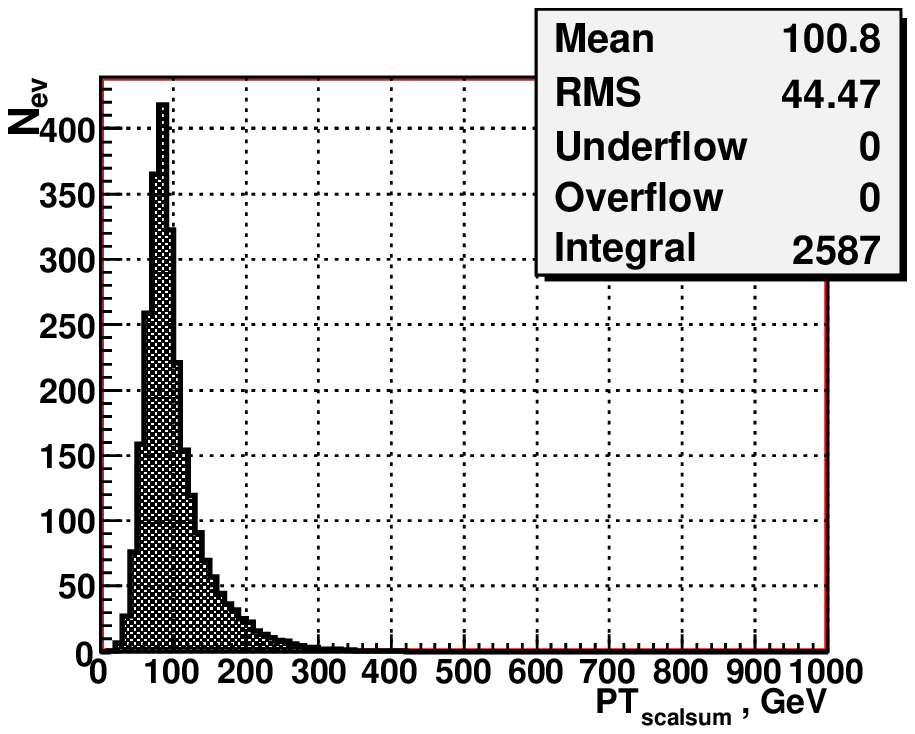}} \\
     \mbox{{\bf c)}\includegraphics[width=7.2cm,  height=4.4cm]{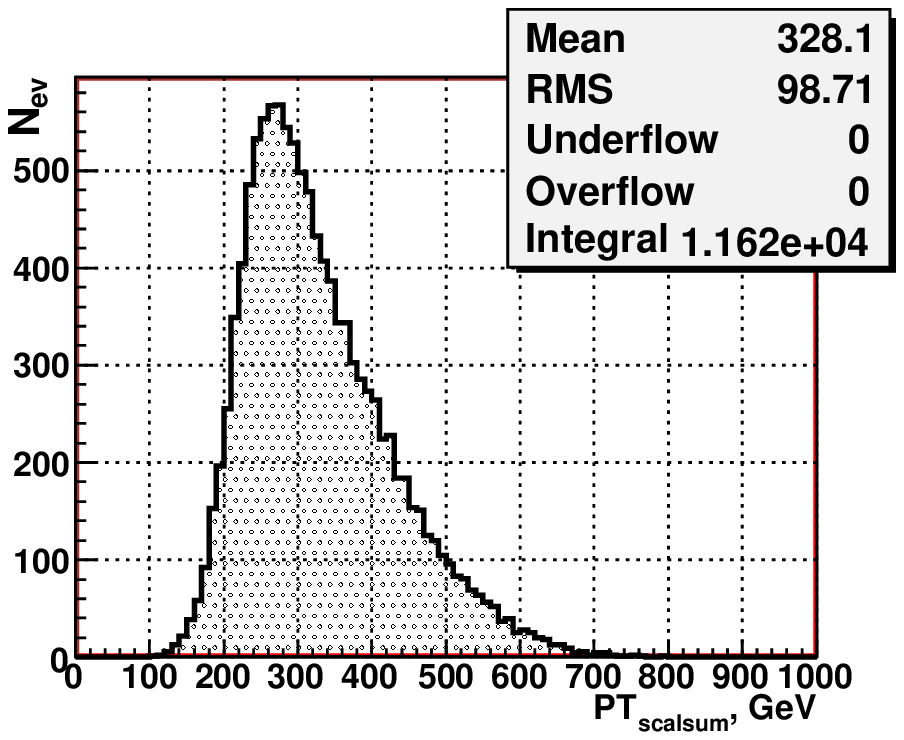}}      
     \mbox{{\bf d)}\includegraphics[width=7.2cm,  height=4.4cm]{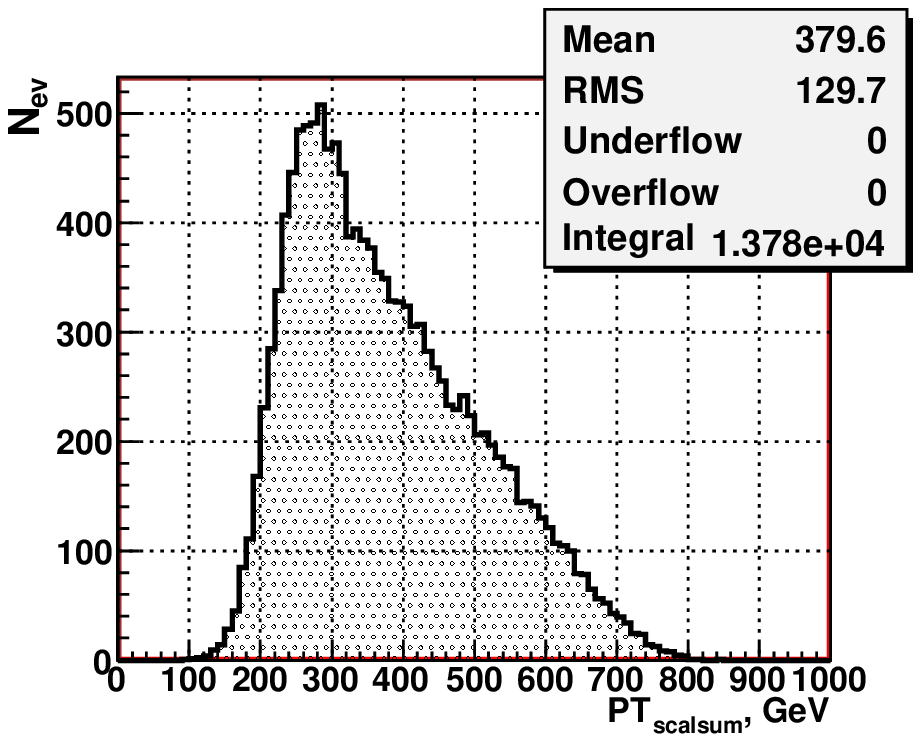}} \\     
    \end{tabular}
     \caption{\small \it $PT_{scalsum}$  distribution.
              {\bf a)} and {\bf b)} are for stop pair production;
	      {\bf c)} and {\bf d)} are for top pair production. 
	  {\bf a)} and {\bf c)} $"+-"$  and $"-+"$ polarizations,
          {\bf b)} and {\bf d)} $"++"$ and $"--"$  polarizations.} 
     \end{center}   
     \end{figure}
     
  We consider also the invariant mass 
 $M_{inv}(All jets, \mu)$  of the 
  system that contains all observable 
  objects in the final state. This invariant
  mass is the modulus of the  vectorial 
  sum  of the  4-momenta $P^{i}_{jet}$ 
  ($N^{jet}=4, i=1,2,3,4$) 
    of all  four  jets in an event plus the 4-momentum of the  
    signal  muon
     \begin{equation}  
         M_{inv}(All jets, \mu) = 
          \sqrt{(\Sigma_{i=1,2,3,4}P^{i}_{jet}+P_{\mu})^{2}}~.
     \end{equation}
     
  The distribution of this invariant mass is shown in Fig.26. 
  Plots {\bf a)} and {\bf b)} show the results  for   stop pair 
  production  while the plots {\bf c)} and {\bf d)} are for 
  top pair production.  As seen from these plots, the cut 
   $M_{inv}(All jets, \mu) \le 230$ GeV will give a 
 good separation of signal stop and top background events.
  
  \begin{figure}[!ht]
    \begin{center}
    \vskip -0.5cm      
    \begin{tabular}{cc}
    \mbox{{\bf a)}\includegraphics[width=7.2cm,  height=4.4cm]{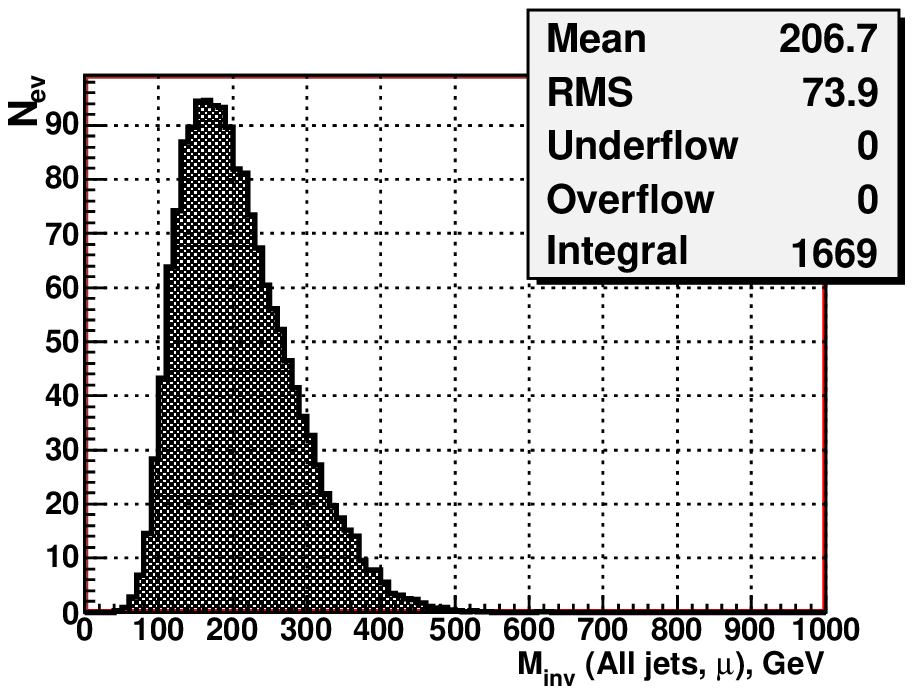}}      
    \mbox{{\bf b)}\includegraphics[width=7.2cm,  height=4.4cm]{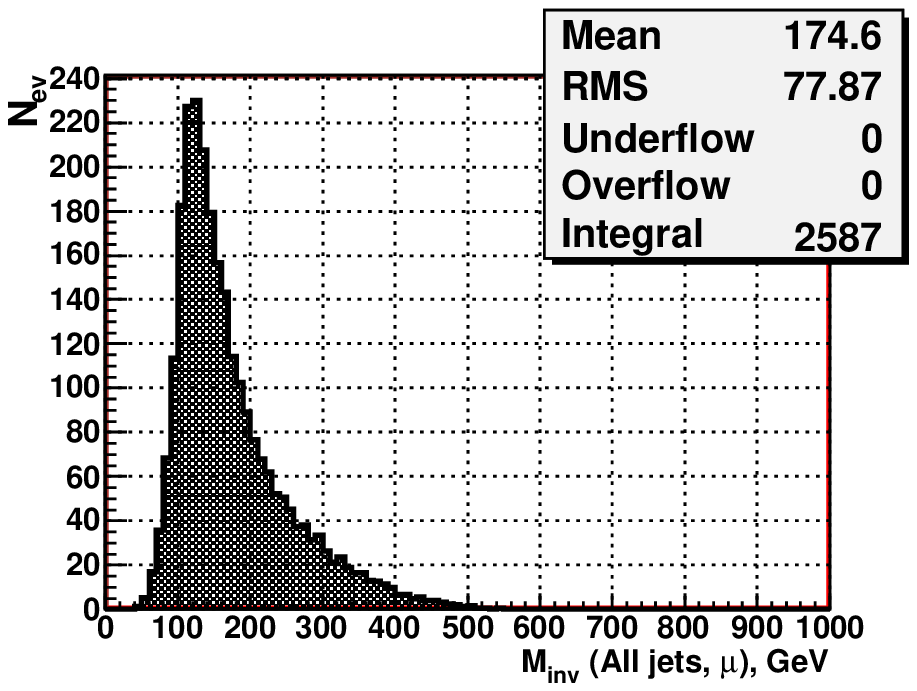}} \\
     \mbox{{\bf c)}\includegraphics[width=7.2cm,  height=4.4cm]{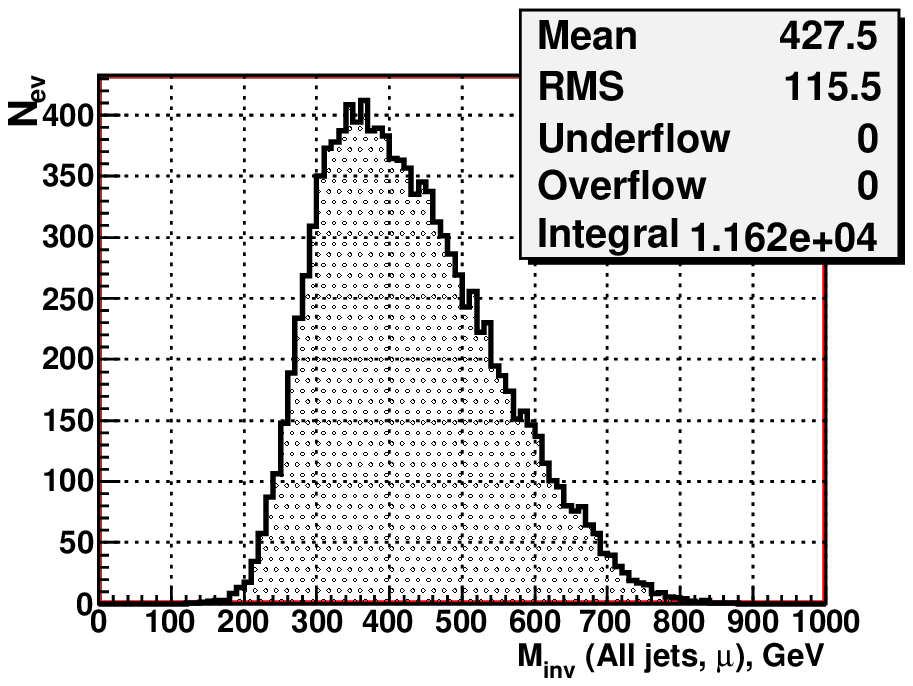}}      
     \mbox{{\bf d)}\includegraphics[width=7.2cm,  height=4.4cm]{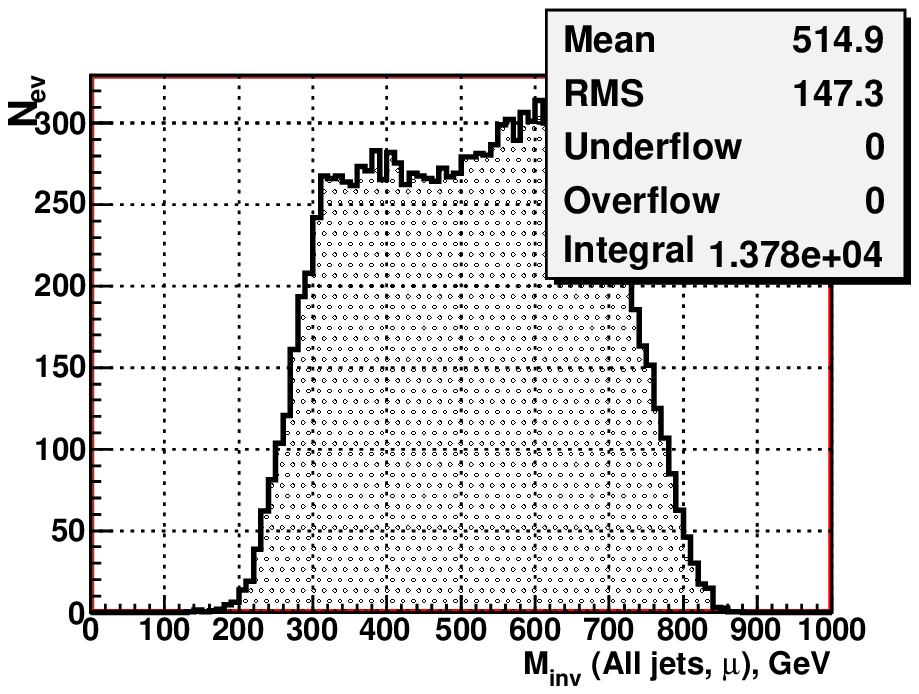}} \\
     \end{tabular}
     \caption{\small \it Distribution of number of events versus 
                                the  reconstructed invariant mass of
			       all jets and signal muon $M_{inv}(All jets, \mu)$.
       	       {\bf a)} and {\bf b)} are for stop production;
               {\bf c)} and {\bf d)} are for top case.
               {\bf a)} and {\bf c)} $"+-"$  and $"-+"$ polarizations,
               {\bf b)} and {\bf d)} $"++"$ and $"--"$  polarizations.}	   
   \vskip -0.5cm 	  
    \end{center}        
 \end{figure}
 
     Another  variable that can also be used 
   for the separation of the signal  and  
   the background is the "missing" mass   $ M_{missing}$ 
   (for $\sqrt {s}=\sqrt {s_{ee}}=1000$ GeV)
 \begin{equation}  
   M_{missing} = 
  \sqrt{ (\sqrt {s} -
   (\sum\nolimits_{i=1}^{N^{jet}}{E^{i}_{jet}+
            E_{\mu}))^{2}} -
   (\sum\nolimits_{i=1}^{N^{jet}}{{\bf P}^{i}_{jet}+
            \bf P_{\mu})^{2}}}   
 \end{equation} 

   This variable takes  into account  the 
   contribution of  those particles that 
   cannot be registered in the detector (neutrinos and neutralinos). 
   The distributions of this invariant
   "missing" mass are given in Fig.27.
   Plot  {\bf a)} and {\bf b)}  show the results  for   
   stop pair  production,  while plots {\bf c)} and {\bf d)}
   are for top pair production.  As seen    from these plots, the 
   cut  $M_{missing} \ge 700$ GeV also allows us to get rid of most
of the background.

\begin{figure}[!ht]
      \begin{center}
    \begin{tabular}{cc}
      \mbox{{\bf a)}\includegraphics[width=7.2cm,  height=4.4cm]{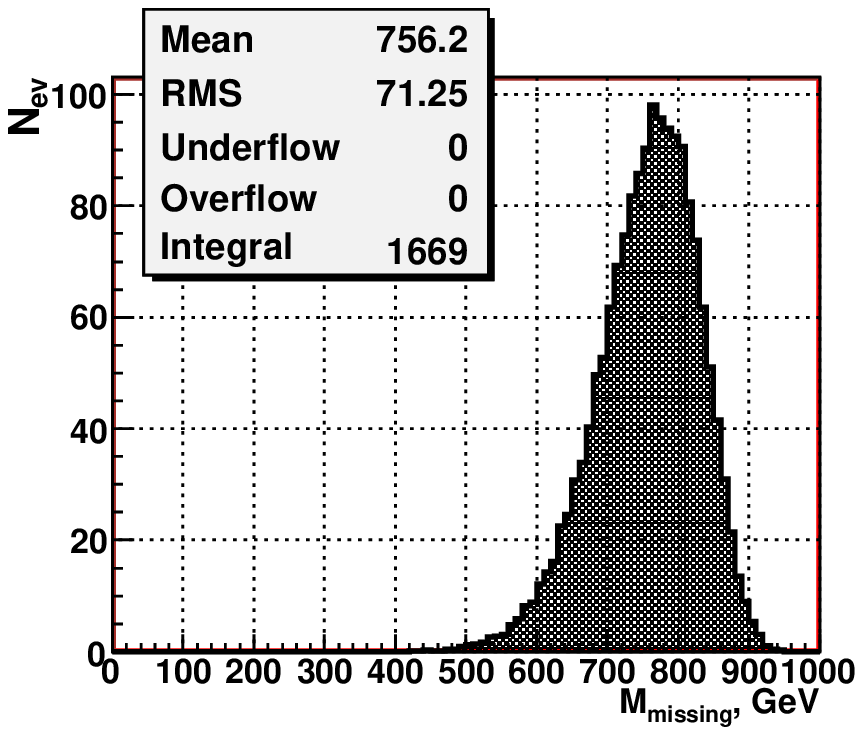}}      
      \mbox{{\bf b)}\includegraphics[width=7.2cm,  height=4.4cm]{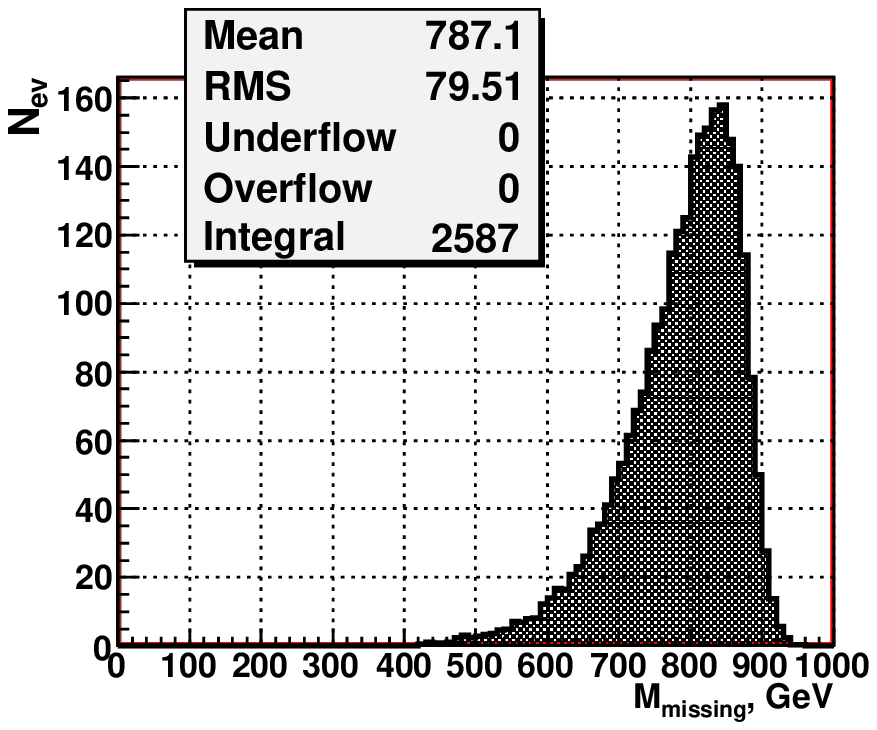}} \\
      \mbox{{\bf c)}\includegraphics[width=7.2cm,  height=4.4cm]{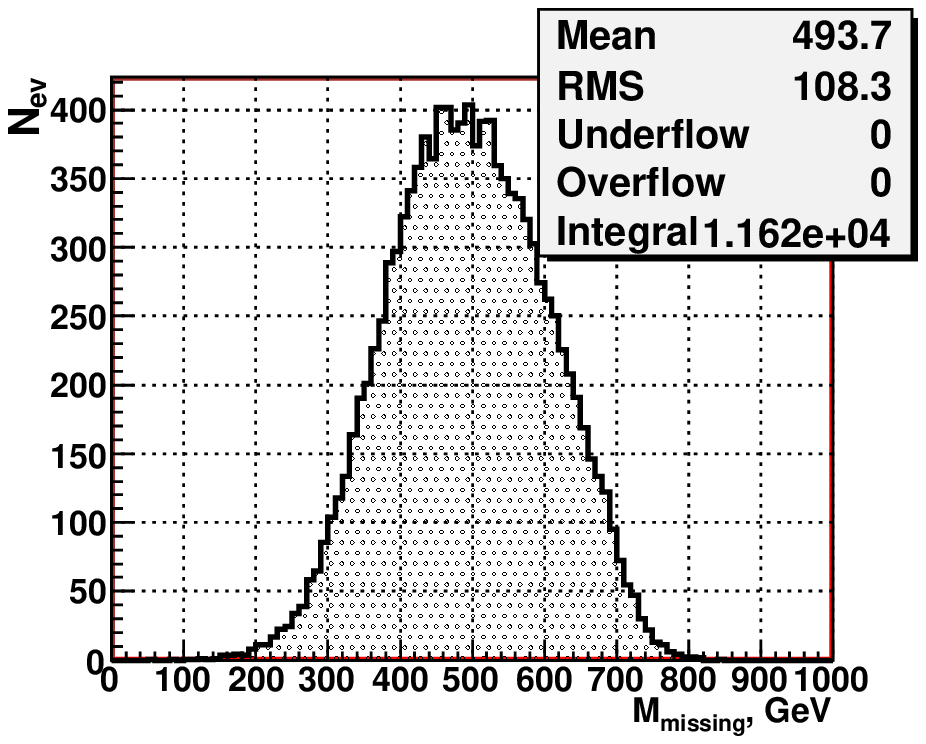}}      
       \mbox{{\bf d)}\includegraphics[width=7.2cm,  height=4.4cm]{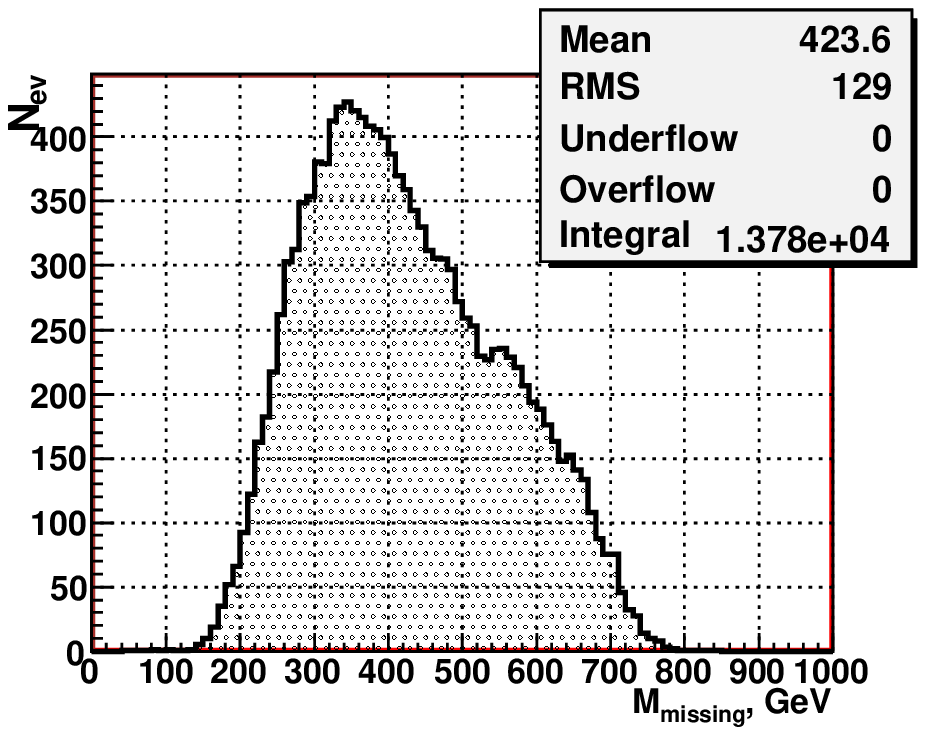}} \\       
    \end{tabular}
     \caption{\small \it Distribution of number  of events,
	    versus the missing mass variable $M_{missing}$.
    	       {\bf a)} and {\bf b)} are for stop pair production;
	       {\bf c)} and {\bf d)} are for top pair production.
	  {\bf a)} and {\bf c)} $"+-"$  and $"-+"$ polarizations,
         {\bf b)} and {\bf d)} $"++"$ and $"--"$  polarizations.}
       \end{center}   
     \vskip -0.5cm                  
 \end{figure}
 
 An even more efficient separation of the signal and 
the background can be obtained by using  the 
 invariant mass   $M_{inv}(All jets)$ 
 of the  system that contains all  jets.
\begin{equation}    
   M_{inv}(All jets)=\sqrt{(\Sigma_{i=1}^{N^{jet}}P^{i}_{jet})^{2}}.
\end{equation}    
   The corresponding distributions for the signal stop events 
   (upper plots) and for the background 
   top events (lower plots) are shown in Fig.28. It is seen
  that the application of  the cut  $M_{inv}(All jets) \le 180$ GeV 
   leads to a  practically complete separation of signal stop
    and top background events.
  
 \begin{figure}[!ht]
     \begin{center}
    \begin{tabular}{cc}
\mbox{{\bf a)}\includegraphics[width=7.2cm,  height=4.4cm]{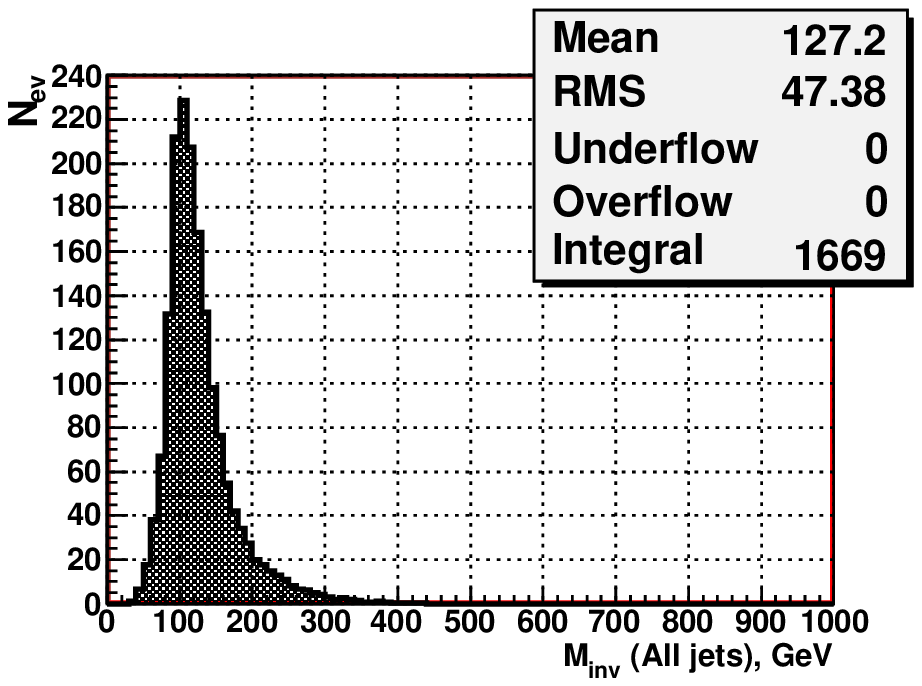}}      
\mbox{{\bf b)}\includegraphics[width=7.2cm,  height=4.4cm]{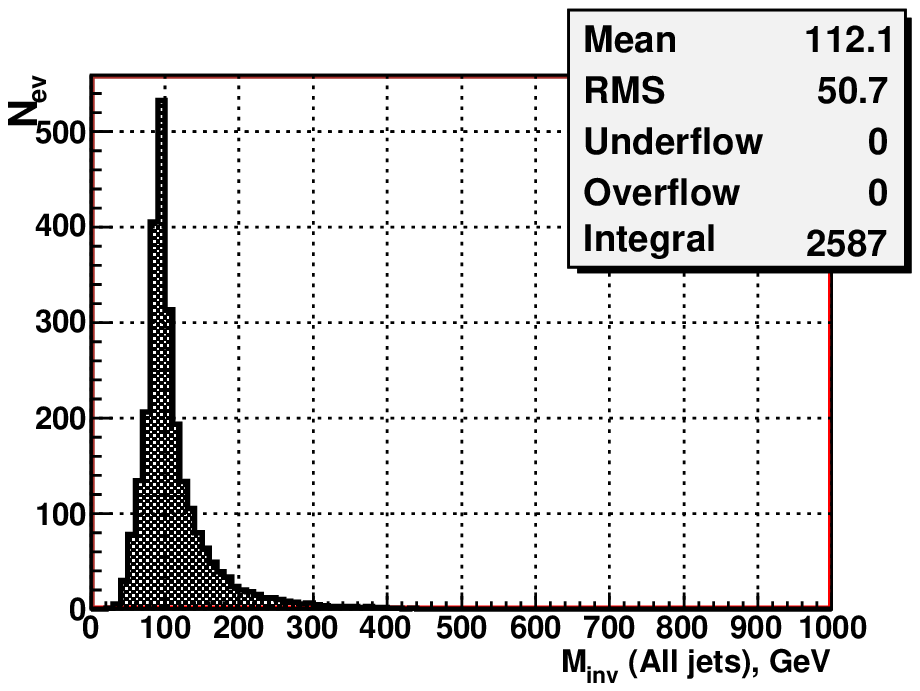}} \\
\mbox{{\bf c)}\includegraphics[width=7.2cm,  height=4.4cm]{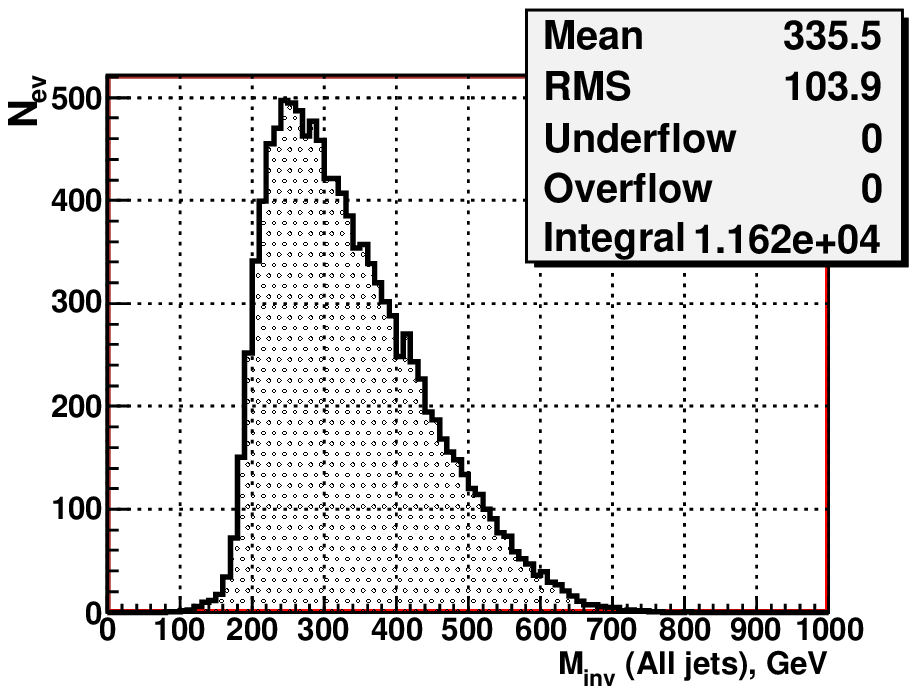}}      
\mbox{{\bf d)}\includegraphics[width=7.2cm,  height=4.4cm]{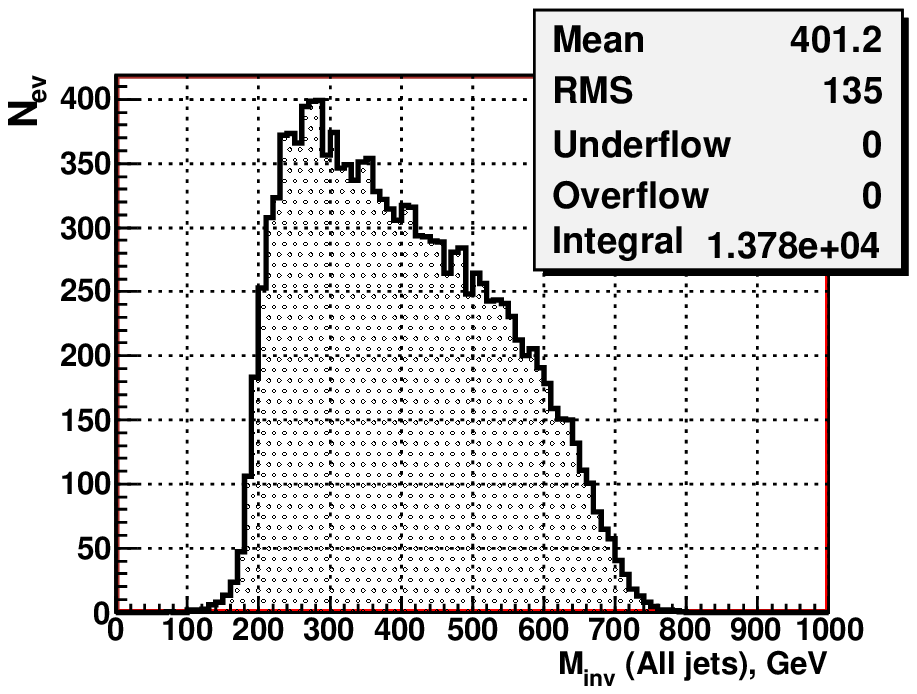}} \\         
     \end{tabular}
     \caption{\small \it Distribution of number of events versus 
      the reconstructed invariant mass of all jets
        $M_{inv}(All jets)$. 
      	       {\bf a)} and {\bf b)} are for stop pair production;
	       {\bf c)} and {\bf d)} are for top pair production.
	  {\bf a)} and {\bf c)} $"+-"$  and $"-+"$ polarizations,
          {\bf b)} and {\bf d)} $"++"$ and $"--"$  polarizations.}	  
	  \vskip -0.5 cm       
     \end{center}        
     \end{figure}

\section{ Cuts and signal-to-background ratio.}

~~~~ To diminish the influence of the jet energy
redistribution effect, discussed in subsections 4.2 and 4.3, we shall use 
  the cuts considered above for  the $E_{vis-tot}$  and 
   $M_{inv}(All jets)$. These variables, by
   definition, include the total 4-momentum of
   all jets, defined as the  vectorial 
   sum  of the  4-momenta of all jets. Therefore
   they do not suffer on energy redistribution
   between jets. Based on our results above,
   we will use  the   following three cuts   to separate the 
   signal and background events: \\

$\bullet$ there must be at least two $b$-jets in an event:

\begin{equation} 
             N_{b-jets} \geq 2 ; 
\end{equation}

$\bullet$ the invariant mass of  all jets  must
             be less  than  $180$ GeV:
\begin{equation}   
            M_{inv}(All jets) \leq 180 ~GeV.
\end{equation}
     
$\bullet$ the detected energy $E_{vis-tot}$ must
               be less than $250$ GeV:
\begin{equation}   
              E_{vis-tot} \leq  250 ~GeV;
\end{equation}

 All the figures presented in this paper are obtained
 after applying the first cut in order to get the right
 picture of jets when the $b$-jets are clearly determined. 
    
These three cuts for the case with  $J=0$ enhanced state considered
here  improve the   signal--to--background  ratio  in the case of $"+-"$ and    $"-+"$ polarizations  from $S/B=0.15$ to $S/B \approx 60$, losing about $23.7\%$  (from 1903 to 1453) of the signal stop events and reduction of  background top events  from 1.227$\cdotp 10^{4}$ to 24. In the case of   $"++"$  and $"--"$ polarizations an improvement of the   signal--to--background  ratio is from $S/B=0.222$ to $S/B \approx 123$, with a loss about $27.6\%$ (from 3233 to 2338) of the signal stop events and a reduction of the background top events from 1.441$\cdotp 10^{4}$ to 19.

     Finally, we present  the efficiency values
    for the three cuts (13)-(15).  We define them  as
    the summary efficiencies. It means that
     if $\varepsilon_{1}$  is the     efficiency 
     of the first cut (13),   $\varepsilon_{12}$ 
    is  the efficiency of applying the first cut (13) 
    and then the second cut (14). Analogously,
    $\varepsilon_{123}$ is the efficiency of  the successive 
    application of the cuts (13), (14) and (15).\\
    
$\bullet$   For SIGNAL STOP events :  \\

$"+-" \& "-+"$ polarizations  - ~~~~$\varepsilon_{1}=0.88$ ; ~~~~ $\varepsilon_{12}=0.78$ ; ~~~~~$\varepsilon_{123} = 0.78$ ;

$"++" \& "--"$ polarizations -  ~~~~$\varepsilon_{1}=0.80$ ; ~~~~ $\varepsilon_{12}=0.73$ ; ~~~~~$\varepsilon_{123} = 0.72$ ;\\ 

 $\bullet$ For BACKGROUND TOP events :  \\

$"+-" \& "-+"$ polarizations - ~~~~ $\varepsilon_{1}=0.94$ ; ~~~~ $\varepsilon_{12}=0.011$ ; ~~~~~$\varepsilon_{123} = 0.002$ ;

$"++" \& "--"$ polarizations - ~~~~ $\varepsilon_{1}=0.94$ ; ~~~~ $\varepsilon_{12}=0.007$ ; ~~~~~$\varepsilon_{123} = 0.001$ .

%
    \section{Determination of the scalar top quark mass.}  
%
 
 ~~~ Another variable of interest is  the 
  invariant   mass   $M_{inv}$($b-jet$, $Jets_{W}$):
     \footnote{ We follow here the notations 
                of subsections  4.2 and 4.3} 
\begin{eqnarray}
   M_{inv}(b-jet,Jets_{W}) \equiv  
    \sqrt{(P_{b-jet} +P_{Jets_{W}})^{2}},
\end{eqnarray}
   which is constructed as the modulus of the
   vectorial sum of the 4-momentum   $P_{b-jet}$  
    of the $b$-jet, plus the total
    4-momentum of $Jets_{W}$ system,  i.e., non-$b$-jets 
    stemming from the W decay
    ($P_{Jets_{W}}= P_{jet1_{W}} + P_{jet2_{W}}$, 
    as there are only two  jets allowed  to be produced in W decay). 
    More precisely, if the signal event 
    contains a $\mu^{-}$ as the signal  muon  (see Fig.1), we 
    have to take the  $b$-jet ($\bar{b}$-jet in the case of $\mu^{+}$ 
    as the signal muon). This     is only possible if one
    can discriminate between the $b$- and 
    $\bar{b}$-jets experimentally.
    Methods of experimental determination    of the  charge of the
    $b$-jet ($\bar{b}$-jet) were developed in  
     \cite{Damerell}. In this paper we 
    do not use any b-tagging  procedure. 
    The PYTHIA information  about
    quark flavor  is taken for  choosing the $b$- and $\bar{b}$-jets.
    In reality,  according to  \cite{Damerell}, a  $50\%$
    efficiency of the separation of $b$-jets and 
    $80\%$ of the corresponding  purity can be  expected.

    The distributions of the  invariant masses  of the
    "$b$-jet+$Jets_{W}$"  system in the case of stop pair
     production are shown in plots {\bf a)} and {\bf b)}
     of Fig.29 for the  two polarization combinations, 
     as well as in  the  case of top pair production in
     plots {\bf c)} and {\bf d)}.  Their analogs
     $M_{inv}(b, 2~quarks_W)$,  obtained at quark 
     level, are presented in Fig.30.  The distributions
     shown in both Figures were obtained without 
     use of cuts (15) and (16).
  
    \begin{figure}[!ht]
     \begin{center}
    \begin{tabular}{cc}
     \mbox{{\bf a)}\includegraphics[width=7.2cm,  height=4.4cm]{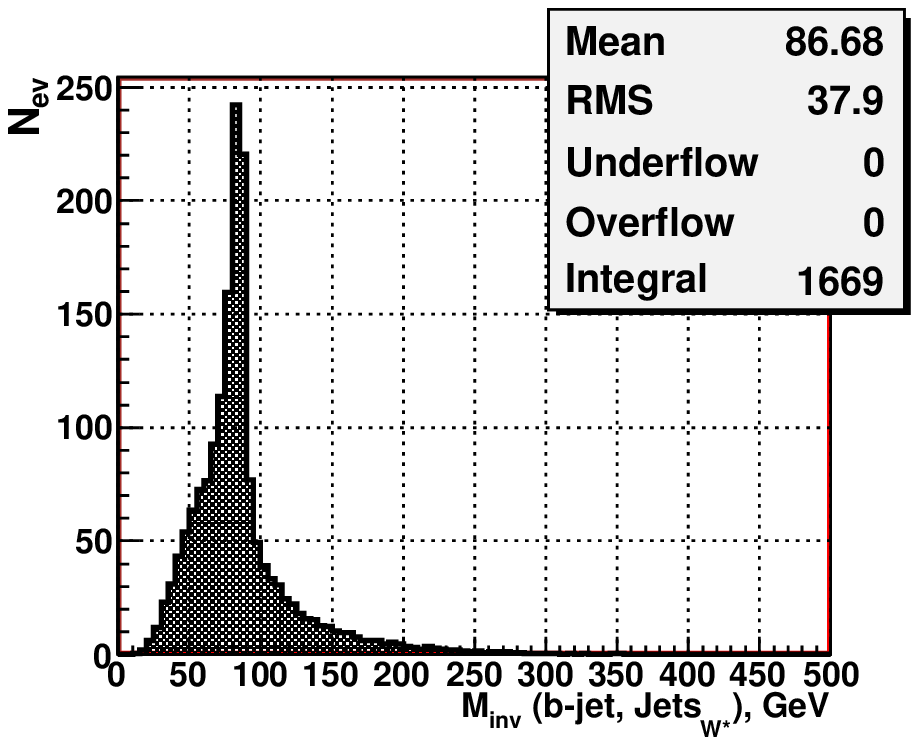}}      
     \mbox{{\bf b)}\includegraphics[width=7.2cm,  height=4.4cm]{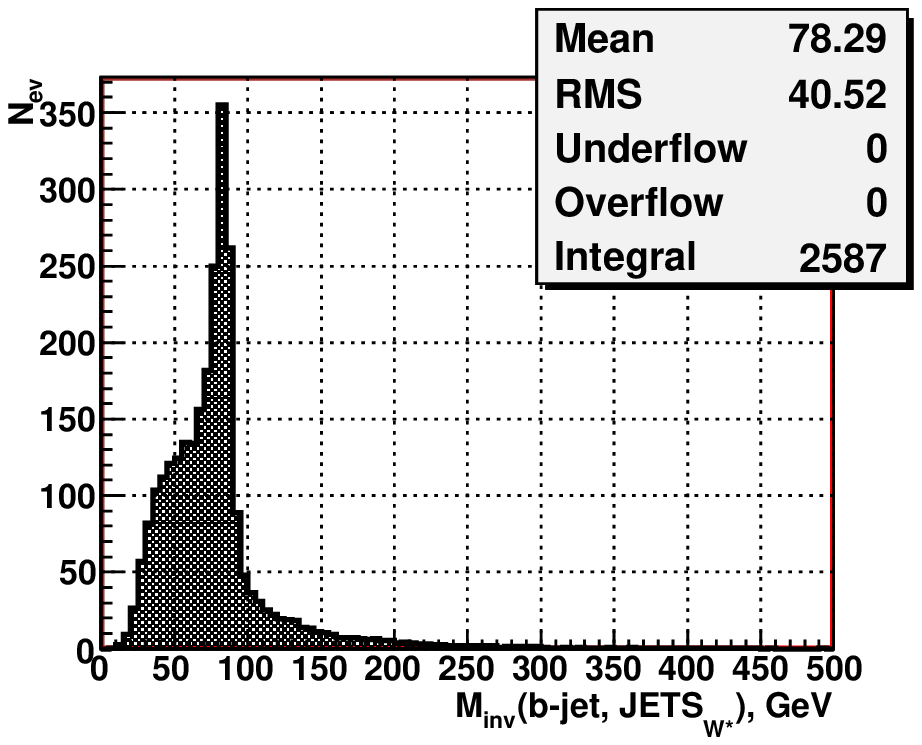}} \\
     \mbox{{\bf c)}\includegraphics[width=7.2cm,  height=4.4cm]{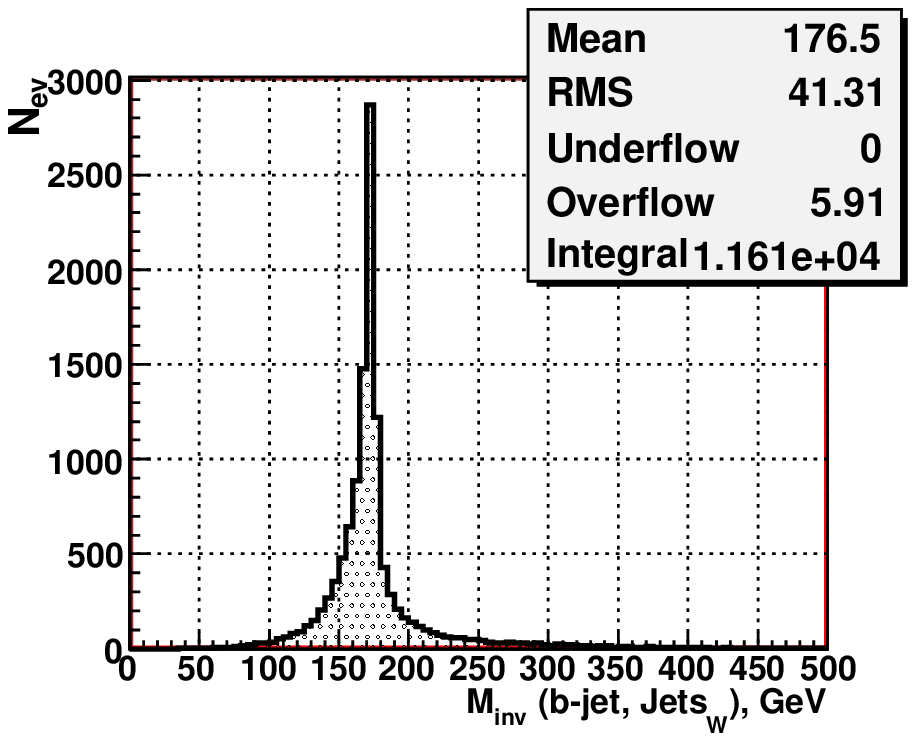}}      
     \mbox{{\bf d)}\includegraphics[width=7.2cm,  height=4.4cm]{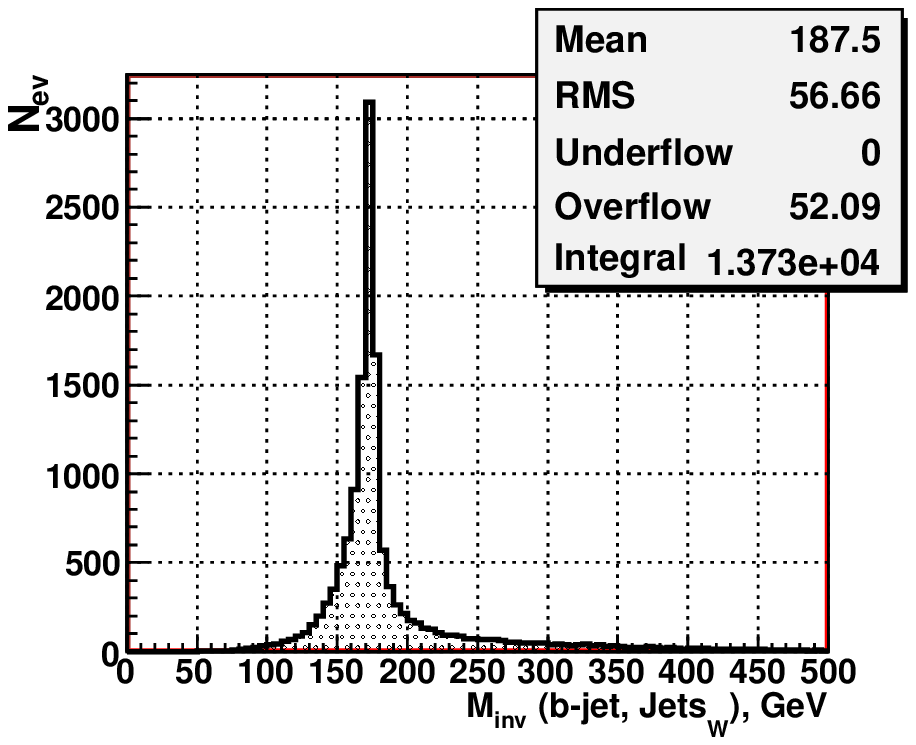}}      
    \end{tabular}
     \caption{\small \it The spectra of the
         invariant masses $M_{inv}$($b-jet$, $Jets_{W}$)
	 obtained without use of cuts (15) and (16).
      	       {\bf a)} and {\bf b)} are for stop pair production;
	       {\bf c)} and {\bf d)} are for top pair production.
	  {\bf a)} and {\bf c)} $"+-"$  and $"-+"$ polarizations,
          {\bf b)} and {\bf d)} $"++"$ and $"--"$  polarizations.}	
    \end{center}
 \vskip -0.5 cm           
    \end{figure}

  \begin{figure}[!ht]
     \begin{center}
\vskip -0.5 cm     
    \begin{tabular}{cc}
     \mbox{{\bf a)}\includegraphics[width=7.2cm,  height=4.4cm]{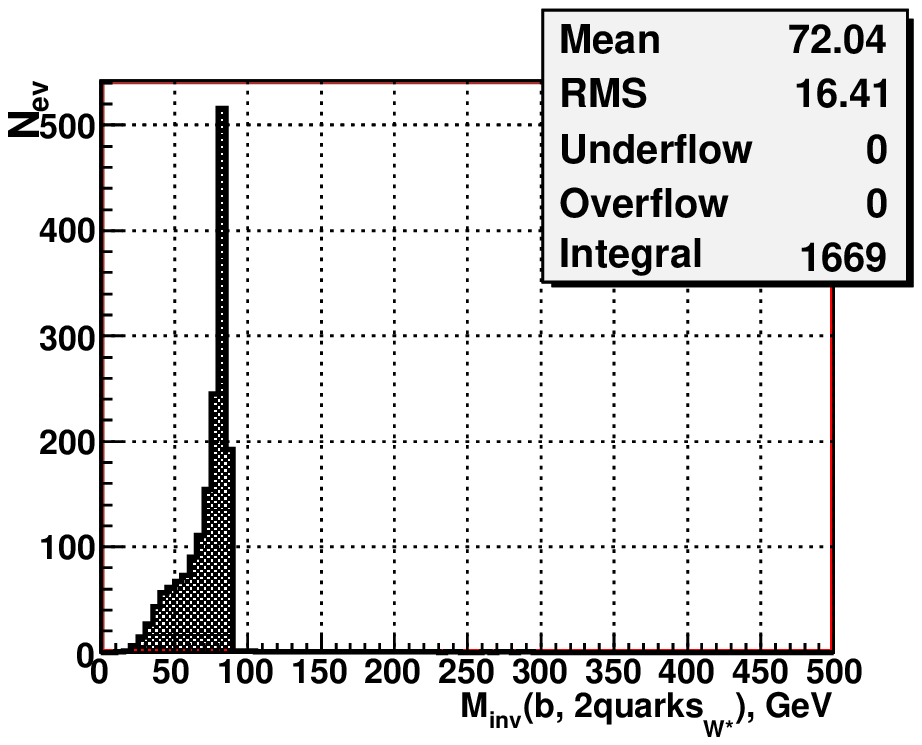}}      
     \mbox{{\bf b)}\includegraphics[width=7.2cm,  height=4.4cm]{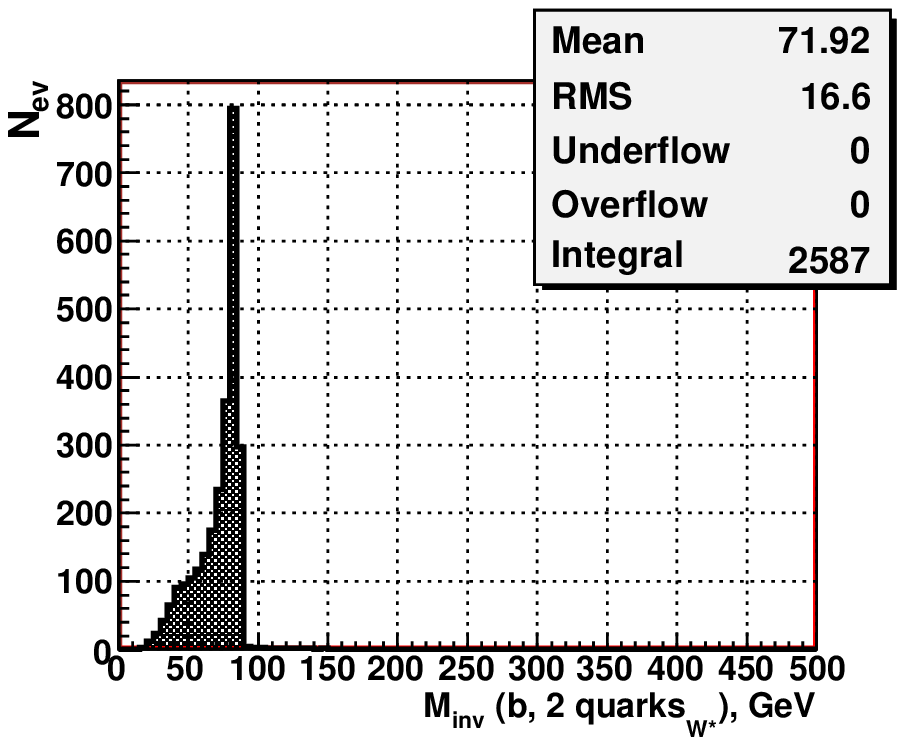}} \\
     \mbox{{\bf c)}\includegraphics[width=7.2cm,  height=4.4cm]{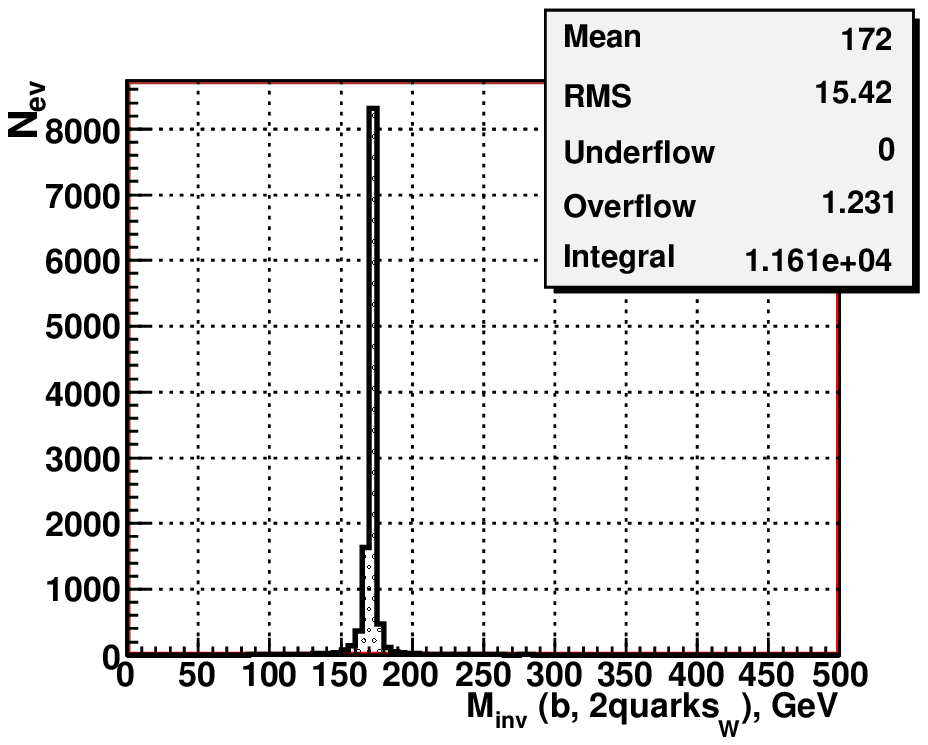}}      
     \mbox{{\bf d)}\includegraphics[width=7.2cm,  height=4.4cm]{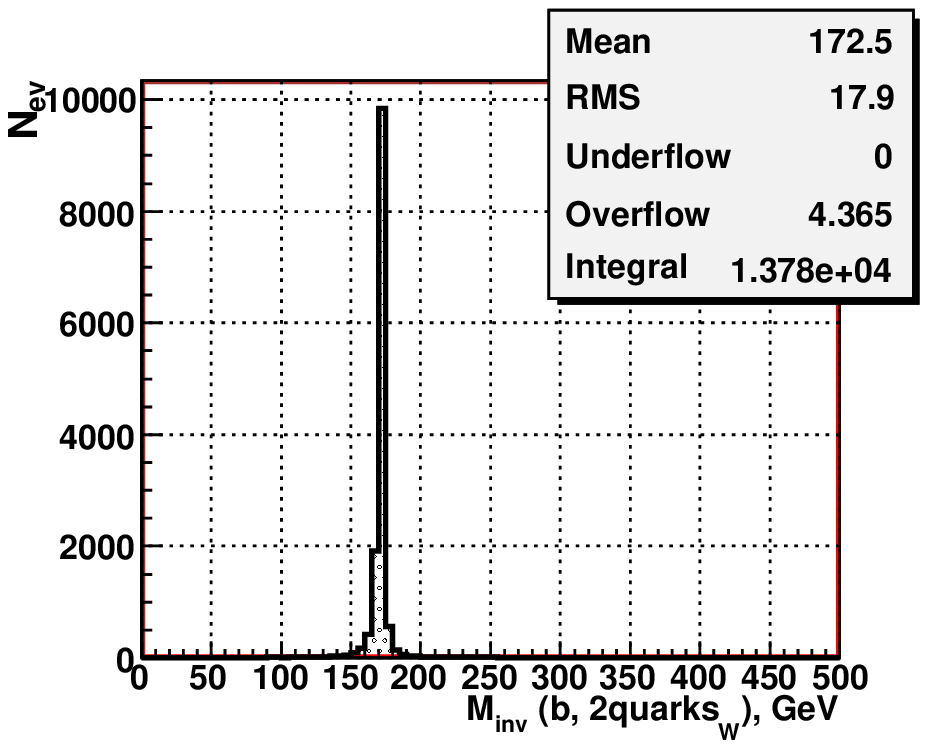}}      
    \end{tabular}
     \caption{\small \it The spectra of the invariant masses
                     $M_{inv}$($b$, $2~quarks_{W}$)  obtained 
		     without use of cuts (15) and (16).
      	       {\bf a)} and {\bf b)} are for stop pair production;
	       {\bf c)} and {\bf d)} are for top pair production.
	  {\bf a)} and {\bf c)} $"+-"$  and $"-+"$ polarizations,
          {\bf b)} and {\bf d)} $"++"$ and $"--"$  polarizations.}	
    \end{center} 
  \vskip -0.5 cm          
    \end{figure}  

In the top case the invariant mass 
$M_{inv}(b, 2~quarks_W)$  of the system 
 composed of a $b$-quark and two quarks from W decay 
 should  reproduce the mass of their parent top quark 
 (see Fig.1). The distributions of   events 
 $dN^{event}/dM_{inv}/5$ GeV   expected  in each
  bin of 5 GeV  versus the  invariant mass 
  $M_{inv}(b, 2~quarks_{W})$ of  the parent three 
  quarks as well as the  invariant mass of jets produced 
  by these quarks, i.e.  $M_{inv}(b-jet, Jets_{W})$, 
  are shown for jet  and quark  levels in plots
  {\bf c)} and {\bf d)} of Fig.29 and 30,  respectively, 
  for both polarizations. These distributions have
  an important common feature. Namely, they  show  
  that the peak positions at  jet  level  and 
  at  quark level, practically coincide to
  a good accuracy with each other  as  well as 
  with the  input value of   the top quark mass  
  $M_{top}=170.9 (\pm 1.8)$ GeV.   It is also 
  seen from Fig.29 that the quark hadronization 
  into jets leads to a broadening of very small tails 
  which are seen in the invariant  mass distribution 
  at quark level (Fig.30). The right tails, which 
  appeared at jet level (see  Fig.29),  is a bit lower 
  than the left tails and are longer than the left ones.
  One may say that the peak  shape  at jet level
  still looks more or less  symmetric. The main
   message from these plots is that the appearance 
  of  tails due to  quark fragmentation into jets does 
  not  change the position of the  peak,
  which allows us to reconstruct  the input top 
  mass both at quark and jet levels. 
 
 An analogous stability of  the  peak position at the jet and 
 quark levels for the stop case can be seen in the  plots {\bf a)}  
 and {\bf b)} of Figs.29  and 30. Note that, 
 according to the  stop decay chain (2), the  
 right edge of the peak of  the invariant mass distribution 
 of the   "$b + 2 quarks_{W}$" system 
 corresponds to the mass difference 
 $M_{\widetilde{t}_1}- M_{\tilde \chi^{0}_{1}}$.

 The distributions of the invariant mass 
 of the "$b + 2 quarks_{W}$"  system  (plots {\bf a)} and {\bf b)}) 
 and  of  the invariant mass of the "$b$-jet+$JETS_W$"  
  system  in the case of stop pair
  production are shown in  Fig.31. Thereby   
   only those stop  events are taken that pass the cuts (14)--(16).  

 \begin{figure}[!ht]
     \begin{center}
   \vskip -0.5 cm     
    \begin{tabular}{cc}
     \mbox{{\bf a)}\includegraphics[width=7.2cm,  height=4.4cm]{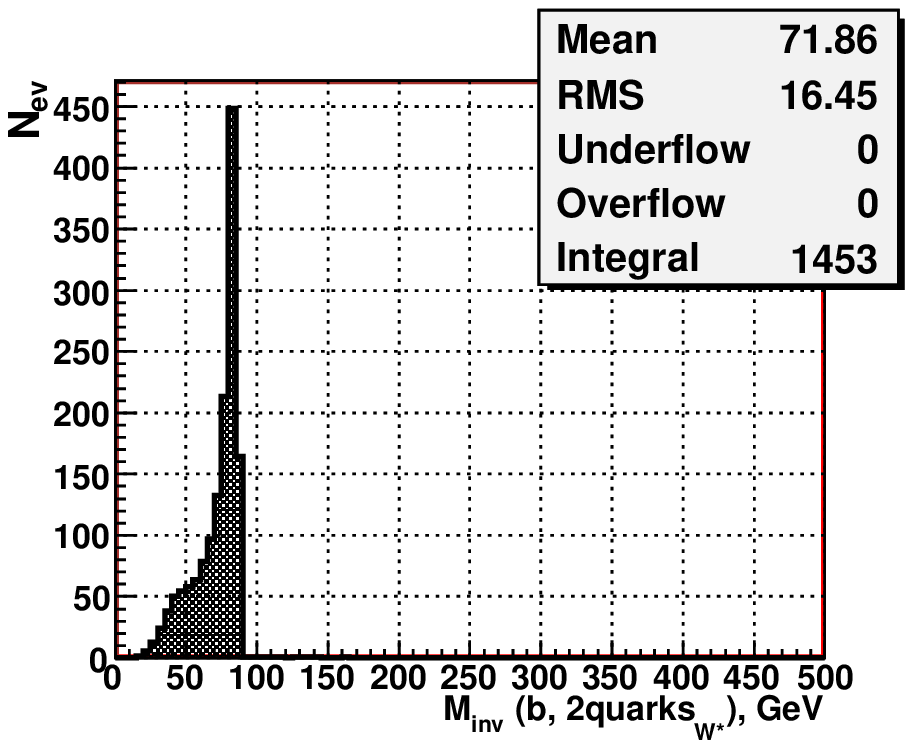}}      
     \mbox{{\bf b)}\includegraphics[width=7.2cm,  height=4.4cm]{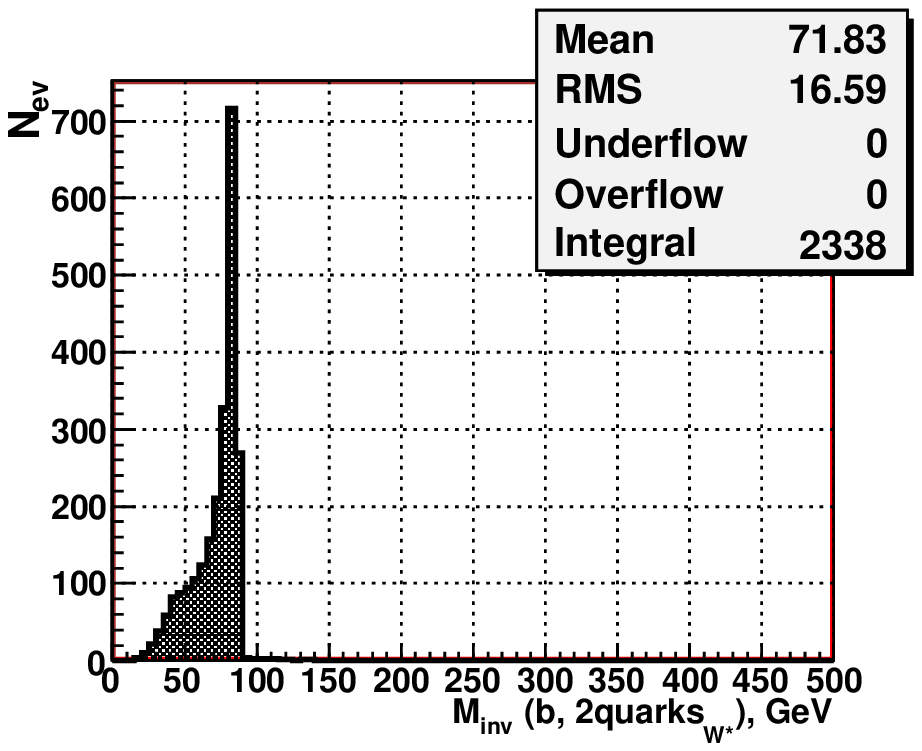}} \\
     \mbox{{\bf c)}\includegraphics[width=7.2cm,  height=4.4cm]{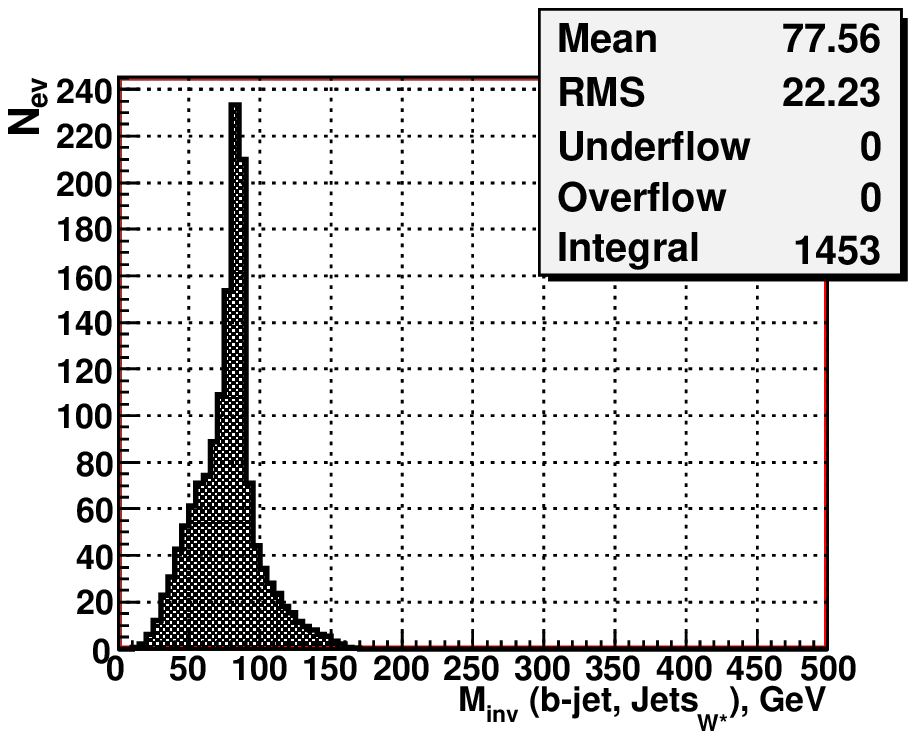}}      
     \mbox{{\bf d)}\includegraphics[width=7.2cm,  height=4.4cm]{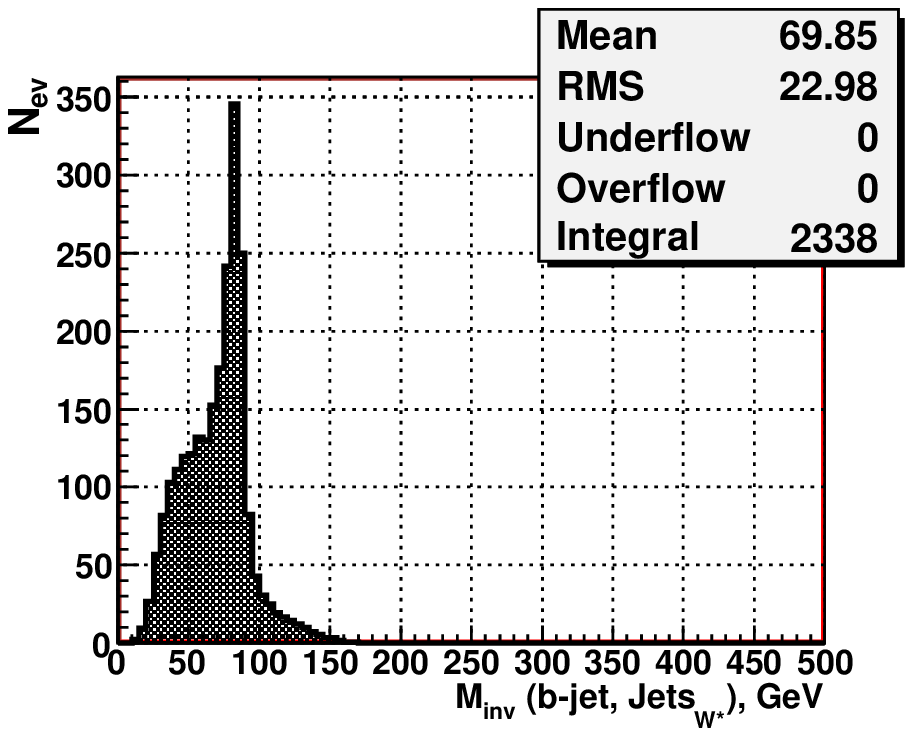}}      
    \end{tabular}
     \caption{\small \it The spectra of the stop invariant 
               masses after the  cuts (14)--(16):
      	       {\bf a)} and {\bf b)} $M_{inv}$($b$, $2~quarks_{W}$);
	       {\bf c)} and {\bf d)} $M_{inv}$($b-jet$, $Jets_{W}$).
	  {\bf a)} and {\bf c)} $"+-"$  and $"-+"$ polarizations,
          {\bf b)} and {\bf d)} $"++"$ and $"--"$  polarizations.}	
    \end{center} 
   \vskip -0.5 cm          
 \end{figure}

 Let us recall that according to 
  Section 6   the application of the   cuts 
  (14)--(16) leaves only  24--19  background top events
(respectively,  for $"+-"$, $"-+"$ and  $"++"$, $"--"$ combinations of photon polarizations) and saves about  76.3 $\%$ of signal stop events. 
 It means that the  distributions shown in plots  {\bf c)} and {\bf d)} 
 of Fig.29 would change drastically and resemble a random distribution of 
  the 24--19 top events  in a rather wide interval. 
 The corresponding plots  {\bf c)} and {\bf d)}
 of top production events which pass the
cuts (14)--(16) are are made with 
a much larger simulated statistics and are 
shown in Fig.32. One sees that the surviving background top events will be 
 mostly distributed  in the region 
 $30 \leq M_{inv}$($b$-jet, $ Jets_{W}$)$ \leq 180$ GeV. 
 This region is by  more than twenty times  wider than the 5 GeV width  
of the  peak  intervals in the $M_{inv}$($b$-jet, $ Jets_{W}$) 
distributions which are shown in the stop 
 plots {\bf c)}  and {\bf d)} of Fig.31
(at jet level) and  which contain  
about  240 (for $``+-``$ and $``-+``$ polarization)  
and 350 (for $``++``$ and $``--``$ polarization)
signal stop events  left  after the cuts. 
 Based on the shape of the
distributions shown in the plots  {\bf c)} and
 {\bf d)} of Fig.32  we can expect that  in future 
measurements the contribution of  these 24-19 
remaining top background events will not 
influence the  position of the peak of the
$M_{inv}$($b$-jet, $ Jets_{W}$) distributions 
(shown in plots {\bf c)}  and {\bf d)} of 
Fig.31) which  allow one to reconstruct 
the input  value of the stop  mass  at
 jet level by adding the mass of the neutralino.
    
   \begin{figure}[!ht]
     \begin{center}
\vskip -0.5 cm
    \begin{tabular}{cc}
     \mbox{{\bf c)} \includegraphics[width=7.2cm,  height=4.4cm]{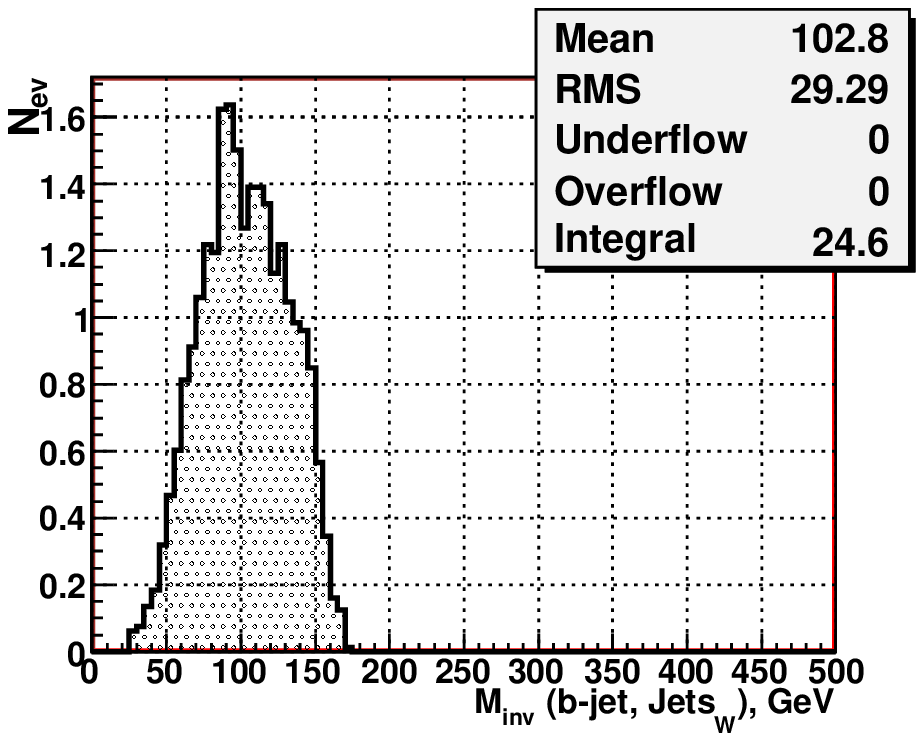}}  
      \mbox{{\bf d)} \includegraphics[width=7.2cm,  height=4.4cm]{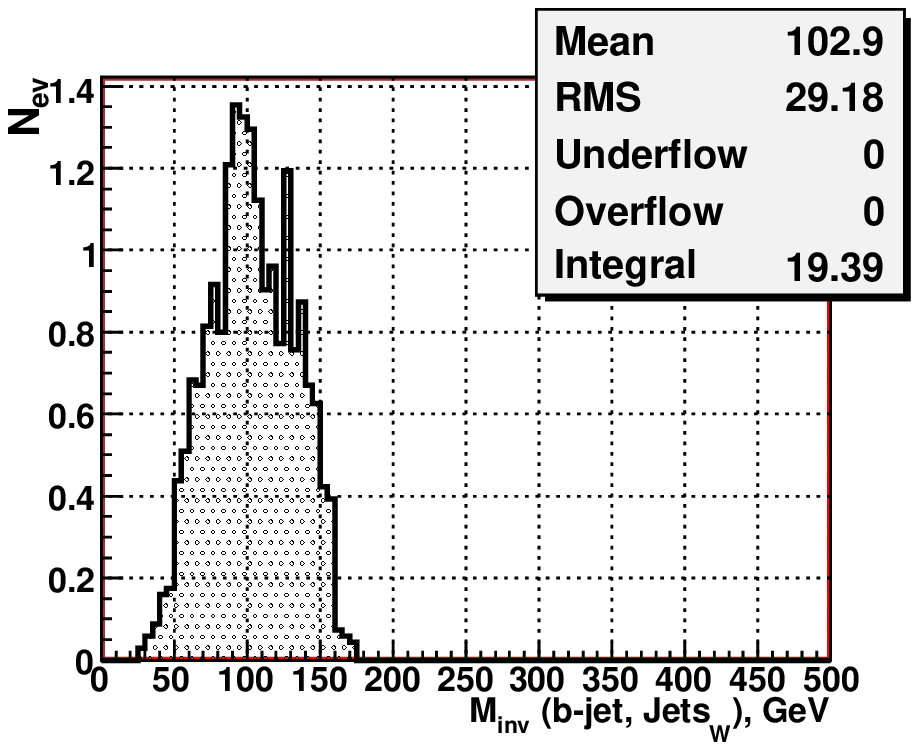}}  \\ 
     \end{tabular}
     \caption{\small \it The spectra of the invariant mass
  $M_{inv}$($b$, $2~quarks_{W}$) for top pair production 
  events  after the  cuts (14)--(16).
              {\bf a)} $"+-"$  and $"-+"$ polarizations,
              {\bf b)}$"++"$ and $"--"$  polarizations.}    
\vskip -0.5 cm	          
     \end{center}        
     \end{figure}   
     
   It is seen that the peak positions of   the  stop distribution
   at jet level  $M_{inv}(b$-jet, $ Jets_{W^{*}})$,
obtained after the cuts (14)--(16) (plots {\bf c)} and {\bf d)} of Fig.31),
   coincide  with the peak  positions at quark 
   level (plots {\bf a)} and {\bf b)} of Fig.31) as well as
   with the peak positions in  plots {\bf a)} 
   and {\bf b)} of  Figs.29 and 30  obtained   without  any cuts.
  Let us note  that the observed  stability of the 
peak position in both  of Figs.29 {\bf a)}, {\bf b)} and 31 {\bf c)}, {\bf d)} is  due to the rather moderate  loss of the number 
of events   in the peak region  (this loss is about 200 events) after
cuts. The cuts lead  (as it can be seen by comparing the mentioned  plots) to a reduction  of the right hand side tails of 
   $M_{inv}(b$-jet, $ Jets_{W^{*}})$ distributions.
\footnote{   The interval 150--350 GeV in the  plot {\bf b)} of
   Fig.31  can be used to calculate the width
   between the grid dots in this plot. It is found to be 
   about 7.4 GeV. This number allows us to estimate the 
   position of the right edge of the  peak  of the
   $M_{inv}(b$-jet, $ Jets_{W^{*}})$  distribution, 
   which seems to be shifted to the  left side from
   100 GeV  by a distance  which of about 
   two dot intervals, i.e., by less than 
   14.8 GeV. Thus, we can estimate that the right
    edge of the peak of the $M_{inv}(b$-jet, $ Jets_{W^{*}})$ 
    distribution lies a little higher than 85.2 GeV.}

  Some additional remarks about the tails in  the stop 
   distributions are in 
  order now. The origin  of the right and left tails
  of the  distributions, shown in the plots {\bf a)} and 
  {\bf b)} of Fig.32, can be clarified  by the results  
  of the stop mass reconstruction by calculating its
  invariant mass at quark level
  $M_{inv}$($b$, $2~quarks_{W^*}, \tilde \chi^{0}_{1})$
  as the modulus of  the sum of the  4-momenta
  of all three quarks and the neutralino (see Fig.1)
  in stop decay. These results are given 
  in plots {\bf a)} and {\bf b)} of Fig.33 which shows
  a very precise reconstruction of the input stop 
  mass at quark level withing the 5 GeV width of
   the  bin containing  the peak. Comparing plots 
  {\bf a)} and {\bf b)} of Fig.31 with  plots  {\bf a)}
  and {\bf b)} of Fig.33  one  can conclude that 
  at quark level the long left tail as well as the 
  very small right  tail in the distribution of
$M_{inv}$($b$, $2~quarks_{W^*})$  disappear when the neutralino  4-momentum is added to the 4-momentum of   the "$b + 2 quarks_{W}$" system.
             
   \begin{figure}[!ht]
     \begin{center}
\vskip -0.5 cm     
    \begin{tabular}{cc}
     \mbox{{\bf a)}\includegraphics[width=7.2cm,  height=4.4cm]{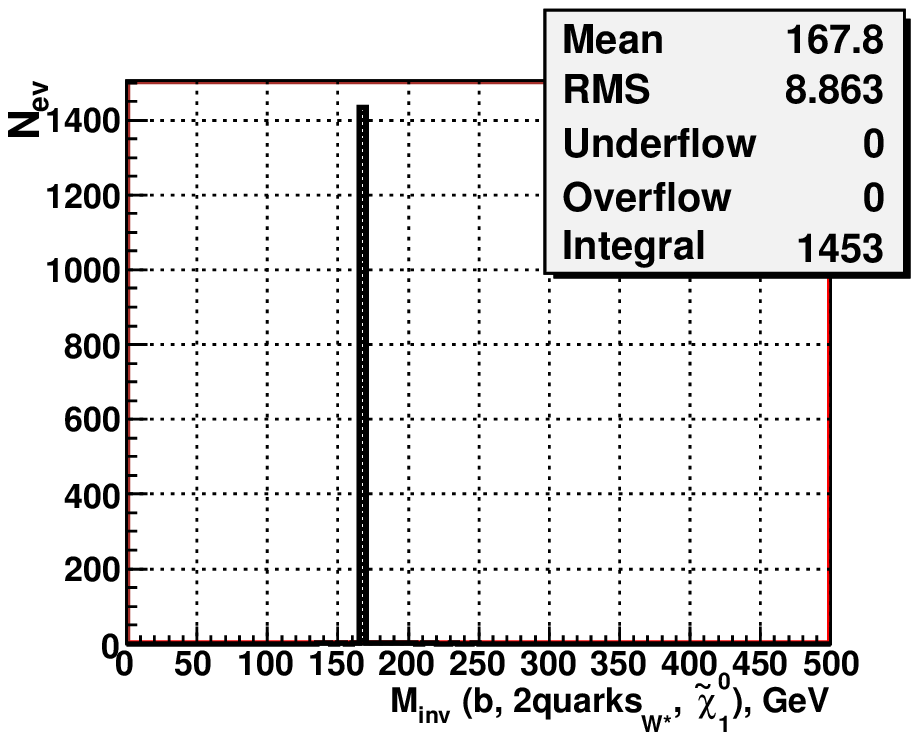}}      
     \mbox{{\bf b)} \includegraphics[width=7.2cm,  height=4.4cm]{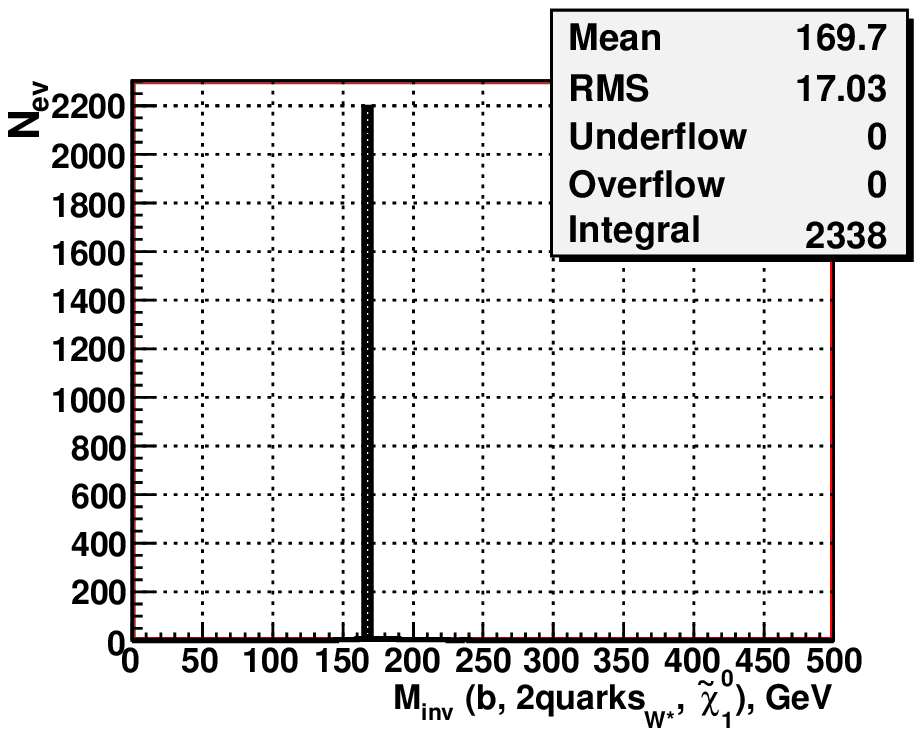}} \\
     \mbox{{\bf c)} \includegraphics[width=7.2cm,  height=4.4cm]{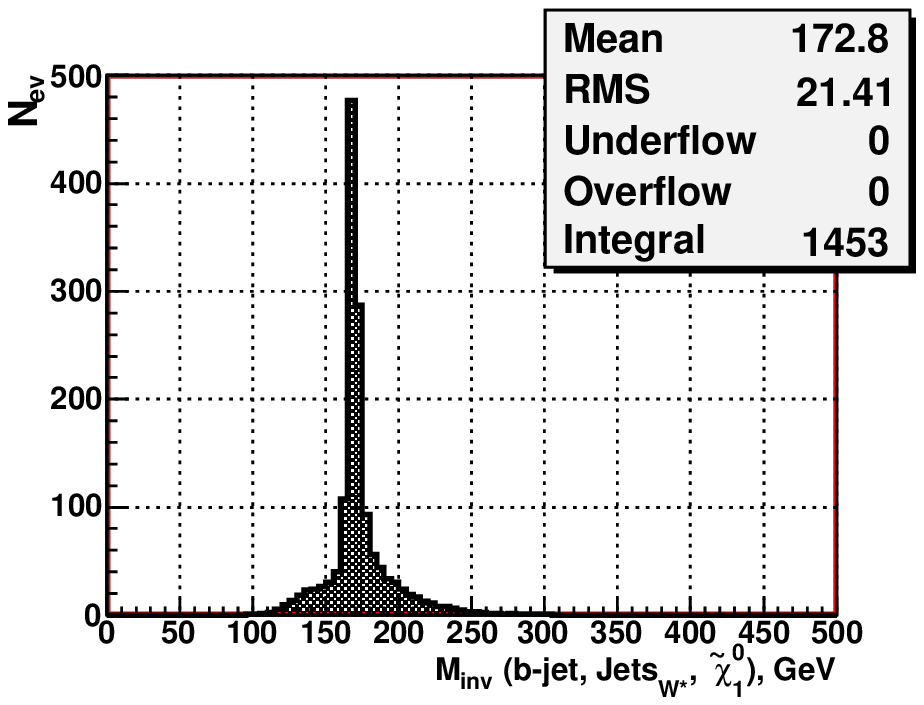}}      
     \mbox{{\bf d)} ~\includegraphics[width=7.2cm,  height=4.4cm]{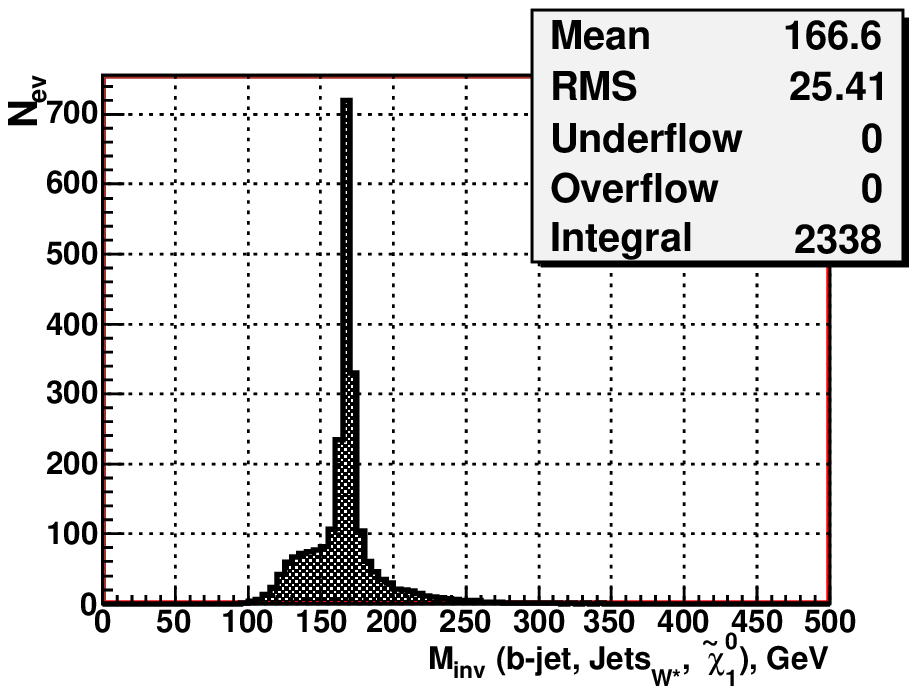}}      
    \end{tabular}
     \caption{\small \it The spectra of the  invariant masses  
 $M_{inv}$($b$-quark, $2~quarks_{W^*}, \tilde \chi^{0}_{1})$, and 
	  $M_{inv}$($b$-jet, $Jets_{W}, \tilde \chi^{0}_{1} $).
      	       {\bf a)} and {\bf b)} are for quarks level;
	       {\bf c)} and {\bf d)} are for jets level.
	  {\bf a)} and {\bf c)} $"+-"$  and $"-+"$ polarizations,
          {\bf b)} and {\bf d)} $"++"$ and $"--"$  polarizations.}	
\vskip -0.5 cm	  
    \end{center}        
    \end{figure}  
   
The influence of the effect of  the  hadronization of the
 $b$-quarks and  of the quarks from W decay  into 
 jets is shown in  plots  {\bf c)} and {\bf d)} of  Fig.33.  
 These plots demonstrate  that the hadronization of
 quarks into jets  practically does not  change the positions
 of  the stop mass peak, which  practically concides with the input value  $M_{\widetilde{t}_1}=167.9$ GeV.
    It is also seen that the hadronization results in  the 
    appearance of  more  or less symmetrical and 
    rather  suppressed short tails  around the peak 
    position.  The mean values of the distributions in 
    plots {\bf c)} and {\bf d)} of Fig.33  are 
    slightly different from the mean values of the 
    quark level distributions shown in plots {\bf a)} 
     and {\bf b)} of Fig.33  but the peak positions remain
     the same.
    It is easy to see from  plots {\bf c)} and {\bf d)} of Fig.31
     that  adding the  mass of the neutralino 
     $M_{\tilde \chi^{0}_{1}} = 80.9$ GeV to the value of 
     the right edge point of the peak 
     $M_{inv}$($b$-jet, $Jets_{W*}) \approx$ 85.2 GeV 
     one can get the left lower limit for the 
     reconstructed stop  mass 
     $M^{reco-low}_{\widetilde{t}_1}  \approx 166.2$~GeV
     which reproduces well  the input value
      $M_{\widetilde{t}_1}=167.9$ GeV. 

Taking into account the bin width of 5 GeV used in the invariant mass distributions we may conclude that the method of the stop mass reconstruction based on the peak positions will be quite useful.

\section{~ Results for top squark mass  $M_{\widetilde t_1}$ = 200 GeV.}
%

 ~~~~ In this section we want to discuss 
  what will change if the mass of the top squark
  is different from the one we have  used before.
  In the present paper we have chosen a 
  rather low scalar top quark mass
  (one of the lowest stop quark's masses 
  that is  allowed for the case of
  $\tilde t_{1} \to b \tilde \chi_{1}^{\pm}$
  decay channel). With increase of the   stop mass the cross section
  for its production is  decreasing. So, for example, for the
 case of $M_{\widetilde{t}_1}=200$ GeV  the number of events  per year  is
 decreasing to 329 for the case of
  $"+-"$  and $"-+"$ polarizations and 1333 for
   the case of $"++"$  and $"--"$ polarizations (after the cuts
 (14)--(16)).  The mass  $M_{\widetilde{t}_1}=200$ GeV
 is still below the highest allowed  stop mass for the 
 $\tilde t_{1} \to b \tilde \chi_{1}^{\pm}$
 decay channel (which is about 255 GeV) corresponding  to
 $M_{\chi^{+}_{1}}=159.2$ GeV  and  $M_{\chi^{0}_{1}}=80.9$ GeV.
 For stop masses below and above the described 
 region, the stop will decay to other
 channels which we do not consider in this paper.

 The distribution of the  invariant  mass $M_{inv}$($b-jet$, $Jets_{W*}$)
   of the  "$b$-jet+$Jets_{W^*}$"  system 
   for events which have passed the  cuts
   (14)--(16) is shown in Fig.34. Plot  
 {\bf a)} is for $"+-"$  and $"-+"$ polarization, 
{\bf b)} is for $"++"$  and $"--"$ polarization. 
 The top background also remains the  same  as it was given in Fig.32. 
                 
   \begin{figure}[!ht]
     \begin{center}
\vskip -0.5cm      
    \begin{tabular}{cc}
     \mbox{a)\includegraphics[width=7.2cm,  height=4.4cm]{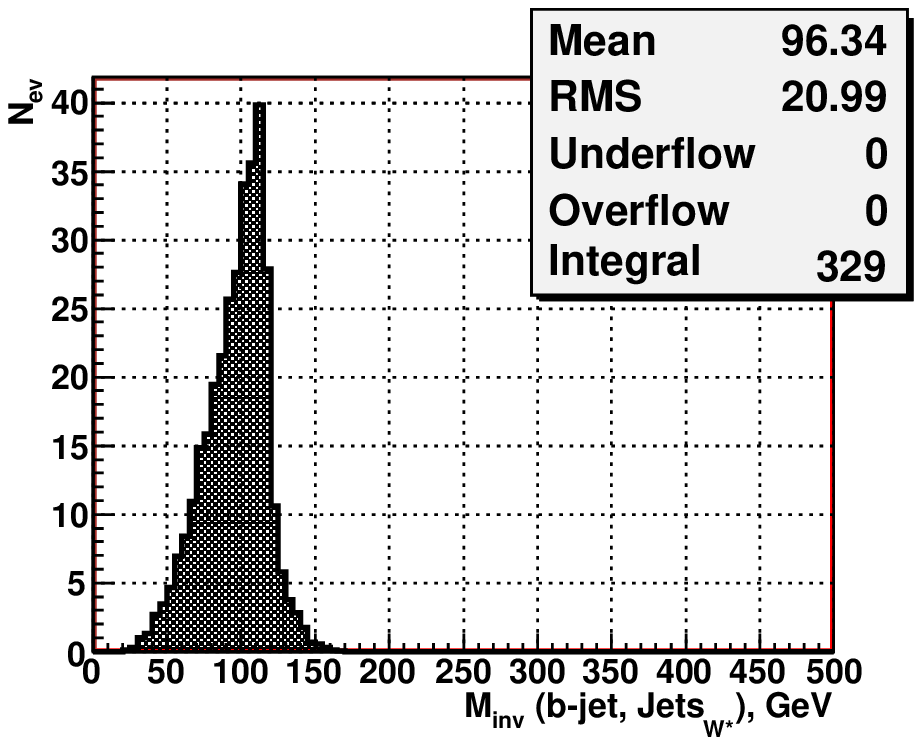}}      
     \mbox{b)\includegraphics[width=7.2cm,  height=4.4cm]{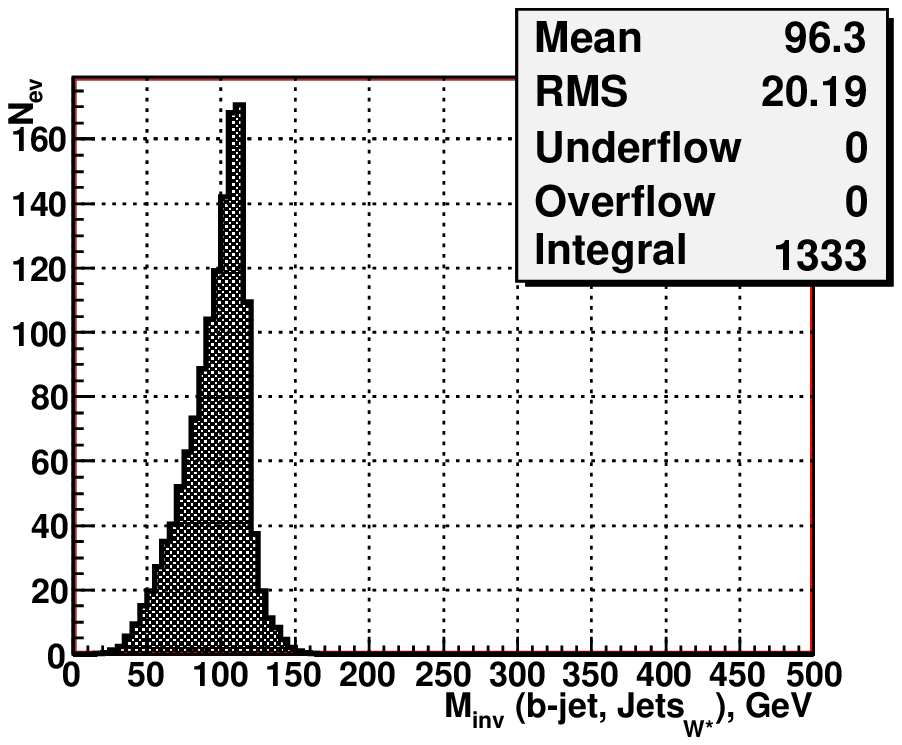}}      
    \end{tabular}
     \caption{\small \it The spectra of the   invariant masses
	  $M_{inv}$($b_{jet}$,$Jets_{W}$) of the
	 "$b$-jet+(all- non-$b$-jets)" system after cuts for $M_{\widetilde{t}_1}=200$ GeV.
	               {\bf a)}  stop pair  production;
	               {\bf b)} top  pair production.}
     \end{center} 
\vskip -0.5cm             
    \end{figure}

   The distributions  in Fig.34 have  peaks 
   at $M_{inv}$($b-{jet}$,$Jets_{W^{*}})\approx$
   110 GeV.  One can also determine the mass
   of the stop quark following the procedure
   described in Section  7, but with  less accuracy 
   than in the case of the lower stop mass used in 
   previous Sections.

\section{Conclusion.}

~~~~ We have studied  stop pair production 
   in photon-photon collisions within the framework 
   of the  MSSM for the total energy of  the 
   $e^{-}e^{-}$  system  $E_{e-e-}^{tot}=\sqrt {s_{ee}}=1000$ GeV.
   We assume that  the 
   stop  quark   decays dominantly into a chargino and a 
   $b$-quark,  $\tilde t_{1} \to b \tilde \chi_{1}^{\pm}$, 
   and the  chargino decays  into a neutralino and 
   a  W boson,
   $\tilde \chi_{1}^{\pm} \to  \tilde \chi_{1}^{0} W^{\pm}$,
   where the W boson is virtual. 
   One of the two W's decays hadronically,
   $W^{+} \to q \bar q $, the other one  leptonically,
    $W^{-} \to \mu^{-} \nu$.
     
      The study is based  on a Monte Carlo simulation
   with two programs. 
   First we have  used the program CIRCE2  which
   gives the luminosity and the energy spectrum of the 
   colliding backscattered photon beams. The results 
   of CIRCE2 are taken as  input for PYTHIA6.4. This 
   event generator is used  to simulate  stop 
   ($M_{\widetilde t_1} = 167.9$ GeV)
   pair production  and decay as well as top pair
   production being the main background.

  Three cuts (14)-(16)  have been proposed. 
  The second (15) and the third (16) cut are the most 
  important  for the separation of the signal stop events 
  from the background  top events. 
   They  restrict the value of the invariant 
   mass of all four jets (produced in
   $\gamma\gamma \to \tilde t_{1} \bar{\tilde t_{1}} \to 
   b\bar{b}q_{i}\bar{q_{j}}\mu\nu_{\mu}\tilde \chi^{0}_{1}\tilde\chi^{0}_{1}$
   process) and the value of the detected energy.
    This set of cuts leads to a 
   signal--to--background ratio as large as
   $S/B=60$ in the case of the $"+-"$  and $"-+"$
   polarizations  and  $S/B=123$  in the case of the 
    $"++"$  and $"--"$ polarizations.
    Thus, we expect about $1-2\%$   admixture of top 
    events  to the stop signal.
      This is different from the more complicated 
   situation  in stop pair production at LHC 
	      (see, for instance, \cite{U.Dydak}).

   We have shown that  determining  the end point
   of the  peak in the distribution of the 
   invariant  mass  $M_{inv}$($b$-jet,$Jets_{W^{*}}$) of the
   "$b$-jet + two jets from W decay" system 
   allow us to reconstruct  the mass of the stop quark
   with a good accuracy based on the statistics of  about
   two years running.  For this the mass of $\tilde \chi_{1}^{0}$ has to be known.
 
   We discussed the  difference in  the main invariant mass distributions 
   for a  mass $M_{\tilde t_{1}}$ = 200 GeV.
       
    In conclusion,  we can say that the $\gamma  \gamma$ 
   channel is very well suited for the study of stop pair 
   production. 

\section{Acknowledgements.}

~~~ This work is supported by the JINR-BMBF 
 project and 
 by the "Fonds zur F$\ddot o$rderung der
 wissenschaftlichen Forschung" (FWF) of
 Austria, project No.P18959-N16.
 The authors acknowledge support from 
 EU under the MRTN-CT-2006-035505 
 and MRTN-CT-2004-503369 network programmes.
 A.B. was supported by the Spanish 
 grants SAB 2006-0072, FPA 2005-01269
 and FPA 2005-25348-E of the
 Ministero de Educacion y Ciencia.


\end{document}